\documentclass{book}
\usepackage{roboto}

\usepackage[english]{babel}
\usepackage[letterpaper,top=2cm,bottom=2cm,left=3cm,right=3cm,marginparwidth=1.75cm]{geometry}

\usepackage{amsfonts}
\UseRawInputEncoding

\usepackage{pdflscape}
\usepackage{rotating}
\usepackage{longtable}
\usepackage{array}
\usepackage{float}
\usepackage{amsmath}
\usepackage{graphicx}
\usepackage{inconsolata}
\usepackage[colorlinks=true, allcolors=blue]{hyperref}
\usepackage{enumitem}
\usepackage{tikz}
\usepackage{listings}
\usepackage{xcolor}
\usepackage[T1]{fontenc}
\usepackage{IEEEtrantools}  
\usepackage{epigraph}  

\definecolor{bgcolor}{rgb}{0.97,0.97,0.97}
\definecolor{codeblue}{rgb}{0.1,0.1,0.8}
\definecolor{codegreen}{rgb}{0,0.4,0}
\definecolor{codegray}{rgb}{0.4,0.4,0.4}
\definecolor{codepurple}{rgb}{0.5,0,0.5}
\definecolor{codered}{rgb}{0.6,0.2,0.2}
\definecolor{lightgray}{rgb}{0.9,0.9,0.9}
\definecolor{darkgray}{rgb}{0.6,0.6,0.6}  

\makeatletter
\renewcommand{\paragraph}{
  \@startsection{paragraph}{4}{\z@}{1ex}{-1em}{\normalfont\normalsize\bfseries\color{gray}}}
\makeatother

\lstdefinestyle{python}{
    language=Python,
    basicstyle=\ttfamily\small\color{black}\usefont{T1}{zi4}{m}{n},  
    keywordstyle=\bfseries\color{codeblue},  
    stringstyle=\color{codegreen},  
    commentstyle=\slshape\color{codegray},  
    showstringspaces=false,
    numbers=left,
    numberstyle=\tiny\color{codegray},  
    stepnumber=1,
    numbersep=8pt,
    frame=single,
    rulecolor=\color{darkgray},  
    breaklines=true,
    backgroundcolor=\color{bgcolor},
    tabsize=4,
    captionpos=b,
    morekeywords={self}, 
}

\lstdefinestyle{matlab}{
    language=Matlab,
    basicstyle=\ttfamily\small\color{black}\usefont{T1}{zi4}{m}{n},  
    keywordstyle=\bfseries\color{codeblue},  
    stringstyle=\color{codegreen},  
    commentstyle=\slshape\color{codegray},  
    showstringspaces=false,
    numbers=left,
    numberstyle=\tiny\color{codegray},  
    stepnumber=1,
    numbersep=8pt,
    frame=single,
    rulecolor=\color{darkgray},  
    breaklines=true,
    backgroundcolor=\color{bgcolor},
    tabsize=4,
    captionpos=b,
    morekeywords={if, elseif, else, end, for, while, function}, 
}

\lstdefinestyle{cmd}{
    language=bash,
    basicstyle=\ttfamily\small\color{black}\usefont{T1}{zi4}{m}{n},  
    keywordstyle=\bfseries\color{blue},
    stringstyle=\color{codegreen},
    commentstyle=\itshape\color{gray},
    showstringspaces=false,
    numbers=none,
    frame=single,
    rulecolor=\color{darkgray},  
    breaklines=true,
    backgroundcolor=\color{bgcolor},
    tabsize=4,
    captionpos=b,
}

\title{FMCW Radar Principles and Human Activity Recognition Systems: Foundations, Techniques, and Applications}

\author{
    Ziqian Bi\textsuperscript{*} \\
    \textit{Purdue University} \\
    bi32@purdue.edu
    \and
    Jiawei Xu\textsuperscript{*} \\ 
    \textit{Purdue University} \\
    xu1644@purdue.edu
    \and
    Xinyuan Song \\ 
    \textit{Emory University} \\
    xsong30@emory.edu
    \and
    Ming Liu\textsuperscript{$\dagger$} \\ 
    \textit{Purdue University} \\
    liu3183@purdue.edu
}

\date{}  

\begin{document}

\maketitle

\begingroup
\renewcommand\thefootnote{}\footnote{
    \textsuperscript{*} Equal contribution \\
    \textsuperscript{$\dagger$} Corresponding author
}
\addtocounter{footnote}{0}
\endgroup

\epigraph{"It may be that our role on this planet is not to worship God but to create him."}{\textit{Arthur C. Clarke}}

\tableofcontents  

\part{Python Foundations for FMCW Radar}

  \chapter{Introduction to Python for Radar Applications}

    \section{Why Python for FMCW Radar?}

Python has become one of the most popular programming languages for scientific and engineering applications \cite{van2007python}, including radar signal processing. Its strengths, such as ease of use, an extensive collection of libraries, and a strong community of users, make it an ideal choice for FMCW (Frequency Modulated Continuous Wave) radar development \cite{marshall2008fmcw}.

Here are several reasons why Python is especially suited for FMCW radar signal processing:

\begin{itemize}
  \item \textbf{Open-source and Free:} Python is free to use and open-source, which means it has a vibrant community constantly improving its functionality. For engineers working on radar systems, the ability to use an open-source language means no additional costs for software licenses.
  
  \item \textbf{Extensive Libraries:} Python boasts a wide range of libraries for scientific computing, such as \texttt{NumPy} for numerical operations, \texttt{SciPy} for scientific computation, \texttt{Matplotlib} for plotting, and \texttt{Pandas} for data manipulation. In particular, for radar signal processing, libraries like \texttt{PyQtGraph} and \texttt{SciPy}'s signal processing tools are highly valuable.

  \item \textbf{Rapid Prototyping:} Python allows engineers to rapidly prototype and experiment with algorithms due to its simple syntax and dynamic typing. This speed is crucial for iterative development, which is common in radar algorithm design.

  \item \textbf{Interoperability:} Python can easily interface with other programming languages such as C, C++, and Fortran. This is particularly useful when high-performance implementations are required for real-time radar processing.
  
  \item \textbf{Visualization Capabilities:} Visualizing radar data is a key aspect of developing FMCW radar systems. Python, through libraries like \texttt{Matplotlib}, \texttt{Seaborn}, and \texttt{Plotly}, provides extensive support for creating high-quality plots and visualizations. This makes Python suitable for analyzing complex radar signals and observing system behavior in real-time.

  \item \textbf{Community and Resources:} Python has a vast user base and a wealth of online resources. For those new to radar signal processing or Python, there are many tutorials, forums, and books available to help you get started.

\end{itemize}

    \section{Setting Up the Python Environment}

Setting up a Python environment suitable for FMCW radar signal processing is crucial for ensuring a smooth development process. The recommended way to set up your environment is by using \textbf{Anaconda}, a powerful distribution that simplifies package management and deployment \cite{clark2022standardizing}. This section will guide you through installing the necessary tools and libraries for FMCW radar development.

\subsection{Installing Anaconda}

Anaconda is a free, open-source distribution of Python and R that is specifically designed for scientific computing. It simplifies package management by including popular libraries like \texttt{NumPy}, \texttt{SciPy}, and \texttt{Matplotlib}, which are essential for radar signal processing. Here's how to install Anaconda:

\begin{enumerate}
    \item Go to the official Anaconda website: \url{https://www.anaconda.com/products/individual}
    \item Download the version of Anaconda for your operating system (Windows, macOS, or Linux).
    \item Follow the installation instructions for your specific operating system:
        \begin{itemize}
            \item \textbf{Windows:} Double-click the downloaded .exe file and follow the on-screen instructions.
            \item \textbf{macOS:} Open the downloaded .pkg file and follow the installation guide.
            \item \textbf{Linux:} Open a terminal and run the following command:
            \begin{lstlisting}[style=cmd]
bash ~/Downloads/Anaconda3-2024.03-Linux-x86_64.sh
            \end{lstlisting}
            Follow the instructions to complete the installation.
        \end{itemize}
\end{enumerate}

Once installed, you can verify that Anaconda has been installed correctly by opening a terminal (or Anaconda Prompt on Windows) and typing:

\begin{lstlisting}[style=cmd]
conda --version
\end{lstlisting}

If the installation was successful, you should see the version of Conda (Anaconda's package manager) printed in the terminal.

\subsection{Creating a Python Environment for Radar Development}

A best practice when working with Python projects is to create isolated environments. These environments allow you to manage dependencies without affecting other projects. You can create a Python environment for radar signal processing as follows:

\begin{enumerate}
    \item Open a terminal or Anaconda Prompt and type:
    \begin{lstlisting}[style=cmd]
conda create --name radar-env python=3.10
    \end{lstlisting}
    This creates a new environment named \texttt{radar-env} with Python version 3.10.

    \item Activate the environment:
    \begin{lstlisting}[style=cmd]
conda activate radar-env
    \end{lstlisting}

    \item Install the essential libraries for radar development:
    \begin{lstlisting}[style=cmd]
conda install numpy scipy matplotlib jupyter pandas
    \end{lstlisting}
    This command installs \texttt{NumPy}, \texttt{SciPy}, \texttt{Matplotlib}, \texttt{Jupyter}, and \texttt{Pandas}, which are essential for radar signal processing and analysis.

\end{enumerate}

\subsection{Installing Jupyter Notebook}

\textbf{Jupyter Notebook} is an interactive environment where you can write and execute Python code, visualize radar data, and document your work. It is widely used in scientific computing and data analysis. To install Jupyter Notebook:

\begin{lstlisting}[style=cmd]
conda install jupyter
\end{lstlisting}

After installation, you can start the Jupyter Notebook server by typing:

\begin{lstlisting}[style=cmd]
jupyter notebook
\end{lstlisting}

This will open the Jupyter interface in your web browser. From here, you can create new notebooks and start coding. 

\subsection{Example: Testing Your Environment}

Let's write a simple script to ensure that everything is working correctly. We will use \texttt{NumPy} to create an array and \texttt{Matplotlib} to plot it. Open a new Jupyter Notebook and enter the following code:

\begin{lstlisting}[style=python]
import numpy as np
import matplotlib.pyplot as plt

# Generate a sine wave
x = np.linspace(0, 2 * np.pi, 100)
y = np.sin(x)

# Plot the sine wave
plt.plot(x, y)
plt.title("Sine Wave")
plt.xlabel("x")
plt.ylabel("sin(x)")
plt.grid(True)
plt.show()
\end{lstlisting}

\begin{figure}[H]
    \centering
    \includegraphics[width=1.0\textwidth]{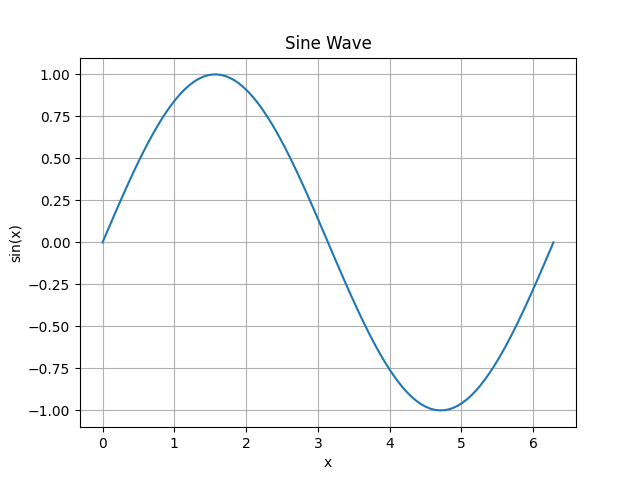}
    \caption{Sine wave}
\end{figure}

This code generates a simple sine wave and plots it. If you see the plot, your Python environment is ready for radar development.

\subsection{Troubleshooting}

If you encounter any issues during installation or setup, here are a few common solutions:

\begin{itemize}
    \item \textbf{Missing Package:} If you receive an error about a missing package, you can install it using the \texttt{conda install} or \texttt{pip install} command.
    \item \textbf{Kernel Not Found in Jupyter Notebook:} If Jupyter Notebook doesn't recognize your environment, you can add it as a kernel by running:
    \begin{lstlisting}[style=cmd]
python -m ipykernel install --user --name=radar-env
    \end{lstlisting}
\end{itemize}

Now that you have set up your Python environment, you are ready to dive into radar signal processing with Python.

\chapter{Data Structures and Basic Operations}

    \section{Lists, Tuples, and Dictionaries: Basic Containers}
    
    Python provides several core data structures that are commonly used to store and manipulate data, particularly when dealing with large datasets like radar signals. These data structures include \textbf{lists}, \textbf{tuples}, and \textbf{dictionaries}, each of which has unique properties and use cases.

    \subsection{Lists}
    A list is a mutable collection of items, which means you can modify the list after it's created. Lists are ordered, meaning the items have a defined sequence and can be accessed by their index. This makes them an excellent choice for storing sequences of radar data, such as a series of pulse measurements or a range of frequency sweeps in an FMCW radar system.

    \paragraph{Example of a Python List:}
    In radar processing, you may want to store a series of signal strength measurements in a list, allowing you to manipulate the data, perform calculations, or analyze the pattern of signal variation over time.
    
    \begin{lstlisting}[style=python]
    # List of radar signal strength values
    signal_strength = [0.1, 0.15, 0.2, 0.18, 0.22, 0.3, 0.4]
    
    # Accessing elements in the list
    first_signal = signal_strength[0]
    last_signal = signal_strength[-1]
    
    # Modifying the list
    signal_strength.append(0.45)  # Add new signal strength value
    signal_strength[2] = 0.25     # Modify the third signal
    \end{lstlisting}

    In this example, we have a list of radar signal strength values. Using indexing, we can access the first signal, modify any value, or append a new signal strength to the list.

    \subsection{Tuples}
    A tuple is similar to a list, but unlike lists, tuples are \textbf{immutable}, meaning once they are created, they cannot be changed. Tuples are useful for storing data that should not be modified, such as constant parameters in a radar system (e.g., radar system settings or fixed coordinates).

    \paragraph{Example of a Python Tuple:}
    In an FMCW radar system, you might store configuration parameters in a tuple to ensure that these values are not accidentally changed during processing.

    \begin{lstlisting}[style=python]
    # Tuple of radar configuration parameters (frequency start, frequency end, sweep time)
    radar_config = (24e9, 24.25e9, 1e-3)
    
    # Accessing elements in the tuple
    freq_start = radar_config[0]
    sweep_time = radar_config[2]
    
    # Trying to modify a tuple will raise an error
    # radar_config[0] = 23e9  # This will result in an error
    \end{lstlisting}

    The tuple \texttt{radar\_config} stores the start and end frequencies of an FMCW radar sweep, as well as the sweep time. These values remain fixed throughout the radar's operation.

    \subsection{Dictionaries}
    A dictionary is a collection of key-value pairs. Dictionaries are \textbf{unordered} but allow you to access, modify, and remove data based on unique keys. In radar processing, dictionaries can be used to store parameters and their associated values, such as mappings of range bins to distances or storing metadata about signal collections.

    \paragraph{Example of a Python Dictionary:}
    Consider storing radar signal data where each signal is associated with its corresponding timestamp.

    \begin{lstlisting}[style=python]
    # Dictionary mapping timestamps to radar signals
    radar_data = {
        "2023-01-01 10:00:00": [0.1, 0.15, 0.2],
        "2023-01-01 10:00:01": [0.12, 0.18, 0.22],
        "2023-01-01 10:00:02": [0.14, 0.2, 0.25]
    }
    
    # Accessing radar signal by timestamp
    first_signal = radar_data["2023-01-01 10:00:00"]
    
    # Modifying a dictionary
    radar_data["2023-01-01 10:00:03"] = [0.16, 0.21, 0.28]  # Adding new signal
    \end{lstlisting}

    Here, each key in the dictionary corresponds to a timestamp, and the value associated with that key is a list of radar signal strength measurements at that moment. Dictionaries are particularly useful when you need to access data based on a unique identifier (in this case, the timestamp).

    \section{Control Structures: Loops and Conditionals}

    Control structures in Python such as \textbf{loops} and \textbf{conditionals} allow you to automate the processing of radar data by executing specific code repeatedly (loops) or based on conditions (conditionals). This section will explore how these constructs can be used in radar signal processing, especially in handling large amounts of data and performing basic preprocessing tasks.

    \subsection{For Loops}
    A \texttt{for} loop allows you to iterate over a sequence (such as a list or dictionary) and perform operations on each item. For radar signal data, you can use a \texttt{for} loop to process each signal measurement, apply filtering, or calculate derived values.

    \paragraph{Example of a Python \texttt{for} Loop:}
    Suppose you have a list of radar signal strength values and you want to normalize each signal (i.e., scale them so that the maximum value is 1).

    \begin{lstlisting}[style=python]
    # List of radar signal strengths
    signal_strength = [0.1, 0.15, 0.2, 0.18, 0.22, 0.3, 0.4]
    
    # Find the maximum value
    max_value = max(signal_strength)
    
    # Normalize each signal strength
    normalized_signal = []
    for signal in signal_strength:
        normalized_signal.append(signal / max_value)
    
    print(normalized_signal)
    # Output: [0.25, 0.375, 0.5, 0.45, 0.55, 0.75, 1.0]
    \end{lstlisting}

    In this example, the \texttt{for} loop iterates over each value in the list \texttt{signal\_strength}, divides it by the maximum value, and appends the normalized value to a new list.

    \subsection{While Loops}
    A \texttt{while} loop continues to execute as long as a certain condition is true. This is useful when processing radar data until a specific criterion is met, such as when a signal crosses a certain threshold or when a specific number of iterations are completed.

    \paragraph{Example of a Python \texttt{while} Loop:}
    Let's say you want to stop processing radar signals once the signal strength exceeds a certain threshold.

    \begin{lstlisting}[style=python]
    # Radar signal strength values
    signal_strength = [0.1, 0.15, 0.2, 0.18, 0.22, 0.3, 0.4]
    
    # Set a threshold
    threshold = 0.25
    i = 0
    
    # Process the signal until a value exceeds the threshold
    while i < len(signal_strength) and signal_strength[i] <= threshold:
        print(f"Processing signal: {signal_strength[i]}")
        i += 1
    \end{lstlisting}

    Here, the \texttt{while} loop will process the radar signals one by one, printing each value until the signal strength exceeds the threshold of 0.25.

    \subsection{Conditionals (If Statements)}
    Conditionals allow you to execute code based on whether a condition is true or false. In radar signal processing, this might be used to check if a signal exceeds a noise threshold, or to apply different preprocessing steps depending on the signal quality.

    \paragraph{Example of a Python Conditional:}
    You may want to classify radar signals into categories based on their strength.

    \begin{lstlisting}[style=python]
    # Classify radar signal strength
    signal_strength = 0.3
    
    if signal_strength < 0.1:
        print("Weak signal")
    elif 0.1 <= signal_strength < 0.3:
        print("Moderate signal")
    else:
        print("Strong signal")
    \end{lstlisting}

    In this example, the radar signal is classified as ``strong'' because its value exceeds the threshold for a ``moderate signal''.

\chapter{Numerical Computation with NumPy}

\section{Introduction to NumPy Arrays}

NumPy is a fundamental package for scientific computing \cite{mehta2015mastering} in Python and plays a crucial role in radar signal processing. Its core feature is the powerful N-dimensional array object, known as the \texttt{ndarray}, which is efficient for handling large-scale, multi-dimensional radar data.

While Python lists can hold multiple elements, NumPy arrays are preferred for large-scale data processing, especially in signal processing applications like FMCW radar. NumPy arrays offer:
\begin{itemize}
    \item Faster processing for numerical data.
    \item Memory efficiency due to homogeneous data types.
    \item Built-in support for mathematical operations on entire arrays.
    \item Broadcasting and vectorized operations for performance improvements.
\end{itemize}

To start working with NumPy, you can import the library and create an array as follows:

\begin{lstlisting}[style=python]
import numpy as np
# Creating a simple 1D array
radar_signal = np.array([1.5, 3.2, 5.1, 2.6])
\end{lstlisting}

In this example, \texttt{radar\_signal} is a NumPy array that holds a sequence of radar signal data points.

\section{Array Indexing and Slicing for Signal Data}

\subsection{Basic Indexing and Slicing}

Indexing and slicing are essential techniques for extracting meaningful portions of radar data from arrays. In NumPy, you can access and manipulate elements in arrays similarly to how you would with Python lists, but with additional features.

Consider a one-dimensional array of radar signal data:
\begin{lstlisting}[style=python]
radar_signal = np.array([1.5, 3.2, 5.1, 2.6, 4.8])
# Accessing single element
first_signal = radar_signal[0]  # Returns 1.5
# Slicing the array (from index 1 to 3, exclusive)
subset_signal = radar_signal[1:3]  # Returns array([3.2, 5.1])
# Slicing with a step
every_second_signal = radar_signal[::2]  # Returns array([1.5, 5.1, 4.8])
\end{lstlisting}

In multi-dimensional arrays, such as a radar signal matrix where each row represents a different channel or sensor, indexing follows a similar pattern:
\begin{lstlisting}[style=python]
radar_matrix = np.array([[1.5, 3.2, 5.1], [2.6, 4.8, 1.9]])
# Accessing element in second row, first column
element = radar_matrix[1, 0]  # Returns 2.6
# Slicing the first two rows and the first two columns
subset_matrix = radar_matrix[:2, :2]  # Returns array([[1.5, 3.2], [2.6, 4.8]])
\end{lstlisting}

\subsection{Advanced Indexing Techniques}

Advanced indexing techniques in NumPy allow for more sophisticated data selection, including boolean and condition-based indexing, which are useful when working with large radar datasets where you may need to filter out corrupted or anomalous signals.

For example, to extract radar signal values greater than 3:
\begin{lstlisting}[style=python]
# Boolean indexing based on a condition
anomalous_signals = radar_signal[radar_signal > 3]  # Returns array([3.2, 5.1, 4.8])
\end{lstlisting}

This technique can be used to identify and isolate specific signal behaviors or anomalies detected in radar data.

\section{Array Shape and Reshaping Techniques}

\subsection{Understanding Array Shapes}

The shape of an array represents the number of elements along each dimension. Understanding array shapes is crucial for efficiently handling radar data, which is often multi-dimensional.

You can view the shape of an array using the \texttt{shape} attribute:
\begin{lstlisting}[style=python]
# Creating a 2D radar data matrix
radar_matrix = np.array([[1.5, 3.2, 5.1], [2.6, 4.8, 1.9]])
print(radar_matrix.shape)  # Output: (2, 3)
\end{lstlisting}
This radar matrix has 2 rows and 3 columns, corresponding to the number of channels and signal samples.

\subsection{Reshaping Arrays for Signal Processing}

In radar signal processing, you may need to reshape arrays to suit the required format for computations. For instance, a one-dimensional time-series radar signal may be transformed into a multi-dimensional matrix:

\begin{lstlisting}[style=python]
# Original 1D radar signal
radar_signal = np.array([1.5, 3.2, 5.1, 2.6, 4.8, 6.0])
# Reshaping to 2D matrix with 2 rows and 3 columns
reshaped_signal = radar_signal.reshape(2, 3)
print(reshaped_signal)
# Output: array([[1.5, 3.2, 5.1],
#                [2.6, 4.8, 6.0]])
\end{lstlisting}

This reshaping can be particularly useful when organizing radar data into matrices for further operations, such as signal correlation or frequency analysis.

\subsection{Flattening Arrays}

Flattening an array is the process of converting a multi-dimensional array into a one-dimensional array. This is useful when performing operations such as statistical analysis, where a single list of data points is needed.

\begin{lstlisting}[style=python]
# Flattening a 2D radar matrix into 1D
flattened_signal = reshaped_signal.flatten()
print(flattened_signal)
# Output: array([1.5, 3.2, 5.1, 2.6, 4.8, 6.0])
\end{lstlisting}

The \texttt{flatten()} function returns a copy of the array, whereas \texttt{ravel()} returns a flattened view, sharing memory with the original array.

\section{Array Transposition and Dimension Swapping}

\subsection{Basic Transpose Operation}

In radar signal processing, transposing an array means rearranging its rows and columns, which can be essential when switching between time-domain and frequency-domain representations.

\begin{lstlisting}[style=python]
# Transposing a 2D radar matrix
transposed_signal = radar_matrix.transpose()
print(transposed_signal)
# Output: array([[1.5, 2.6],
#                [3.2, 4.8],
#                [5.1, 1.9]])
\end{lstlisting}

Transposing is often used when working with frequency-domain data in FMCW radar systems, where the matrix needs to be reordered for further processing.

\subsection{Swapping Axes in Multi-dimensional Arrays}

The \texttt{swapaxes()} function allows you to swap any two axes in a multi-dimensional array, which is useful when reorganizing data structures for specific processing tasks in radar systems.

\begin{lstlisting}[style=python]
# Swapping axes in a 3D radar data array
radar_3d = np.random.rand(3, 4, 5)  # Example 3D radar signal data
swapped_axes = radar_3d.swapaxes(0, 2)
print(swapped_axes.shape)  # Output: (5, 4, 3)
\end{lstlisting}
This can be applied to rearrange data for real-time signal processing or analysis in FMCW radar applications.

\chapter{Data Visualization for Radar Signals}

    \section{Introduction to Matplotlib}

Matplotlib is a powerful plotting library \cite{sial2021comparative} in Python that allows users to create various types of static, interactive, and animated visualizations. For radar signal processing, visualization is an essential part of analyzing and interpreting signals at different stages, such as in the time domain and frequency domain. This section introduces Matplotlib and demonstrates how to use it for radar signal visualization.

Matplotlib is commonly used through its \texttt{pyplot} module, which provides functions for creating different types of plots. To start using Matplotlib in your radar signal processing projects, you need to install and import it as follows:

\begin{lstlisting}[style=python]
import matplotlib.pyplot as plt
\end{lstlisting}

Matplotlib supports various plot types, including line plots, scatter plots, bar plots, histograms, and more. In radar signal processing, the most common visualizations include time-domain plots, frequency-domain plots (spectra), and point cloud plots for radar targets.

    \section{Basic Operations in plt}

When using Matplotlib to create plots, the basic operation starts with creating a figure and adding charts, titles, and labels to it. Below, we will explore the fundamental operations in Matplotlib step by step.

\subsection{Creating Figures and Axes}
In Matplotlib, a plot starts with creating a figure object. We can use the \texttt{plt.figure()} function to create a new figure and then use the \texttt{plt.plot()} function to draw the graph within the figure. For example:

\begin{lstlisting}[style=python]
import matplotlib.pyplot as plt
import numpy as np

# Create a new figure
plt.figure()

# Plot a simple sine wave
x = np.linspace(0, 2 * np.pi, 100)
y = np.sin(x)
plt.plot(x, y)

# Show the plot
plt.show()
\end{lstlisting}

\begin{figure}[H]
    \centering
    \includegraphics[width=1.0\textwidth]{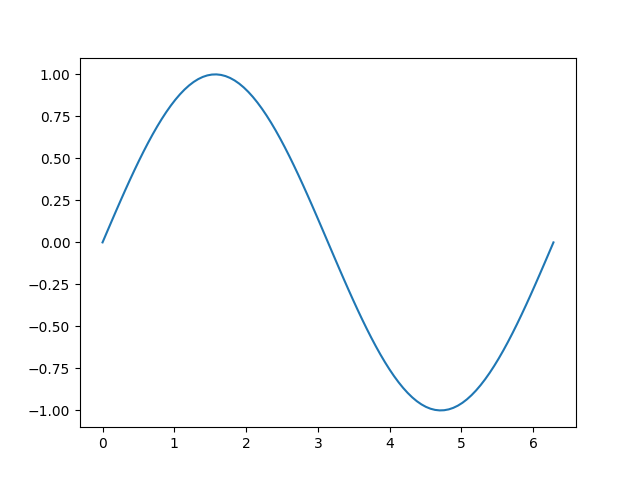}
    \caption{Creating Figures and Axes}
\end{figure}

This code creates a line plot of a sine wave and displays it within the figure.

\subsection{Adding Titles, Labels, and Grids}
When plotting radar signals, it is important to add titles, x-axis and y-axis labels to clearly explain the signal. Additionally, adding grid lines can help in visualizing the trends in the signal. Below is an example of how to add these elements:

\begin{lstlisting}[style=python]
# Create the figure
plt.figure()

# Plot the sine wave
plt.plot(x, y)

# Add title and labels
plt.title("Sine Wave")
plt.xlabel("Time (seconds)")
plt.ylabel("Amplitude")

# Add grid lines
plt.grid(True)

# Show the plot
plt.show()
\end{lstlisting}

\begin{figure}[H]
    \centering
    \includegraphics[width=1.0\textwidth]{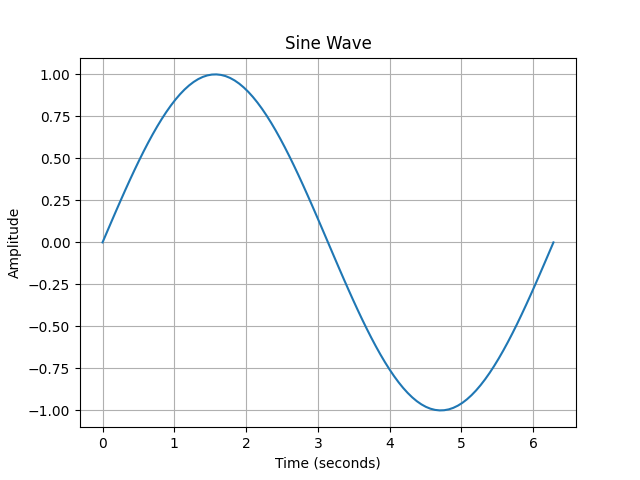}
    \caption{Adding Titles, Labels, and Grids}
\end{figure}

In this code, the \texttt{plt.title()} function adds the title to the plot, \texttt{plt.xlabel()} and \texttt{plt.ylabel()} add labels to the x and y axes, respectively, and \texttt{plt.grid(True)} enables grid lines.

\subsection{Saving Plots}
In radar signal processing, it is often necessary to save plots for further analysis or documentation. The \texttt{plt.savefig()} function allows you to save the current figure to a file. For example:

\begin{lstlisting}[style=python]
# Save the plot to a file
plt.savefig("sine_wave.png")
\end{lstlisting}

    \section{Creating Line Plots with plt}

Line plots are one of the most common types of visualizations in radar signal processing, especially for displaying time-domain signals. Line plots are typically used to represent the amplitude of the radar return signal as a function of time.

Below is a detailed example of how to create a simple line plot to visualize time-domain radar signals.

\subsection{Example: Plotting Time-Domain Radar Signals}
Suppose we have captured a radar return signal that is modeled as a sine wave with noise. We can plot this signal over time using the following code:

\begin{lstlisting}[style=python]
# Generate a time vector
t = np.linspace(0, 1, 500)

# Generate a sine wave signal with 5 Hz frequency and add noise
signal = np.sin(2 * np.pi * 5 * t) + 0.5 * np.random.randn(500)

# Create the figure and plot the signal
plt.figure()
plt.plot(t, signal)

# Add title and labels
plt.title("Time-Domain Radar Signal")
plt.xlabel("Time (seconds)")
plt.ylabel("Amplitude")

# Show the plot
plt.show()
\end{lstlisting}

\begin{figure}[H]
    \centering
    \includegraphics[width=1.0\textwidth]{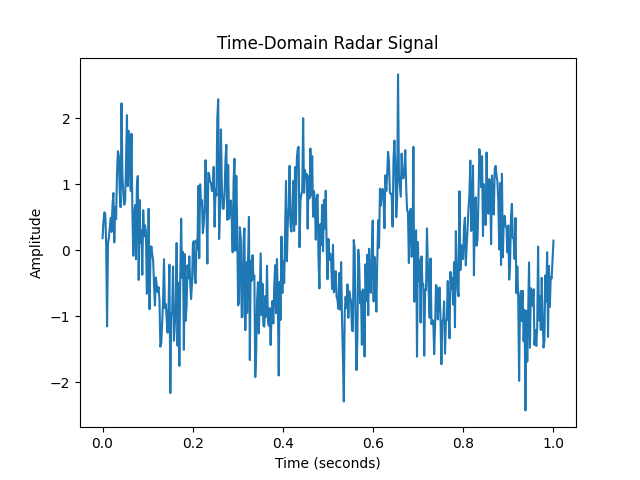}
    \caption{Example: Plotting Time-Domain Radar Signals}
\end{figure}

In this example, we generate a 5 Hz sine wave signal with added random noise to simulate real radar return data. The line plot displays the signal's amplitude as it varies over time.

    \section{Creating Scatter Plots with plt}

Scatter plots are often used to visualize radar target point clouds, showing the location, speed, or other attributes of detected objects. For FMCW radar, point clouds can be used to represent the two-dimensional or three-dimensional positions of multiple targets.

\subsection{Example: Plotting a Radar Target Point Cloud}
The following code demonstrates how to use the \texttt{scatter()} function in Matplotlib to create a scatter plot, simulating the two-dimensional coordinates of radar-detected targets:

\begin{lstlisting}[style=python]
# Simulate target positions detected by radar (randomly generated)
x_coords = np.random.rand(50) * 100  # x-coordinates
y_coords = np.random.rand(50) * 100  # y-coordinates

# Create the figure and plot the point cloud
plt.figure()
plt.scatter(x_coords, y_coords)

# Add title and labels
plt.title("Radar Target Point Cloud")
plt.xlabel("X Coordinate (meters)")
plt.ylabel("Y Coordinate (meters)")

# Show the plot
plt.show()
\end{lstlisting}

\begin{figure}[H]
    \centering
    \includegraphics[width=1.0\textwidth]{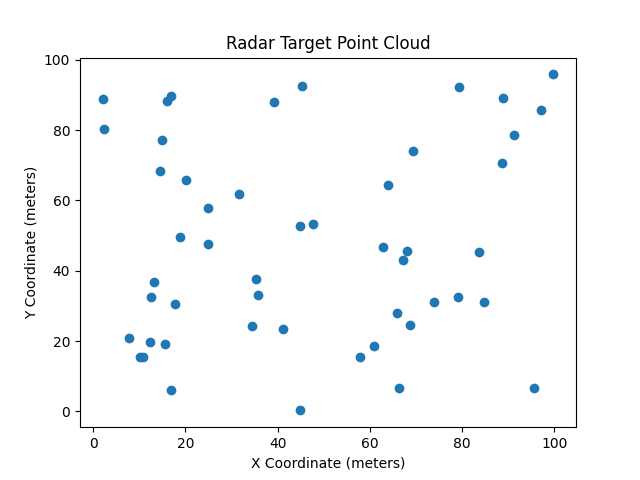}
    \caption{Example: Plotting a Radar Target Point Cloud}
\end{figure}

In this example, we generate the positions of 50 randomly located targets and display them using a scatter plot. Scatter plots are commonly used in radar target tracking and for visualizing range and azimuth angles.

\subsection{Example: Plotting a 3D Radar Target Point Cloud with DBSCAN Clustering}
The following code demonstrates how to use the \texttt{scatter()} function in Matplotlib to create a 3D scatter plot, simulating the three-dimensional coordinates of radar-detected targets, and applying DBSCAN clustering to group the points.

\begin{lstlisting}[style=python]
from sklearn.cluster import DBSCAN
from mpl_toolkits.mplot3d import Axes3D
import numpy as np
import matplotlib.pyplot as plt

# Simulate 3D target positions detected by radar (randomly generated)
np.random.seed(0)  # For reproducibility
x_coords = np.random.rand(100) * 100  # x-coordinates
y_coords = np.random.rand(100) * 100  # y-coordinates
z_coords = np.random.rand(100) * 100  # z-coordinates

# Combine the coordinates into a single dataset
coordinates = np.vstack((x_coords, y_coords, z_coords)).T

# Apply DBSCAN clustering
dbscan = DBSCAN(eps=15, min_samples=5)
labels = dbscan.fit_predict(coordinates)

# Create the figure and 3D plot
fig = plt.figure()
ax = fig.add_subplot(111, projection='3d')

# Plot each cluster with different colors
unique_labels = np.unique(labels)
for label in unique_labels:
    cluster_points = coordinates[labels == label]
    ax.scatter(cluster_points[:, 0], cluster_points[:, 1], cluster_points[:, 2], label=f'Cluster {label}')

# Add title and labels
ax.set_title("3D Radar Target Point Cloud with DBSCAN Clustering")
ax.set_xlabel("X Coordinate (meters)")
ax.set_ylabel("Y Coordinate (meters)")
ax.set_zlabel("Z Coordinate (meters)")

# Show legend and plot
ax.legend()
plt.show()
\end{lstlisting}

\begin{figure}[H]
    \centering
    \includegraphics[width=1.0\textwidth]{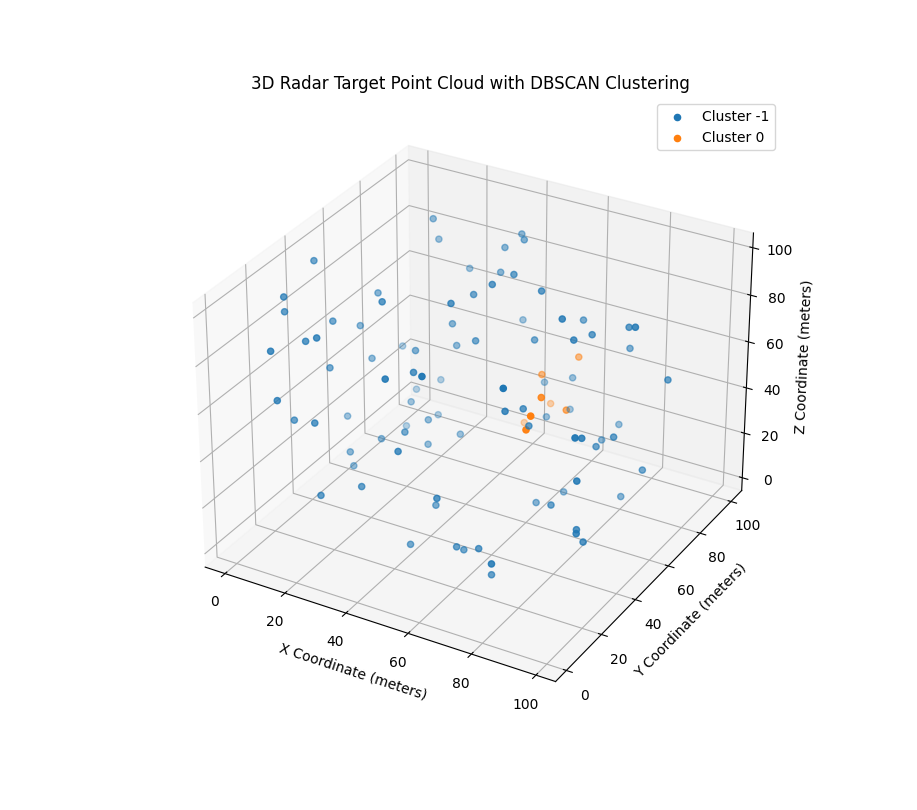}
    \caption{Example: Plotting a 3D Radar Target Point Cloud with DBSCAN Clustering}
\end{figure}

In this example, we generate the positions of 100 randomly located 3D targets and apply DBSCAN clustering to group them. The clusters are visualized in a 3D scatter plot, which is a useful technique in radar signal processing to analyze spatial distribution and clustering of targets.

    \section{Plotting Time-Domain and Frequency-Domain Signals}

In radar signal processing, analyzing the signal in both the time and frequency domains is essential. The time-domain signal describes how the signal varies over time, while the frequency-domain signal, obtained via the Fourier transform, reveals the frequency components of the signal.

\subsection{Example: Plotting Time-Domain and Frequency-Domain Signals}
The following example demonstrates how to plot both the time-domain and frequency-domain representations of a radar signal. We generate a signal with multiple frequency components and use the Fourier transform to convert it into the frequency domain:

\begin{lstlisting}[style=python]
# Generate a time vector
t = np.linspace(0, 1, 500)

# Generate a composite signal with 5 Hz and 20 Hz sine waves
signal = np.sin(2 * np.pi * 5 * t) + 0.5 * np.sin(2 * np.pi * 20 * t)

# Perform the Fourier Transform of the signal
signal_fft = np.fft.fft(signal)
frequencies = np.fft.fftfreq(len(signal), t[1] - t[0])

# Create the figure
plt.figure()

# Plot the time-domain signal
plt.subplot(2, 1, 1)
plt.plot(t, signal)
plt.title("Time-Domain Signal")
plt.xlabel("Time (seconds)")
plt.ylabel("Amplitude")

# Plot the frequency-domain signal (show only positive frequencies)
plt.subplot(2, 1, 2)
plt.plot(frequencies[:len(frequencies)//2], np.abs(signal_fft[:len(frequencies)//2]))
plt.title("Frequency-Domain Signal")
plt.xlabel("Frequency (Hz)")
plt.ylabel("Magnitude")

# Display the plots
plt.tight_layout()
plt.show()
\end{lstlisting}

\begin{figure}[H]
    \centering
    \includegraphics[width=1.0\textwidth]{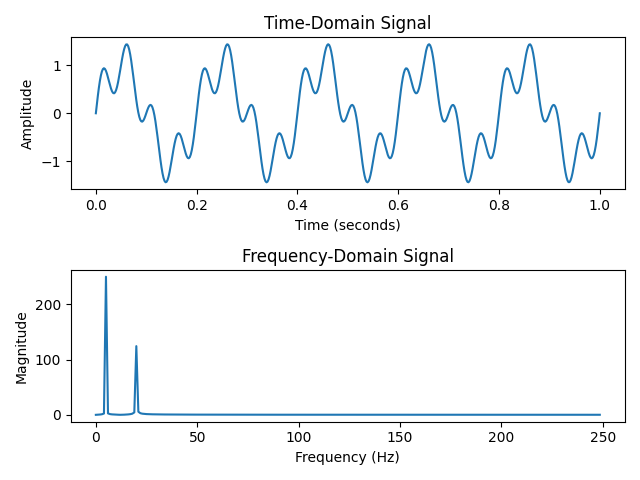}
    \caption{Example: Plotting Time-Domain and Frequency-Domain Signals}
\end{figure}

This example demonstrates how to plot both the time-domain and frequency-domain representations of a radar signal. We use the \texttt{np.fft.fft()} function to perform the Fast Fourier Transform (FFT) on the signal to reveal its frequency components, which are then displayed in the frequency-domain plot.

By using these visualization techniques, engineers can better understand radar signals at different stages of processing and optimize signal processing algorithms accordingly.

\chapter{Basic Concepts of Object-Oriented Programming (OOP)}

Object-Oriented Programming (OOP) is a programming paradigm that allows you to structure your code in a more organized and modular way by grouping related data and functions together in \textit{classes} and creating instances of these classes known as \textit{objects}. In radar systems, particularly in complex ones like FMCW radar, OOP helps in managing various components such as sensors, signal processors, and algorithms, making the system more scalable and easier to maintain.

    \section{Classes and Objects in Python}

    In Python, a \textbf{class} is a blueprint for creating objects \cite{hernandez2021python}. An \textbf{object} is an instance of a class, containing both \textbf{attributes} (data) and \textbf{methods} (functions) that operate on the data. Classes allow you to encapsulate radar-related functionalities into logical units. For example, you might have a class that represents a radar sensor, a signal processor, or even a specific radar waveform.

    \subsection{Defining a Class and Creating Objects}
    Let's begin by defining a simple class for an FMCW radar system. The class will store radar parameters such as frequency range and sweep time and will have methods to perform basic operations like calculating the bandwidth of the radar sweep.

    \paragraph{Example of Defining a Python Class:}

    Code for Example of Defining a Python Class:
    
    \begin{lstlisting}[style=python]
    # Define an FMCW radar class
    class FMCWRadar:
        def __init__(self, start_freq, end_freq, sweep_time):
            self.start_freq = start_freq  # Start frequency (Hz)
            self.end_freq = end_freq      # End frequency (Hz)
            self.sweep_time = sweep_time  # Sweep time (seconds)
        
        def calculate_bandwidth(self):
            # Method to calculate the radar bandwidth
            return self.end_freq - self.start_freq

    # Creating an object of the FMCWRadar class
    radar = FMCWRadar(24e9, 24.25e9, 1e-3)

    # Accessing the object's attributes and methods
    print("Radar Start Frequency:", radar.start_freq)
    print("Radar Bandwidth:", radar.calculate_bandwidth())
    \end{lstlisting}

    In this example, we defined an \texttt{FMCWRadar} class with an \texttt{\_\_init\_\_} method (constructor) that initializes the start frequency, end frequency, and sweep time. The \texttt{calculate\_bandwidth} method computes the radar's bandwidth, which is the difference between the end and start frequencies. We then created an object of the \texttt{FMCWRadar} class and accessed its attributes and methods.

    \subsection{Why Use Classes?}
    By organizing your radar system components into classes, you make it easier to manage different parts of the system. If you later need to add more features or change the way radar parameters are handled, you can update the class rather than modifying code in multiple places. This modular approach simplifies debugging and enhances reusability.

    For example, if you want to add a new method to calculate the chirp rate (the rate at which the radar frequency increases over time), you can easily extend the \texttt{FMCWRadar} class.

    \paragraph{Adding New Methods to a Class:}

    Code for Adding New Methods to a Class:

    \begin{lstlisting}[style=python]
    class FMCWRadar:
        def __init__(self, start_freq, end_freq, sweep_time):
            self.start_freq = start_freq
            self.end_freq = end_freq
            self.sweep_time = sweep_time
        
        def calculate_bandwidth(self):
            return self.end_freq - self.start_freq
        
        def calculate_chirp_rate(self):
            # Chirp rate is bandwidth divided by sweep time
            return self.calculate_bandwidth() / self.sweep_time

    # Creating a radar object and calculating chirp rate
    radar = FMCWRadar(24e9, 24.25e9, 1e-3)
    print("Radar Chirp Rate:", radar.calculate_chirp_rate())
    \end{lstlisting}

    Now, the \texttt{FMCWRadar} class has an additional method \texttt{calculate\_chirp\_rate} that computes the chirp rate. This demonstrates how classes can be easily extended to accommodate more complex radar functionality.

    \section{Inheritance and Polymorphism in Radar Systems}

    \subsection{Inheritance}
    Inheritance allows a class (called a \textbf{child class} or \textbf{subclass}) to inherit attributes and methods from another class (called a \textbf{parent class} or \textbf{superclass}). This is especially useful when creating radar systems that share common functionalities but have specialized behavior. For instance, you might have a generic radar class and then create specialized subclasses for FMCW, pulse-Doppler, or CW (continuous-wave) radars.

    \paragraph{Example of Inheritance:}
    Let's say we want to create a base class for generic radars and then create a specialized class for FMCW radars that inherits from it.

    \begin{lstlisting}[style=python]
    # Base class for a generic radar system
    class Radar:
        def __init__(self, name):
            self.name = name
        
        def display_info(self):
            print(f"Radar Name: {self.name}")
    
    # Derived class for an FMCW radar system
    class FMCWRadar(Radar):
        def __init__(self, name, start_freq, end_freq, sweep_time):
            super().__init__(name)  # Inherit the name attribute from Radar class
            self.start_freq = start_freq
            self.end_freq = end_freq
            self.sweep_time = sweep_time
        
        def calculate_bandwidth(self):
            return self.end_freq - self.start_freq

    # Create an FMCWRadar object and display its information
    fmcw_radar = FMCWRadar("FMCW Radar A", 24e9, 24.25e9, 1e-3)
    fmcw_radar.display_info()  # Calls method from the base Radar class
    print("Radar Bandwidth:", fmcw_radar.calculate_bandwidth())
    \end{lstlisting}

    In this example, the \texttt{FMCWRadar} class inherits from the \texttt{Radar} class, which means it can use the \texttt{display\_info} method defined in the \texttt{Radar} class. This illustrates how inheritance helps in reusing code across different radar systems.

    \subsection{Polymorphism}
    Polymorphism allows methods in different classes to have the same name but behave differently depending on which class the object belongs to. This is useful in radar systems when you want to apply similar operations (such as signal processing) to different types of radars but with specific implementations for each radar type.

    \paragraph{Example of Polymorphism:}

    Code for Example of Polymorphism:

    \begin{lstlisting}[style=python]
    # Base class for radar systems
    class Radar:
        def process_signal(self):
            raise NotImplementedError("This method should be overridden in subclasses")
    
    # Subclass for FMCW radar
    class FMCWRadar(Radar):
        def process_signal(self):
            print("Processing FMCW radar signal...")
    
    # Subclass for Pulse-Doppler radar
    class PulseDopplerRadar(Radar):
        def process_signal(self):
            print("Processing Pulse-Doppler radar signal...")
    
    # Create radar objects
    radars = [FMCWRadar(), PulseDopplerRadar()]
    
    # Loop through each radar and process its signal (polymorphism in action)
    for radar in radars:
        radar.process_signal()
    \end{lstlisting}

    Here, both \texttt{FMCWRadar} and \texttt{PulseDopplerRadar} override the \texttt{process\_signal} method of the \texttt{Radar} base class. When iterating through the list of radar objects, Python automatically calls the correct \texttt{process\_signal} method based on the type of radar, demonstrating polymorphism.

    \section{When and How to Use OOP for Radar Applications}

    Object-oriented programming is useful when your radar system has multiple components that interact with each other and require state (data) to be maintained. Use OOP in the following scenarios:
    
    \begin{itemize}
        \item When modeling complex radar systems with many interacting modules (e.g., multiple types of radar sensors).
        \item When the system needs to be easily extendable (e.g., adding new radar types like FMCW or Pulse-Doppler).
        \item When managing large amounts of data across different parts of the radar system.
    \end{itemize}

    However, OOP might not be necessary for simpler radar applications where the focus is on single functions or scripts. In such cases, \textbf{functional programming} \cite{hughes1989functional}, where you simply define functions and call them, might be more appropriate. You should avoid unnecessary complexity when a function-based approach is sufficient for tasks like signal filtering or generating radar waveforms.

    \paragraph{When to use functional programming:}
    \begin{itemize}
        \item For small, simple radar tasks that do not require managing state.
        \item When the focus is on performing individual, stateless operations (e.g., applying a Fourier Transform to a radar signal).
    \end{itemize}

    By balancing OOP and functional programming, you can design radar systems that are both efficient and scalable.

\chapter{Multithreading and Multiprocessing in Python}

\section{Introduction to Concurrency and Parallelism}

Concurrency and parallelism are two important concepts in modern programming that allow programs to run multiple tasks simultaneously, improving performance and efficiency. In radar signal processing, where large datasets and intensive computations are involved, using concurrency and parallelism can greatly accelerate the processing of radar signals and reduce bottlenecks.

\textbf{Concurrency} refers to the ability of a system to manage multiple tasks at the same time, potentially switching between them. \textbf{Parallelism}, on the other hand, means actually running multiple tasks at the same time on different processors or cores.

Python offers two primary ways to achieve concurrency and parallelism:
\begin{itemize}
    \item \textbf{Multithreading}: Multiple threads share the same memory space and are ideal for I/O-bound tasks such as reading from files or real-time radar data acquisition.
    \item \textbf{Multiprocessing}: Multiple processes run independently with separate memory spaces, making it ideal for CPU-bound tasks like radar signal computation and data analysis.
\end{itemize}

For radar signal processing:
\begin{itemize}
    \item Use \textbf{multiprocessing} for computationally intensive tasks like filtering, Fourier transforms, or radar image processing.
    \item Use \textbf{multithreading} for I/O-bound tasks like real-time data acquisition, data writing, and network communication.
\end{itemize}

\section{Multiprocessing with Pool}

Python's \texttt{multiprocessing} module allows you to create multiple processes that run concurrently, taking advantage of multiple CPU cores to perform tasks in parallel. One of the easiest and most efficient ways to distribute tasks across processes is by using the \texttt{Pool} class from the \texttt{multiprocessing} module.

\subsection{Creating and Managing a Pool of Processes}

A \texttt{Pool} is a group of worker processes that can be used to execute tasks concurrently. You can create a pool of processes and then distribute tasks using various methods such as \texttt{apply()}, \texttt{map()}, and \texttt{apply\_async()}. These methods allow you to perform parallel radar signal processing tasks efficiently.

Here is an example of how to use \texttt{multiprocessing.Pool} to parallelize radar signal processing tasks:

\begin{lstlisting}[style=python]
import numpy as np
from multiprocessing import Pool

# A function to process radar signal data (e.g., filtering)
def process_signal(data_chunk):
    # Perform some radar signal processing, such as filtering
    filtered_data = np.fft.fft(data_chunk)  # Example Fourier transform
    return filtered_data

# Splitting radar signal data into chunks for parallel processing
radar_data = np.random.rand(1000)  # Simulated radar signal data
chunks = np.array_split(radar_data, 4)  # Split data into 4 chunks

# Creating a pool of 4 processes
with Pool(4) as pool:
    # Applying parallel processing
    results = pool.map(process_signal, chunks)

# Combining the processed chunks back into a single array
processed_data = np.concatenate(results)
print(processed_data)
\end{lstlisting}

In this example:
\begin{itemize}
    \item \texttt{radar\_data} is split into 4 chunks using \texttt{np.array\_split()}, which distributes the data evenly for parallel processing.
    \item The \texttt{pool.map()} method applies the \texttt{process\_signal()} function to each chunk in parallel across 4 processes.
    \item After processing, the results are combined back into a single array.
\end{itemize}

\subsection{Parallel Signal Processing Using Pool}

FMCW radar signal processing often involves computationally intensive tasks such as frequency analysis (e.g., Fast Fourier Transforms) or filtering large datasets. By using multiple processes, these tasks can be parallelized, greatly reducing the time required for processing.

For example, if you need to apply frequency analysis to large radar data, the task can be distributed across multiple CPU cores using the \texttt{Pool}:

\begin{lstlisting}[style=python]
# Function to perform frequency analysis on radar signal data
def frequency_analysis(signal_chunk):
    # Perform Fourier Transform to convert from time to frequency domain
    freq_data = np.fft.fft(signal_chunk)
    return freq_data

# Parallelizing frequency analysis on radar data
with Pool(4) as pool:
    frequency_results = pool.map(frequency_analysis, chunks)

# Combining frequency-domain data
frequency_domain_data = np.concatenate(frequency_results)
\end{lstlisting}

This approach can be used to handle real-time signal processing in complex radar systems by parallelizing key operations.

\section{Multithreading in Python}

Multithreading in Python allows you to execute multiple threads concurrently within a single process. Threads are lightweight and share the same memory space, making them ideal for I/O-bound tasks, such as reading radar data from sensors, writing to files, or handling network communication.

For example, if you are reading real-time radar data from a sensor and simultaneously writing it to a file, using threads can help improve efficiency:

\begin{lstlisting}[style=python]
import threading

# Function to read radar data from a sensor
def read_radar_data():
    while True:
        # Simulated radar data acquisition
        data = np.random.rand(100)
        print("Reading radar data:", data)

# Function to write radar data to a file
def write_radar_data():
    while True:
        # Simulated radar data writing
        print("Writing radar data to file...")

# Creating threads for reading and writing radar data
thread1 = threading.Thread(target=read_radar_data)
thread2 = threading.Thread(target=write_radar_data)

# Starting both threads
thread1.start()
thread2.start()

# Joining threads to ensure they finish execution
thread1.join()
thread2.join()
\end{lstlisting}

In this example, two threads are created:
\begin{itemize}
    \item \texttt{thread1} reads radar data in real-time.
    \item \texttt{thread2} writes the data to a file.
\end{itemize}
By using multithreading, both operations can be performed concurrently, improving efficiency.

\subsection{Thread Management and Synchronization}

When working with threads, managing shared resources between threads can lead to race conditions and deadlocks. To prevent these issues, synchronization mechanisms such as locks and semaphores can be used.

A lock ensures that only one thread can access a shared resource at a time:

\begin{lstlisting}[style=python]
import threading

lock = threading.Lock()

# Shared radar data buffer
radar_data_buffer = []

# Function to safely write radar data
def safe_write_data():
    with lock:
        # Write data to the buffer
        radar_data_buffer.append(np.random.rand(100))

# Creating and starting multiple threads
threads = [threading.Thread(target=safe_write_data) for _ in range(5)]
for t in threads:
    t.start()

# Joining all threads
for t in threads:
    t.join()

print("Radar data buffer:", radar_data_buffer)
\end{lstlisting}

In this example, the \texttt{lock} ensures that only one thread at a time can write to the \texttt{radar\_data\_buffer}, preventing data corruption.

\subsection{When to Use Multithreading for Radar Applications}

Multithreading is particularly useful for radar-related tasks that involve I/O operations or tasks that are not CPU-intensive. These include:
\begin{itemize}
    \item \textbf{Real-time radar data acquisition}: Continuously collecting data from radar sensors in real-time.
    \item \textbf{Data writing}: Writing collected radar data to a file or database without interrupting data collection.
    \item \textbf{Network communication}: Sending or receiving radar data over a network in real-time.
\end{itemize}

By using threads, these tasks can be performed concurrently, ensuring smooth operation without I/O bottlenecks.

\section{Asynchronous Programming}

Asynchronous programming \cite{hattingh2020using} is a programming paradigm that allows you to run tasks in a non-blocking manner, meaning that tasks can execute concurrently without waiting for each other to complete. In Python, the \texttt{asyncio} library provides a framework for writing asynchronous code.

Asynchronous programming is ideal for scenarios like real-time data streaming, where radar data is continuously acquired, processed, and transmitted.

\subsection{Asynchronous Tasks with asyncio}

The \texttt{asyncio} library allows you to define asynchronous functions (coroutines) using the \texttt{async def} syntax. These functions can be used to handle non-blocking I/O tasks, such as reading from sensors or sending radar data over a network.

\begin{lstlisting}[style=python]
import asyncio

# Simulating asynchronous radar data acquisition
async def read_radar_data():
    while True:
        # Simulated radar data acquisition
        data = np.random.rand(100)
        print("Asynchronously reading radar data:", data)
        await asyncio.sleep(1)  # Simulating data acquisition delay

# Running the asynchronous task
asyncio.run(read_radar_data())
\end{lstlisting}

In this example, the \texttt{read\_radar\_data()} function continuously reads radar data without blocking other operations.

\subsection{Event Loops and Coroutines}

The \texttt{asyncio} library is built around an event loop, which schedules and runs asynchronous tasks (coroutines). Coroutines are functions that can be paused and resumed, allowing for efficient task management in radar applications.

For example, if you need to acquire radar data while transmitting it over a network, both tasks can be handled asynchronously using the event loop:

\begin{lstlisting}[style=python]
async def transmit_data():
    while True:
        print("Asynchronously transmitting data...")
        await asyncio.sleep(2)

async def main():
    # Running both tasks concurrently
    await asyncio.gather(read_radar_data(), transmit_data())

# Running the event loop
asyncio.run(main())
\end{lstlisting}

In this case, both radar data acquisition and data transmission happen concurrently, improving overall system performance without blocking.

\section{Combining Multithreading and Asynchronous Programming}

In complex radar systems, you may need to combine both multithreading and asynchronous programming. For example, you can use multithreading to handle I/O-bound tasks such as radar data acquisition, while using asynchronous programming to process the data in a non-blocking manner.

Here's an example of how to combine both approaches:

\begin{lstlisting}[style=python]
import threading
import asyncio

# Function to collect radar data in a separate thread
def radar_data_collection():
    while True:
        print("Collecting radar data in a thread...")
        # Simulated radar data collection
        data = np.random.rand(100)
        asyncio.run(process_data(data))  # Asynchronously process the data

# Asynchronous function to process radar data
async def process_data(data):
    print("Asynchronously processing radar data:", data)
    await asyncio.sleep(1)

# Running radar data collection in a separate thread
collection_thread = threading.Thread(target=radar_data_collection)
collection_thread.start()
collection_thread.join()
\end{lstlisting}

In this example:
\begin{itemize}
    \item Radar data is collected in a separate thread using \texttt{radar\_data\_collection()}.
    \item The collected data is processed asynchronously using the \texttt{process\_data()} coroutine.
\end{itemize}

By combining multithreading for I/O-bound tasks and asynchronous programming for non-blocking processing, radar systems can achieve higher performance and responsiveness.

\part{MATLAB Foundations for FMCW Radar}

  \chapter{Introduction to MATLAB for Radar Applications}

  \section{Why MATLAB for FMCW Radar?}

MATLAB (short for \textbf{MAT}rix \textbf{LAB}oratory) is a powerful programming environment \cite{otto2005introduction} designed for numerical computing and visualization. Its extensive library of built-in functions and toolboxes makes it particularly useful for engineers working in fields such as signal processing, control systems, and radar systems. When applied to FMCW (Frequency Modulated Continuous Wave) radar, MATLAB provides several distinct advantages for both beginners and experienced engineers alike.

Here are some of the key reasons why MATLAB is an excellent choice for FMCW radar signal processing:

\begin{itemize}
    \item \textbf{Rapid Prototyping:} MATLAB allows you to develop algorithms quickly with fewer lines of code compared to traditional programming languages like C or Python. This is crucial in radar signal processing, where fast iteration of designs is required to test and optimize radar performance.
    \item \textbf{Toolboxes:} MATLAB offers specialized toolboxes such as the \texttt{Signal Processing Toolbox}, \texttt{Radar Toolbox}, and \texttt{DSP System Toolbox}, which contain pre-built functions for signal analysis, filtering, and radar-specific algorithms. These toolboxes eliminate the need to code everything from scratch and speed up the development process.
    \item \textbf{Simulation Capabilities:} MATLAB excels in simulating complex systems. For FMCW radar, simulations allow users to model target detection, range-Doppler processing, and clutter suppression in a controlled environment. This aids in testing algorithms before implementing them in real-world hardware.
    \item \textbf{Visualization:} MATLAB's powerful plotting functions provide easy visualization of signals, allowing users to observe time-domain, frequency-domain, and spectrogram representations. Visual feedback is crucial for understanding the effects of signal processing algorithms and making necessary adjustments.
    \item \textbf{Ease of Use:} MATLAB's syntax is intuitive, especially for those familiar with linear algebra and matrix manipulations. It is designed for working with vectors and matrices, which makes it well-suited for handling the complex data structures used in radar signal processing.
\end{itemize}

\section{Setting Up the MATLAB Environment}

Before we dive into radar signal processing, it is essential to have a working MATLAB environment. Follow the steps below to set up MATLAB and install the necessary toolboxes for FMCW radar applications.

\subsection{Installation of MATLAB}

If you have not installed MATLAB yet, follow these steps to get started:

\begin{enumerate}
    \item Visit the \textbf{MathWorks} website at \url{https://www.mathworks.com/} and navigate to the \texttt{Products} section.
    \item Click on \texttt{MATLAB} and select \texttt{Download Trial} if you don't have a license, or log in to your account to download the version that matches your license.
    \item Run the installer and follow the prompts. You will need to create a MathWorks account if you don't already have one.
    \item During installation, you can choose which toolboxes to install. For radar applications, make sure to select:
    \begin{itemize}
        \item \texttt{Signal Processing Toolbox}
        \item \texttt{Radar Toolbox}
        \item \texttt{DSP System Toolbox}
        \item \texttt{Phased Array System Toolbox} (optional, but useful for advanced radar applications)
    \end{itemize}
    \item Once the installation is complete, open MATLAB by clicking the application icon or running the command \texttt{matlab} from the terminal (on Linux or macOS).
\end{enumerate}

\subsection{MATLAB Interface Overview}

When you first open MATLAB, you will be presented with the \texttt{MATLAB Desktop}, which consists of the following key components:

\begin{itemize}
    \item \textbf{Command Window:} This is where you can enter MATLAB commands, run scripts, and interact with variables.
    \item \textbf{Workspace:} The workspace displays all variables that are currently loaded into memory. It allows you to see their values and dimensions.
    \item \textbf{Current Folder:} This section shows the files and directories that are in the working directory. MATLAB executes scripts and functions from the current directory.
    \item \textbf{Editor:} This is where you write and save MATLAB scripts or functions. The editor supports syntax highlighting, debugging, and error-checking.
    \item \textbf{Plots and Figures:} MATLAB will open new windows for visualizing plots, graphs, and radar signal analysis results.
\end{itemize}

Below is a typical MATLAB command that sets up a simple signal for analysis in radar applications:

\begin{lstlisting}[style=matlab]
% Generate a simple FMCW radar signal
fs = 1000; % Sampling frequency in Hz
T = 1; % Signal duration in seconds
t = 0:1/fs:T-1/fs; % Time vector
f_start = 0; % Start frequency of the chirp in Hz
f_end = 100; % End frequency of the chirp in Hz
signal = chirp(t, f_start, T, f_end);

% Plot the signal
figure;
plot(t, signal);
xlabel('Time (s)');
ylabel('Amplitude');
title('FMCW Chirp Signal');
\end{lstlisting}

This code snippet generates a simple FMCW chirp signal and plots it over time. The \texttt{chirp()} function is a convenient way to create frequency-modulated signals for radar applications.

\subsection{Installing Toolboxes Post-Installation}

If you did not install the necessary toolboxes during the initial setup, you can add them later by following these steps:

\begin{enumerate}
    \item Open MATLAB.
    \item Go to the \texttt{Home} tab and click on \texttt{Add-Ons} and select \texttt{Get Add-Ons}.
    \item In the Add-On Explorer, search for the toolboxes you need (e.g., \texttt{Signal Processing Toolbox} or \texttt{Radar Toolbox}).
    \item Click on the toolbox, and then click the \texttt{Install} button.
    \item Once installed, you can verify the installation by entering the command \texttt{ver} in the Command Window, which lists all installed toolboxes.
\end{enumerate}

With MATLAB set up and the key toolboxes installed, you're ready to begin exploring radar signal processing!

\section{Getting Familiar with MATLAB Basics}

MATLAB's strength lies in its ability to work efficiently with matrices and vectors, which are fundamental for radar signal processing. Let's explore some of the key operations and how they apply to FMCW radar.

\subsection{Creating and Manipulating Variables}

In MATLAB, variables are created simply by assigning them a value. For example:

\begin{lstlisting}[style=matlab]
A = 5; % Create a scalar variable
B = [1, 2, 3; 4, 5, 6]; % Create a 2x3 matrix
\end{lstlisting}

In the context of radar processing, you often work with signals stored as vectors (1D arrays) or matrices (2D arrays).

\subsection{Matrix Operations for Radar Applications}

Matrices are essential in radar processing for handling multi-dimensional data, such as signals received from different antenna elements or over time. Below are some fundamental operations:

\begin{lstlisting}[style=matlab]
% Matrix addition
C = B + 2; % Add 2 to each element of matrix B

% Matrix multiplication
D = B * B'; % Multiply matrix B with its transpose

% Element-wise multiplication
E = B .* B; % Multiply element-by-element
\end{lstlisting}

These operations are often used when manipulating radar signals, for instance, when performing Doppler processing or range-Doppler map generation.

\subsection{Basic Signal Processing}

MATLAB provides built-in functions for performing common signal processing tasks, such as the Fourier Transform, which is crucial for analyzing the frequency content of radar signals.

\begin{lstlisting}[style=matlab]
% Compute the Fourier Transform of a radar signal
X = fft(signal); % Fast Fourier Transform of the signal
f = (0:length(X)-1)*fs/length(X); % Frequency axis

% Plot the spectrum
figure;
plot(f, abs(X));
xlabel('Frequency (Hz)');
ylabel('Magnitude');
title('Spectrum of FMCW Radar Signal');
\end{lstlisting}

The \texttt{fft()} function converts a time-domain radar signal into its frequency-domain representation, which is useful for detecting targets and analyzing Doppler shifts.

With these basic operations, you are now equipped to perform essential radar signal processing tasks in MATLAB!

\chapter{MATLAB Basics for Data Handling and Processing}

\section{MATLAB Variables and Data Types}

In MATLAB, data can be represented in several forms, including scalars, vectors, matrices, and higher-dimensional arrays. Radar signal processing often involves the manipulation of data in these forms. Understanding MATLAB's variable types and how to create and manipulate them is crucial for working efficiently with radar data.

\textbf{Scalars} are single data elements, such as a single number:
\begin{lstlisting}[style=matlab]
x = 5;  % scalar value
\end{lstlisting}

\textbf{Vectors} are one-dimensional arrays that hold multiple data elements. For instance, in radar, a vector could represent a set of signal samples over time:
\begin{lstlisting}[style=matlab]
v = [1, 2, 3, 4, 5];  % row vector
w = [1; 2; 3; 4; 5];  % column vector
\end{lstlisting}

\textbf{Matrices} are two-dimensional arrays, commonly used in radar for storing data such as the range-Doppler map or signal covariance matrices:
\begin{lstlisting}[style=matlab]
M = [1, 2, 3; 4, 5, 6; 7, 8, 9];  % 3x3 matrix
\end{lstlisting}

For radar data, we may also encounter \textbf{higher-dimensional arrays}, where data is organized into three or more dimensions, such as 3D radar data cubes.

\section{Matrix and Array Operations}

\textbf{Matrix operations} are fundamental when working with radar signal processing algorithms, particularly in FMCW radar systems. MATLAB provides an extensive range of built-in operations for working with matrices and arrays.

\subsection{Matrix Creation and Initialization}

There are various ways to create and initialize matrices in MATLAB. Below are some of the most common:

\textbf{Zero Matrix:}
\begin{lstlisting}[style=matlab]
Z = zeros(3, 3);  % 3x3 matrix of all zeros
\end{lstlisting}

\textbf{Identity Matrix:}
\begin{lstlisting}[style=matlab]
I = eye(4);  % 4x4 identity matrix
\end{lstlisting}

\textbf{Random Matrix:}
\begin{lstlisting}[style=matlab]
R = rand(2, 5);  % 2x5 matrix of random numbers between 0 and 1
\end{lstlisting}

These matrices are commonly used in initializing filters, covariance matrices, or simulation scenarios in radar processing.

\subsection{Matrix Indexing and Slicing}

Indexing is essential for accessing specific elements of a matrix or extracting a subset of the data. In radar signal processing, you often need to extract a certain range of samples or dimensions from a matrix.

To access individual elements:
\begin{lstlisting}[style=matlab]
val = M(2, 3);  % access element in the 2nd row and 3rd column
\end{lstlisting}

To extract a submatrix:
\begin{lstlisting}[style=matlab]
subM = M(1:2, 2:3);  % extract rows 1 to 2 and columns 2 to 3
\end{lstlisting}

\section{Reshaping and Manipulating Data Dimensions}

In radar data handling, we frequently need to reshape data arrays to match the requirements of algorithms or to make data more accessible for certain types of processing. MATLAB provides functions like \texttt{reshape()} and \texttt{permute()} to handle such operations.

\subsection{Reshape for Signal Processing}

The \texttt{reshape()} function changes the dimensions of a matrix without altering the data values. For instance, if we have radar data arranged in a one-dimensional array, we can reshape it into a 2D or 3D form for further processing:

\begin{lstlisting}[style=matlab]
data = 1:12;  % 1D array
reshapedData = reshape(data, 3, 4);  % 3x4 matrix
\end{lstlisting}

This operation is particularly useful in radar processing, such as reshaping time-domain signals into matrix form for range-Doppler mapping.

\subsection{Transposing and Permuting Dimensions}

For multi-dimensional radar data, it may be necessary to transpose or rearrange data dimensions. The \texttt{transpose()} and \texttt{permute()} functions facilitate this.

\textbf{Transposing a matrix:}
\begin{lstlisting}[style=matlab]
T = M.';  % transpose of matrix M
\end{lstlisting}

\textbf{Permuting dimensions:}
\begin{lstlisting}[style=matlab]
permutedData = permute(data3D, [2, 1, 3]);  % swap 1st and 2nd dimensions of a 3D array
\end{lstlisting}

\section{MATLAB Scripts and Functions}

Modular programming in MATLAB allows users to break down large radar processing tasks into smaller, reusable parts. Scripts and functions are the main tools for implementing modular code.

\subsection{Creating MATLAB Scripts}

A MATLAB script is a simple file containing a sequence of commands. Scripts are ideal for performing data processing, simulations, and visualizations in radar applications.

Example script for visualizing a radar signal:
\begin{lstlisting}[style=matlab]
% Radar signal simulation and visualization
t = 0:0.01:1;  % time vector
signal = cos(2 * pi * 5 * t);  % simulated radar signal

plot(t, signal);  % visualize the signal
title('Simulated Radar Signal');
xlabel('Time (s)');
ylabel('Amplitude');
\end{lstlisting}

\subsection{Writing Functions for Radar Processing}

Functions in MATLAB are essential for structuring more complex radar algorithms. A function can accept inputs, process data, and return outputs. For example, an FMCW radar processing function can be written as follows:

\begin{lstlisting}[style=matlab]
function range = calculateRange(frequencyShift, sweepTime, bandwidth)
    % This function calculates the range of an object in FMCW radar
    c = 3e8;  % speed of light in m/s
    range = (frequencyShift * c * sweepTime) / (2 * bandwidth);
end
\end{lstlisting}

This function modularizes the range calculation, which can be reused across different scripts and radar simulations.

\chapter{Signal Processing with MATLAB}

\section{Time-Domain Signal Analysis}

Time-domain signal analysis focuses on analyzing signals \cite{nguyen2014time} as they evolve over time. For FMCW (Frequency Modulated Continuous Wave) radar systems, the time-domain representation of signals is fundamental to understanding how the radar transmits and receives waveforms. MATLAB offers powerful functions for working with time-domain signals, and this section introduces some of the most important tools and techniques.

\subsection{Basic Time-Domain Operations}

In this subsection, we discuss basic time-domain operations such as signal smoothing, filtering, and statistical analysis (mean, variance), which are essential in processing raw radar signals.

\textbf{1. Signal Smoothing}

Signal smoothing \cite{enke1976signal} is a common operation that removes high-frequency noise from the signal, making the underlying pattern easier to identify. In MATLAB, this can be achieved using moving average filters or other smoothing techniques. Below is an example of how to smooth a noisy signal using a moving average filter:

\begin{lstlisting}[style=matlab]
% Example of Signal Smoothing using a Moving Average Filter
t = 0:0.01:10;          % Time vector
noisy_signal = sin(2*pi*t) + 0.5*randn(size(t));  % Noisy sine wave

% Moving average filter with window size 5
windowSize = 5;
smooth_signal = movmean(noisy_signal, windowSize);

% Plot the original noisy signal and the smoothed signal
figure;
plot(t, noisy_signal, 'r--', 'DisplayName', 'Noisy Signal');
hold on;
plot(t, smooth_signal, 'b-', 'DisplayName', 'Smoothed Signal');
legend;
xlabel('Time (s)');
ylabel('Amplitude');
title('Signal Smoothing with Moving Average Filter');
\end{lstlisting}

\textbf{2. Filtering}

Filtering is crucial in isolating certain frequency components from the signal. For instance, you might want to filter out high-frequency noise from the radar signal. In MATLAB, built-in functions allow you to apply filters such as low-pass, high-pass, and band-pass filters.

\textbf{3. Statistical Analysis}

Basic statistical operations like calculating the mean and variance of a signal help in understanding the signal's overall behavior.

\begin{lstlisting}[style=matlab]
% Example of Statistical Analysis
mean_signal = mean(noisy_signal);   % Calculate the mean of the signal
var_signal = var(noisy_signal);     % Calculate the variance of the signal

fprintf('Mean of the signal: %.2f\n', mean_signal);
fprintf('Variance of the signal: %.2f\n', var_signal);
\end{lstlisting}

\subsection{Signal Plotting and Visualization in Time Domain}

Visualizing time-domain signals helps in identifying key features such as signal amplitude, duration, and noise characteristics. MATLAB provides extensive plotting functions for this purpose. 

Below is an example of how to visualize a time-domain signal:

\begin{lstlisting}[style=matlab]
% Time-Domain Signal Visualization
t = 0:0.001:1;                 % Time vector
signal = cos(2*pi*50*t) + 0.5*randn(size(t));   % Cosine wave with noise

figure;
plot(t, signal, 'b');           % Plot the signal
xlabel('Time (s)');
ylabel('Amplitude');
title('Time-Domain Signal Visualization');
grid on;
\end{lstlisting}

In this example, a cosine wave is plotted along with random noise, which simulates a noisy radar signal. This simple plot can help in identifying whether any noise filtering or smoothing is necessary.

\section{Frequency-Domain Signal Analysis}

Frequency-domain analysis allows us to observe the signal's frequency components, which is crucial for radar systems. MATLAB's Fourier Transform functions enable us to analyze the spectral content of FMCW radar signals.

\subsection{Fourier Transform with MATLAB}

The Fast Fourier Transform \cite{nussbaumer1982fast} (FFT) is a computational tool that converts a time-domain signal into its frequency-domain representation. For radar systems, the FFT helps in analyzing how different frequency components contribute to the overall signal.

\begin{lstlisting}[style=matlab]
% Example of FFT on a Radar Signal
fs = 1000;                     % Sampling frequency (Hz)
t = 0:1/fs:1-1/fs;             % Time vector
signal = cos(2*pi*100*t) + 0.5*cos(2*pi*200*t);   % Signal with two frequencies

% Compute the FFT of the signal
N = length(signal);
signal_fft = fft(signal);

% Frequency vector for plotting
f = (0:N-1)*(fs/N);

% Plot the FFT result
figure;
plot(f, abs(signal_fft));
xlabel('Frequency (Hz)');
ylabel('Amplitude');
title('Frequency Spectrum using FFT');
\end{lstlisting}

In this example, we create a signal with two different frequency components (100 Hz and 200 Hz) and analyze it using FFT. The frequency spectrum reveals the distinct peaks corresponding to the signal's frequency content.

\subsection{Power Spectral Density (PSD) Estimation}

Power Spectral Density (PSD) provides information \cite{czifra2009sensitivity} about how the signal's power is distributed across different frequencies. In radar systems, PSD is useful for identifying signal components and background noise levels.

\begin{lstlisting}[style=matlab]
% Example of Power Spectral Density (PSD) Estimation
fs = 1000;                    % Sampling frequency
t = 0:1/fs:1-1/fs;            % Time vector
signal = cos(2*pi*100*t) + 0.5*randn(size(t));   % Signal with noise

% Estimate the Power Spectral Density
[psd, f] = periodogram(signal, [], [], fs);

% Plot the PSD
figure;
plot(f, 10*log10(psd));
xlabel('Frequency (Hz)');
ylabel('Power/Frequency (dB/Hz)');
title('Power Spectral Density of the Signal');
grid on;
\end{lstlisting}

The `periodogram` function in MATLAB estimates the PSD of the signal, and the plot shows how the power is distributed across different frequencies.

\section{Filtering and Noise Reduction}

Signal filtering and noise reduction are essential steps in processing FMCW radar signals. Filters help remove unwanted components, such as noise, while preserving important signal features.

\subsection{Designing Filters in MATLAB}

MATLAB provides several ways to design filters, including low-pass, high-pass, and band-pass filters, which are useful for isolating specific frequency components of radar signals. The `designfilt` \cite{dong2019developing} function can be used for this purpose.

\begin{lstlisting}[style=matlab]
% Designing a Low-Pass Filter in MATLAB
fs = 1000;                     % Sampling frequency
lowpass_filter = designfilt('lowpassfir', 'PassbandFrequency', 150, ...
                            'StopbandFrequency', 200, 'SampleRate', fs);

% Visualize the filter response
fvtool(lowpass_filter);
\end{lstlisting}

The `fvtool` function provides a graphical representation of the filter's frequency response, making it easy to see the cutoff frequencies.

\subsection{Applying Filters to Radar Signals}

Once a filter is designed, it can be applied to radar signals using the `filter` function. Below is an example of applying a low-pass filter to a noisy radar signal.

\begin{lstlisting}[style=matlab]
% Applying the Low-Pass Filter to a Noisy Radar Signal
noisy_signal = cos(2*pi*100*t) + 0.5*randn(size(t));  % Noisy signal
filtered_signal = filter(lowpass_filter, noisy_signal);

% Plot the original and filtered signals
figure;
plot(t, noisy_signal, 'r--', 'DisplayName', 'Noisy Signal');
hold on;
plot(t, filtered_signal, 'b-', 'DisplayName', 'Filtered Signal');
legend;
xlabel('Time (s)');
ylabel('Amplitude');
title('Filtered Radar Signal using Low-Pass Filter');
\end{lstlisting}

In this example, we apply the low-pass filter designed earlier to remove high-frequency noise, improving the quality of the radar signal.

\chapter{MATLAB for FMCW Radar System Simulation}

\section{Simulating FMCW Waveforms}

Frequency Modulated Continuous Wave \cite{zheng2004analysis} (FMCW) radar transmits signals whose frequency increases linearly over time during a specific duration, known as the sweep period. The key parameters for generating FMCW waveforms include:

\begin{itemize}
    \item \textbf{Start frequency} $(f_0)$: The initial frequency of the transmitted signal.
    \item \textbf{End frequency} $(f_1)$: The maximum frequency of the transmitted signal.
    \item \textbf{Sweep bandwidth} $(B)$: The difference between the start and end frequencies $(B = f_1 - f_0)$.
    \item \textbf{Sweep time} $(T_{sweep})$: The time it takes to sweep from the start frequency to the end frequency.
\end{itemize}

The formula for the frequency of an FMCW signal at any time $t$ is:

\[
f(t) = f_0 + \frac{B}{T_{sweep}} t, \quad 0 \leq t \leq T_{sweep}
\]

Here is how you can generate an FMCW waveform in MATLAB:

\begin{lstlisting}[style=matlab]
% Parameters for FMCW waveform
fs = 1e6; % Sampling frequency in Hz
T_sweep = 1e-3; % Sweep time in seconds (1 ms)
f_start = 77e9; % Start frequency (77 GHz)
B = 200e6; % Sweep bandwidth (200 MHz)
f_end = f_start + B; % End frequency

% Time vector
t = 0:1/fs:T_sweep-1/fs;

% FMCW chirp signal generation using linear frequency modulation
fmcw_signal = chirp(t, f_start, T_sweep, f_end);

% Plot the waveform
figure;
plot(t*1e3, fmcw_signal);
xlabel('Time (ms)');
ylabel('Amplitude');
title('FMCW Chirp Signal');
\end{lstlisting}

In this code:

\begin{itemize}
    \item The \texttt{chirp()} function generates the FMCW signal.
    \item The time vector \texttt{t} is created based on the sampling frequency and the sweep time.
    \item The resulting signal is plotted, showing the time-domain representation of the FMCW waveform.
\end{itemize}

\section{Radar Range and Velocity Estimation}

FMCW radar works by measuring the time delay and frequency shift between the transmitted and received signals. This section will discuss the key concepts for estimating range and velocity from the received signals.

\subsection{Calculating Range Using Beat Frequency}

When a radar signal reflects off a target, the returned signal will have a time delay that corresponds to the distance of the target. In FMCW radar, this time delay manifests as a frequency difference between the transmitted and received signals, known as the \textbf{beat frequency}.

The beat frequency is proportional to the range of the target:

\[
f_b = \frac{2BR}{cT_{sweep}}
\]

Where:
\begin{itemize}
    \item $f_b$ is the beat frequency.
    \item $B$ is the sweep bandwidth.
    \item $R$ is the range to the target.
    \item $c$ is the speed of light ($3 \times 10^8$ m/s).
    \item $T_{sweep}$ is the sweep time.
\end{itemize}

Rearranging the equation to solve for the range:

\[
R = \frac{f_b c T_{sweep}}{2B}
\]

Here is how you can calculate the range of a target using the beat frequency in MATLAB:

\begin{lstlisting}[style=matlab]
% Given parameters
B = 200e6; % Bandwidth (Hz)
T_sweep = 1e-3; % Sweep time (seconds)
c = 3e8; % Speed of light (m/s)
f_b = 50e3; % Beat frequency (Hz)

% Calculate target range
R = (f_b * c * T_sweep) / (2 * B);

% Display the result
fprintf('Estimated target range: %.2f meters\n', R);
\end{lstlisting}

\subsection{Velocity Estimation via Doppler Effect}

The Doppler effect \cite{neipp2003analysis}causes a frequency shift in the received signal due to the relative motion between the radar and the target. For an FMCW radar, the Doppler frequency shift $f_D$ is related to the velocity of the target:

\[
f_D = \frac{2v f_0}{c}
\]

Where:
\begin{itemize}
    \item $f_D$ is the Doppler frequency.
    \item $v$ is the relative velocity between the radar and the target.
    \item $f_0$ is the carrier frequency (start frequency).
    \item $c$ is the speed of light.
\end{itemize}

The velocity of the target can be estimated as:

\[
v = \frac{f_D c}{2 f_0}
\]

Here is how you can calculate the velocity of a target using Doppler frequency in MATLAB:

\begin{lstlisting}[style=matlab]
% Given parameters
f_D = 1e3; % Doppler frequency (Hz)
f_0 = 77e9; % Carrier frequency (Hz)

% Calculate target velocity
v = (f_D * c) / (2 * f_0);

% Display the result
fprintf('Estimated target velocity: %.2f m/s\n', v);
\end{lstlisting}

\section{Simulating Multi-Target Scenarios}

In real-world radar scenarios, there are often multiple targets present. To simulate such situations in MATLAB, we can generate multiple reflected signals corresponding to each target's range and velocity. Each target will have its own time delay (which affects the beat frequency) and Doppler shift (which affects the frequency modulation).

Here's an example where we simulate two targets with different ranges and velocities:

\begin{lstlisting}[style=matlab]
% Define target parameters
ranges = [50, 150]; % Target ranges in meters
velocities = [30, -20]; % Target velocities in m/s
f_b_targets = (2 * B * ranges) / (c * T_sweep); % Beat frequencies
f_D_targets = (2 * velocities * f_0) / c; % Doppler frequencies

% Generate received signals for each target
received_signal = zeros(size(t));
for i = 1:length(ranges)
    % Received signal is delayed and frequency shifted
    received_signal = received_signal + ...
        chirp(t - (2 * ranges(i)) / c, f_start, T_sweep, f_end) .* ...
        exp(1j * 2 * pi * f_D_targets(i) * t);
end

% Plot the simulated received signal
figure;
plot(t*1e3, real(received_signal));
xlabel('Time (ms)');
ylabel('Amplitude');
title('Simulated Multi-Target Received Signal');
\end{lstlisting}

This code generates a signal that simulates the reflections from two targets with different ranges and velocities.

\section{Radar Data Post-Processing and Visualization}

Post-processing of radar data typically involves transforming the received signal into a more interpretable form, such as a range-Doppler map or spectrogram. MATLAB provides several tools for visualizing the results of radar signal processing.

\subsection{Visualizing Radar Spectrograms}

A radar spectrogram displays how the frequency content of the received signal changes over time, which can be used to identify moving targets and distinguish between them.

Here is how to generate a spectrogram in MATLAB:

\begin{lstlisting}[style=matlab]
% Compute the spectrogram of the received signal
window_size = 256; % Size of the window for the STFT
noverlap = 200; % Number of overlapping samples
nfft = 512; % Number of FFT points

figure;
spectrogram(received_signal, window_size, noverlap, nfft, fs, 'yaxis');
title('Radar Signal Spectrogram');
\end{lstlisting}

The \texttt{spectrogram()} function generates a time-frequency plot showing how the frequency content of the signal varies over time.

\subsection{3D Visualization of Radar Data}

For visualizing the position and trajectory of multiple targets in 3D, MATLAB's \texttt{plot3()} function is useful. You can plot the estimated positions of targets over time to see how their positions change in space.

\begin{lstlisting}[style=matlab]
% Define target positions and trajectories
target_positions = [ranges; zeros(1, length(ranges)); velocities];

% 3D plot of target motion over time
figure;
plot3(target_positions(1,:), target_positions(2,:), target_positions(3,:));
xlabel('Range (m)');
ylabel('Cross-Range (m)');
zlabel('Velocity (m/s)');
title('3D Visualization of Target Trajectories');
\end{lstlisting}

This provides a 3D view of the motion of the targets as detected by the radar system.
\chapter{Advanced MATLAB Techniques for Radar Processing}

As radar signal processing becomes more advanced and complex, there is a growing need for efficient computation and real-time processing capabilities. In this chapter, we will explore advanced MATLAB techniques that can be applied to large-scale radar data and real-time systems, helping you handle these demands effectively.

\section{Parallel Computing in MATLAB}

When dealing with large-scale radar data, processing can become computationally expensive. MATLAB provides powerful tools for parallel computing \cite{sharma2009matlab}that allow users to take advantage of multi-core processors, GPUs, and distributed computing environments to speed up data processing tasks.

\subsection{Using Parallel for Loops}

In radar systems, you might need to perform repeated calculations over large datasets. Using MATLAB's \texttt{parfor} loop (parallel for loop) is a simple yet effective way to parallelize these operations, which can drastically reduce computation time.

For example, if you are processing multiple radar returns in parallel, the traditional \texttt{for} loop can be replaced with a \texttt{parfor} loop. Below is an example where we apply \texttt{parfor} to process radar signals in parallel:

\begin{lstlisting}[style=matlab]
% Serial version
for i = 1:1000
    processed_signal = processRadarSignal(raw_signal(i, :));
end

% Parallel version using parfor
parfor i = 1:1000
    processed_signal = processRadarSignal(raw_signal(i, :));
end
\end{lstlisting}

By using \texttt{parfor}, MATLAB distributes the iterations of the loop across multiple workers (cores), allowing the computations to be executed concurrently, thus speeding up the overall process.

It is important to ensure that the operations inside the \texttt{parfor} loop are independent, as parallel loops cannot handle interdependent operations efficiently.

\subsection{Working with Distributed Arrays}

For very large radar datasets that may not fit into the memory of a single machine, MATLAB's distributed arrays allow data to be spread across multiple machines or clusters. This is particularly useful in scenarios such as processing high-resolution radar data over long time durations.

To create a distributed array, you can use the \texttt{distributed()} function. Here is an example of how to initialize and work with distributed arrays for radar data:

\begin{lstlisting}[style=matlab]
% Create a large random matrix
large_data = rand(10000, 10000);

% Distribute the matrix across a computing cluster
D = distributed(large_data);

% Perform an operation on the distributed array
result = sum(D, 2);  % Summing over each row in parallel
\end{lstlisting}

By distributing the array, each worker holds a portion of the data, allowing for more efficient memory usage and faster computation times for large-scale radar processing tasks. MATLAB automatically handles the communication between workers and the aggregation of results.

\section{MATLAB Optimization Toolbox}

Optimization plays a critical role in radar system design and operation. MATLAB's Optimization Toolbox offers a wide range of algorithms to solve problems such as radar waveform design, parameter estimation, and resource allocation.

Here is an example of optimizing a simple objective function in the context of waveform design for radar:

\begin{lstlisting}[style=matlab]
% Define the objective function for waveform optimization
objFun = @(x) -performanceMetric(x);  % Minimizing negative performance metric

% Define initial guess for parameters
x0 = [0.5, 0.5, 0.5];

% Set optimization options
options = optimoptions('fmincon', 'Display', 'iter');

% Run the optimization
[x_opt, fval] = fmincon(objFun, x0, [], [], [], [], lb, ub, [], options);

% Display the optimized waveform parameters
disp(x_opt);
\end{lstlisting}

In this example, \texttt{fmincon} is used to optimize radar waveform parameters to maximize a performance metric (e.g., detection probability, range resolution). The Optimization Toolbox provides several algorithms, including gradient-based methods and global optimization techniques, making it versatile for different radar system design challenges.

\section{Real-time Data Acquisition and Processing}

Real-time radar data acquisition and processing are critical in applications such as automotive radar, surveillance, and target tracking. MATLAB provides interfaces to connect with radar hardware and perform real-time processing using the Data Acquisition Toolbox and Instrument Control Toolbox.

\subsection{Connecting to Radar Hardware}
To acquire real-time data from radar hardware, you can use MATLAB's data acquisition tools. Here is an example of setting up a real-time data acquisition session:

\begin{lstlisting}[style=matlab]
% Create a data acquisition session
daqSession = daq.createSession('ni');

% Add analog input channels to acquire radar data
addAnalogInputChannel(daqSession, 'Dev1', 'ai0', 'Voltage');

% Set the acquisition rate
daqSession.Rate = 10000;

% Acquire data in real-time
data = startForeground(daqSession);

% Plot the acquired radar data
plot(data);
xlabel('Time (s)');
ylabel('Amplitude');
title('Real-time Radar Signal Acquisition');
\end{lstlisting}

In this example, we set up a session to acquire analog radar signal data from a National Instruments (NI) device. The data is then processed in real-time and visualized.

\subsection{Real-time Signal Processing}
Once the data is acquired in real-time, you can process it immediately using various signal processing algorithms. For example, applying an FFT (Fast Fourier Transform) to the real-time radar data:

\begin{lstlisting}[style=matlab]
% Perform real-time FFT on the acquired data
fftData = fft(data);

% Plot the frequency spectrum
f = (0:length(fftData)-1)*daqSession.Rate/length(fftData);
plot(f, abs(fftData));
xlabel('Frequency (Hz)');
ylabel('Magnitude');
title('Frequency Spectrum of Real-time Radar Signal');
\end{lstlisting}

This allows you to visualize the frequency components of the real-time radar signal, which is crucial in many radar applications such as Doppler processing and range estimation.

By combining data acquisition and signal processing in real-time, MATLAB enables rapid prototyping and testing of radar systems, ensuring that you can analyze and process data as it is being collected.

\part{Deep Learning for Radar Signal Processing}

\chapter{Introduction to Deep Learning in Radar Systems}

Deep learning introduces new methodologies and applications in radar systems. Compared to traditional signal processing algorithms, deep learning techniques, particularly Convolutional Neural Networks (CNN), Long Short-Term Memory (LSTM) networks, and Transformer models, offer significant advantages when handling the complex non-linear and high-dimensional features of radar data. By leveraging neural networks' ability to automatically extract features, deep learning can efficiently process large-scale data without the need for manual feature engineering, often outperforming traditional methods in tasks such as target detection, tracking, and classification.

Deep learning also provides new approaches for point cloud data analysis, with models like PointNet and PointNet++ directly processing point cloud data. These models have shown remarkable success in radar signal processing, particularly for 3D spatial localization and object classification tasks. Additionally, Transformer models, known for their superiority in handling sequence data and understanding global context, are increasingly being applied to radar and point cloud data analysis, offering more flexible and efficient solutions for complex radar scenarios.

\section{Why and How Use Deep Learning for Radar Signal Processing?}

Radar systems, particularly Frequency Modulated Continuous Wave (FMCW) radars, are essential in various applications such as automotive safety, aerospace, maritime navigation, and surveillance. Traditional radar signal processing techniques have served these applications well, but they have limitations when dealing with complex environments and large volumes of data. Deep learning offers powerful tools to overcome these challenges by automatically extracting features and learning intricate patterns from data \cite{lang2020comprehensive}.

In the field of deep learning, while MATLAB provides some toolboxes and functions, it is relatively limited compared to Python, especially in terms of community support, framework diversity (such as TensorFlow, PyTorch), and model deployment. Python offers broader applications and flexibility, so it is recommended to use Python for deep learning development in real projects. However, in many cases, the data might initially be stored in the MATLAB environment, so we need to export the data from MATLAB to a format that Python can read, commonly saving it as a `.mat` file, and then reading that file in Python for processing.

MATLAB `.mat` files come in two formats, `v7.3` and earlier versions (such as `v7`), and the choice depends on the data size:

\begin{itemize}
    \item \textbf{v7.3 and later versions} use the HDF5 format and are suitable for files larger than 2GB. This format is more complex but is ideal for handling large-scale data.
    \item \textbf{v7 and earlier versions} are suitable for data smaller than 2GB, and the files saved in this format are simpler and can be directly read by Python using \texttt{scipy.io.loadmat}.
\end{itemize}

Therefore, it is recommended to save the data in the \textbf{v7} format if the size allows, as it can be easily read using \texttt{scipy}. If the data exceeds 2GB, \texttt{h5py} can be used to read the \textbf{v7.3} format files.

Below is an example of how to save the data in MATLAB in different formats and read it in Python accordingly:

\begin{lstlisting}[style=matlab]
% MATLAB code - Saving data as a .mat file in v7 format
data = rand(100, 10);  % Example data
labels = randi([0, 1], 100, 1);  % Example labels
save('dataset_v7.mat', 'data', 'labels');  % Save in v7 format (default, <2GB)

% MATLAB code - Saving data as a .mat file in v7.3 format (for files >2GB)
save('dataset_v7_3.mat', 'data', 'labels', '-v7.3');  % Specify saving in v7.3 format
\end{lstlisting}

Next, we can read the corresponding `.mat` file in Python:

\begin{lstlisting}[style=python]
# Python code - Reading a .mat file in v7 format
import scipy.io as sio

# Load the v7 format .mat file
mat_contents = sio.loadmat('dataset_v7.mat')

# Extract the data and labels
data = mat_contents['data']
labels = mat_contents['labels']

print(data.shape)  # Output the shape of the data
print(labels.shape)  # Output the shape of the labels
\end{lstlisting}

If the file is saved in \textbf{v7.3} format (for data larger than 2GB), \texttt{h5py} can be used to read the file:

\begin{lstlisting}[style=python]
# Python code - Reading a .mat file in v7.3 format
import h5py

# Load the v7.3 format .mat file
with h5py.File('dataset_v7_3.mat', 'r') as mat_file:
    data = mat_file['data'][:]
    labels = mat_file['labels'][:]

print(data.shape)  # Output the shape of the data
print(labels.shape)  # Output the shape of the labels
\end{lstlisting}

By using this approach, we can seamlessly import data from MATLAB into Python and choose the appropriate reading method based on the data size, making full use of Python's powerful deep learning frameworks for model training and evaluation. For data smaller than 2GB, it is recommended to use \texttt{scipy.io.loadmat} due to its simplicity and efficiency.

\subsection{Limitations of Traditional Radar Signal Processing}

Traditional radar signal processing relies on well-established methods like the Fourier Transform, matched filtering, and Doppler processing. While effective, these methods have several limitations:

\begin{itemize}
    \item \textbf{Manual Feature Extraction}: Requires domain expertise to design features that capture relevant information from radar signals.
    \item \textbf{Difficulty with Non-linear Patterns}: Struggles to model complex, non-linear relationships inherent in real-world scenarios.
    \item \textbf{Sensitivity to Noise and Clutter}: Performance degrades in environments with high noise levels or cluttered backgrounds.
    \item \textbf{Scalability Issues}: Inefficient when processing large-scale data or real-time applications due to computational constraints.
\end{itemize}

\subsection{Advantages of Deep Learning in Radar Systems}

Deep learning addresses these limitations by offering the following advantages:

\begin{itemize}
    \item \textbf{Automatic Feature Learning}: Neural networks can automatically learn relevant features from raw data, reducing the need for manual feature engineering.
    \item \textbf{Handling Complex and Non-linear Relationships}: Capable of modeling complex patterns and relationships within the data.
    \item \textbf{Robustness to Noise and Clutter}: Improves detection and classification performance in challenging environments.
    \item \textbf{Scalability and Real-time Processing}: Optimized architectures can handle large datasets and meet real-time processing requirements.
\end{itemize}

\subsection{Applications of Deep Learning in FMCW Radar}

Deep learning techniques have been successfully applied to various FMCW radar signal processing tasks:

\begin{itemize}
    \item \textbf{Target Detection}: Identifying objects within the radar's field of view.
    \item \textbf{Classification}: Determining the type of detected objects (e.g., vehicles, pedestrians, cyclists).
    \item \textbf{Human Activity Recognition (HAR)}: Understanding human movements and activities for applications like security monitoring and human-computer interaction.
    \item \textbf{Gesture Recognition}: Interpreting specific gestures for control systems or interactive applications.
\end{itemize}

\subsection{Deep Learning Techniques for Radar Signal Processing}

Several deep learning models are particularly well-suited for radar signal processing:

\begin{itemize}
    \item \textbf{Convolutional Neural Networks (CNNs)}: Excellent for processing two-dimensional radar data like range-Doppler maps.
    \item \textbf{Recurrent Neural Networks (RNNs) and LSTMs}: Effective for sequential radar data and time-series analysis.
    \item \textbf{Transformers}: Provide advanced capabilities for processing sequential data without the limitations of traditional RNNs.
    \item \textbf{PointNet and PointNet++}: Specialized for processing point cloud data generated by radar systems.
\end{itemize}

\subsection{Example: Using a CNN for Target Classification}

Let's walk through an example of using a CNN to classify targets detected by an FMCW radar.

\subsubsection{Data Preparation}

Assume we have a dataset of radar signals represented as range-Doppler maps. Each map is associated with a label indicating the type of target.

\begin{lstlisting}[style=python]
import numpy as np
import torch
from torch.utils.data import Dataset, DataLoader

class RadarDataset(Dataset):
    def __init__(self, data_files, labels):
        self.data = [np.load(file) for file in data_files]
        self.labels = labels

    def __len__(self):
        return len(self.data)

    def __getitem__(self, idx):
        x = torch.tensor(self.data[idx], dtype=torch.float32)
        y = torch.tensor(self.labels[idx], dtype=torch.long)
        return x.unsqueeze(0), y  # Add channel dimension
\end{lstlisting}

\subsubsection{Defining the CNN Model}

\begin{lstlisting}[style=python]
import torch.nn as nn
import torch.nn.functional as F

class RadarCNN(nn.Module):
    def __init__(self):
        super(RadarCNN, self).__init__()
        self.conv1 = nn.Conv2d(1, 16, 3)  # Input channels, output channels, kernel size
        self.conv2 = nn.Conv2d(16, 32, 3)
        self.fc1 = nn.Linear(32 * 6 * 6, 128)  # Adjust dimensions accordingly
        self.fc2 = nn.Linear(128, num_classes)

    def forward(self, x):
        x = F.relu(self.conv1(x))      # First convolutional layer
        x = F.max_pool2d(x, 2)         # First pooling layer
        x = F.relu(self.conv2(x))      # Second convolutional layer
        x = F.max_pool2d(x, 2)         # Second pooling layer
        x = x.view(-1, 32 * 6 * 6)     # Flatten the tensor
        x = F.relu(self.fc1(x))        # First fully connected layer
        x = self.fc2(x)                # Output layer
        return x
\end{lstlisting}

\subsubsection{Training the Model}

\begin{lstlisting}[style=python]
model = RadarCNN()
criterion = nn.CrossEntropyLoss()
optimizer = torch.optim.Adam(model.parameters(), lr=0.001)

train_loader = DataLoader(radar_dataset, batch_size=32, shuffle=True)

num_epochs = 10
for epoch in range(num_epochs):
    total_loss = 0
    for inputs, labels in train_loader:
        optimizer.zero_grad()
        outputs = model(inputs)
        loss = criterion(outputs, labels)
        loss.backward()
        optimizer.step()
        total_loss += loss.item()
    print(f"Epoch {epoch+1}/{num_epochs}, Loss: {total_loss/len(train_loader)}")
\end{lstlisting}

\subsection{Example: Human Activity Recognition with LSTM Networks}

For tasks involving temporal dynamics, such as Human Activity Recognition, LSTM networks are more suitable.

\subsubsection{Implementing an LSTM Model}

\begin{lstlisting}[style=python]
class RadarLSTM(nn.Module):
    def __init__(self, input_size, hidden_size, num_layers, num_classes):
        super(RadarLSTM, self).__init__()
        self.lstm = nn.LSTM(input_size, hidden_size, num_layers, batch_first=True)
        self.fc = nn.Linear(hidden_size, num_classes)

    def forward(self, x):
        h0 = torch.zeros(num_layers, x.size(0), hidden_size)
        c0 = torch.zeros(num_layers, x.size(0), hidden_size)
        out, _ = self.lstm(x, (h0, c0))
        out = self.fc(out[:, -1, :])  # Use output from the last time step
        return out
\end{lstlisting}

\subsubsection{Training the LSTM Model}

\begin{lstlisting}[style=python]
model = RadarLSTM(input_size, hidden_size, num_layers, num_classes)
criterion = nn.CrossEntropyLoss()
optimizer = torch.optim.Adam(model.parameters(), lr=0.001)

train_loader = DataLoader(radar_dataset, batch_size=32, shuffle=True)

num_epochs = 15
for epoch in range(num_epochs):
    total_loss = 0
    for sequences, labels in train_loader:
        optimizer.zero_grad()
        outputs = model(sequences)
        loss = criterion(outputs, labels)
        loss.backward()
        optimizer.step()
        total_loss += loss.item()
    print(f"Epoch {epoch+1}/{num_epochs}, Loss: {total_loss/len(train_loader)}")
\end{lstlisting}

\subsection{Advanced Techniques: Transformers and PointNet}

\subsubsection{Transformers for Sequential Data}

Transformers can process sequential data without the need for recurrent connections, making them efficient for radar signal processing.

\begin{lstlisting}[style=python]
from torch.nn import Transformer

class RadarTransformer(nn.Module):
    def __init__(self, input_size, num_classes):
        super(RadarTransformer, self).__init__()
        self.transformer = Transformer(d_model=input_size, nhead=8, num_encoder_layers=6)
        self.fc = nn.Linear(input_size, num_classes)

    def forward(self, x):
        x = self.transformer(x)
        x = self.fc(x[:, -1, :])  # Use the output of the last time step
        return x
\end{lstlisting}

\subsubsection{PointNet for Point Cloud Data}

PointNet is designed to handle point cloud data, which can be generated from radar signals for 3D representation.

\begin{lstlisting}[style=python]
class PointNet(nn.Module):
    def __init__(self, num_classes):
        super(PointNet, self).__init__()
        self.fc1 = nn.Linear(3, 64)
        self.fc2 = nn.Linear(64, 128)
        self.fc3 = nn.Linear(128, 1024)
        self.fc4 = nn.Linear(1024, num_classes)

    def forward(self, x):
        x = F.relu(self.fc1(x))    # Input is (batch_size, num_points, 3)
        x = F.relu(self.fc2(x))
        x = F.relu(self.fc3(x))
        x = F.max_pool1d(x, x.size(-1))
        x = x.view(-1, 1024)
        x = self.fc4(x)
        return x
\end{lstlisting}

\subsection{Comparing Traditional Methods with Deep Learning}

\begin{table}[h]
\centering
\begin{tabular}{|l|l|l|}
\hline
\textbf{Aspect} & \textbf{Traditional Methods} & \textbf{Deep Learning} \\ \hline
Feature Extraction & Manual & Automatic \\ \hline
Non-linear Modeling & Limited & Strong \\ \hline
Scalability & Less Scalable & Highly Scalable \\ \hline
Noise Robustness & Sensitive & Robust \\ \hline
\end{tabular}
\caption{Comparison between Traditional Methods and Deep Learning in Radar Signal Processing}
\end{table}

\subsection{Tree Diagram: Deep Learning Techniques in Radar Signal Processing}

\begin{itemize}
    \item \textbf{Deep Learning Techniques}
    \begin{itemize}
        \item \textbf{Convolutional Neural Networks}
        \begin{itemize}
            \item Target Detection
            \item Classification
        \end{itemize}
        \item \textbf{Recurrent Neural Networks}
        \begin{itemize}
            \item LSTM Networks
            \begin{itemize}
                \item Human Activity Recognition
            \end{itemize}
        \end{itemize}
        \item \textbf{Transformers}
        \begin{itemize}
            \item Sequential Data Processing
        \end{itemize}
        \item \textbf{PointNet Models}
        \begin{itemize}
            \item Point Cloud Analysis
        \end{itemize}
    \end{itemize}
\end{itemize}

\subsection{Conclusion}

Integrating deep learning into FMCW radar signal processing unlocks new possibilities for enhanced performance and functionality. By overcoming the limitations of traditional methods, deep learning enables:

\begin{itemize}
    \item Improved accuracy in target detection and classification.
    \item Real-time processing capabilities for dynamic environments.
    \item The ability to learn and adapt to new patterns and scenarios.
\end{itemize}

As we continue exploring deep learning techniques, their applications in radar systems will expand, leading to more intelligent and autonomous systems.

\chapter{Convolutional Neural Networks (CNN) for Radar Data}

Radar data, such as Range-Doppler Maps and radar images, can be efficiently processed using Convolutional Neural Networks (CNNs). CNNs, with their ability to handle high-dimensional data and automatically learn features, are particularly well-suited for radar signal processing tasks, including target detection, classification, and human activity recognition.

\section{Principles of Convolutional Neural Networks}

Convolutional Neural Networks (CNNs) are a type of deep learning model specifically designed for processing grid-like data structures, such as images. Radar data, when transformed into a two-dimensional form, such as a Range-Doppler Map, can be processed similarly to image data. CNNs consist of three main types of layers: convolutional layers, pooling layers, and fully connected layers.

\subsection{Convolutional Layers}

The core component of a CNN is the convolutional layer. This layer applies a set of learnable filters (or kernels) to the input data, generating feature maps. The filters slide over the input image (or radar map), performing a dot product between the filter weights and the input values. This process is known as convolution. The output of a convolutional layer highlights important features in the data, such as edges in an image or patterns in a radar map.

\begin{lstlisting}[style=python]
import torch
import torch.nn as nn

class BasicCNN(nn.Module):
    def __init__(self):
        super(BasicCNN, self).__init__()
        self.conv1 = nn.Conv2d(1, 16, kernel_size=3, stride=1, padding=1)
        self.conv2 = nn.Conv2d(16, 32, kernel_size=3, stride=1, padding=1)
        self.fc1 = nn.Linear(32 * 8 * 8, 128)
        self.fc2 = nn.Linear(128, 2)  # Binary classification

    def forward(self, x):
        x = nn.functional.relu(self.conv1(x))
        x = nn.functional.max_pool2d(x, 2)
        x = nn.functional.relu(self.conv2(x))
        x = nn.functional.max_pool2d(x, 2)
        x = x.view(-1, 32 * 8 * 8)  # Flatten the tensor
        x = nn.functional.relu(self.fc1(x))
        x = self.fc2(x)
        return x
\end{lstlisting}

\subsection{Pooling Layers}

After a convolutional layer, pooling layers are typically added to reduce the spatial dimensions of the feature maps, which decreases the computational complexity and helps prevent overfitting. The most common pooling operation is max-pooling, where the maximum value is selected from a set of neighboring values.

\subsection{Fully Connected Layers}

The final layers in a CNN are fully connected (FC) layers. These layers take the flattened feature maps from the last convolutional or pooling layer and perform classification. The output of the fully connected layer corresponds to the classification categories, such as "vehicle" or "pedestrian" in a radar system.

\section{CNN for Range-Doppler Maps and Radar Images}

Range-Doppler Maps are two-dimensional representations of radar data, where one axis represents the distance (range) to the target, and the other represents the velocity (Doppler shift) of the target. CNNs are highly effective for processing these two-dimensional radar maps, much like they are used for image processing.

\subsection{Weight Sharing in CNN for Radar Data}

One of the key advantages of CNNs is the weight-sharing mechanism. In traditional neural networks, each connection between neurons has its own weight, which can lead to a very large number of parameters for high-dimensional inputs like images or radar data. In CNNs, the same set of weights (filters) is applied across the entire input, significantly reducing the number of parameters and making the model more efficient.

\begin{itemize}
    \item \textbf{Reduction in Parameters}: By sharing weights across different regions of the input, CNNs are able to capture spatial hierarchies in data (e.g., patterns in radar images) without requiring a large number of parameters.
    \item \textbf{Efficient Processing}: This mechanism allows CNNs to efficiently process high-dimensional radar data, such as Range-Doppler Maps, where the relationships between neighboring pixels (or points in the map) are important.
\end{itemize}

\section{Pre-trained CNN for Radar Applications}

Pre-trained CNN models, such as ResNet, VGG, or MobileNet, can be leveraged for radar data processing through a technique called transfer learning. Transfer learning involves taking a model that has been pre-trained on a large dataset (such as ImageNet) and fine-tuning it on a specific task, such as radar-based target detection. This approach is particularly useful when the available radar dataset is limited.

\subsection{Transfer Learning for Radar Applications}

Transfer learning works by freezing the early layers of a pre-trained CNN (which typically learn general features like edges and textures) and retraining the later layers on the new radar dataset. This allows the model to learn task-specific features (such as radar reflections from vehicles or pedestrians) without the need for a large dataset.

\begin{lstlisting}[style=python]
from torchvision import models

# Load a pre-trained ResNet model
resnet = models.resnet18(pretrained=True)

# Freeze all layers except the last one
for param in resnet.parameters():
    param.requires_grad = False

# Replace the final layer to fit radar data classification
resnet.fc = nn.Linear(resnet.fc.in_features, num_classes)

# Example radar dataset (binary classification)
train_loader = DataLoader(radar_dataset, batch_size=32, shuffle=True)

# Optimizer and loss function
optimizer = torch.optim.Adam(resnet.fc.parameters(), lr=0.001)
criterion = nn.CrossEntropyLoss()

# Training loop
for epoch in range(num_epochs):
    total_loss = 0
    for inputs, labels in train_loader:
        optimizer.zero_grad()
        outputs = resnet(inputs)
        loss = criterion(outputs, labels)
        loss.backward()
        optimizer.step()
        total_loss += loss.item()
    print(f'Epoch {epoch+1}/{num_epochs}, Loss: {total_loss/len(train_loader)}')
\end{lstlisting}

\subsection{Fine-tuning a Pre-trained Model for Radar Classification}

Fine-tuning involves not only training the last few layers of the pre-trained model but also adjusting some of the earlier layers to better adapt to the specific characteristics of radar data.

\begin{itemize}
    \item \textbf{Step 1}: Load a pre-trained CNN model (e.g., ResNet, VGG).
    \item \textbf{Step 2}: Replace the final fully connected layer to match the number of classes in the radar classification task.
    \item \textbf{Step 3}: Freeze early layers to preserve general features learned from the pre-trained model.
    \item \textbf{Step 4}: Fine-tune the final layers using radar data to adapt the model to radar-specific tasks, such as human activity recognition (HAR) or object classification.
\end{itemize}

\subsection{Tree Diagram: CNN in Radar Applications}

\begin{itemize}
    \item \textbf{CNN Applications in Radar}
    \begin{itemize}
        \item \textbf{Range-Doppler Maps}
        \begin{itemize}
            \item Target Detection
            \item Classification
        \end{itemize}
        \item \textbf{Radar Images}
        \begin{itemize}
            \item Gesture Recognition
            \item Human Activity Recognition
        \end{itemize}
        \item \textbf{Pre-trained CNN Models}
        \begin{itemize}
            \item Transfer Learning
        \end{itemize}
    \end{itemize}
\end{itemize}

\subsection{Conclusion}

Convolutional Neural Networks (CNNs) are highly effective for processing radar data, particularly in applications involving Range-Doppler Maps and radar images. By leveraging weight sharing, CNNs are able to efficiently process high-dimensional radar data with a reduced number of parameters. Furthermore, pre-trained CNN models can be fine-tuned using transfer learning techniques, enabling accurate radar signal processing even with limited data.

\chapter{Long Short-Term Memory (LSTM) Networks for Temporal Radar Data}

Radar data is often collected over time, capturing dynamic changes in the environment. Traditional neural networks struggle to retain information over long sequences, which is crucial for understanding temporal dependencies in radar data. Long Short-Term Memory (LSTM) networks address this problem by introducing memory cells that can retain information for long durations, making them suitable for radar signal processing tasks involving time-series data, such as waveforms or sequences of Range-Doppler Maps.

\section{Principles of LSTM Networks}

LSTM networks are a type of Recurrent Neural Network (RNN) designed to overcome the limitations of standard RNNs in capturing long-term dependencies. In standard RNNs, as information passes through the network, it tends to fade, making it difficult to retain information across long sequences. LSTMs solve this issue by using special structures called \textbf{gates}, which control the flow of information.

\subsection{The Structure of an LSTM Cell}

Each LSTM cell has three gates:
\begin{itemize}
    \item \textbf{Input Gate}: Controls which part of the new information should be added to the memory cell.
    \item \textbf{Forget Gate}: Decides which part of the old information should be discarded from the memory cell.
    \item \textbf{Output Gate}: Determines which part of the memory cell's content will be passed to the next time step.
\end{itemize}

These gates work together to retain long-term information while selectively forgetting irrelevant information. This ability to control memory makes LSTMs highly suitable for processing time-series radar data, which often contains both short-term and long-term dependencies.

\subsection{LSTM for Radar Signal Processing}

Radar systems, such as those using Frequency Modulated Continuous Wave (FMCW), generate time-series data that includes the motion and distance of objects over time. LSTM networks are ideal for analyzing this temporal data because they can learn patterns that evolve over time, such as human activity, object tracking, or signal modulation.

\begin{lstlisting}[style=python]
import torch
import torch.nn as nn

class RadarLSTM(nn.Module):
    def __init__(self, input_size, hidden_size, num_layers, num_classes):
        super(RadarLSTM, self).__init__()
        self.lstm = nn.LSTM(input_size, hidden_size, num_layers, batch_first=True)
        self.fc = nn.Linear(hidden_size, num_classes)

    def forward(self, x):
        h0 = torch.zeros(num_layers, x.size(0), hidden_size)  # Initial hidden state
        c0 = torch.zeros(num_layers, x.size(0), hidden_size)  # Initial cell state
        out, _ = self.lstm(x, (h0, c0))  # LSTM forward pass
        out = self.fc(out[:, -1, :])  # Output from the last time step
        return out

# Example: Processing time-series radar data
input_size = 64  # Assume 64 features per time step
hidden_size = 128  # Number of LSTM units
num_layers = 2  # Number of LSTM layers
num_classes = 2  # Binary classification (e.g., object or no object)
model = RadarLSTM(input_size, hidden_size, num_layers, num_classes)

# Example input: batch_size=10, sequence_length=100, input_size=64
radar_data = torch.randn(10, 100, input_size)
output = model(radar_data)
print(output.shape)  # Output shape: (10, 2)
\end{lstlisting}

In this example, the LSTM network is designed to process time-series radar data with 64 features at each time step, and it outputs a classification decision after processing the entire sequence.

\section{Combining CNN and LSTM for Spatio-Temporal Radar Data}

In many radar applications, the data has both spatial and temporal components. For instance, a sequence of Range-Doppler Maps captures the spatial distribution of targets, while the sequence over time shows how these targets move. To handle such spatio-temporal data, a common approach is to combine CNNs, which are effective at processing spatial information, with LSTMs, which excel at handling temporal dependencies.

\subsection{CNN+LSTM Model for Human Activity Recognition (HAR)}

Human Activity Recognition (HAR) is a common radar application where CNNs and LSTMs are combined. The CNN extracts spatial features from each radar frame (e.g., a Range-Doppler Map), and the LSTM processes the temporal sequence of these frames to recognize activities such as walking, running, or waving.

\begin{lstlisting}[style=python]
class CNN_LSTM(nn.Module):
    def __init__(self, input_size, hidden_size, num_layers, num_classes):
        super(CNN_LSTM, self).__init__()
        self.conv1 = nn.Conv2d(1, 16, kernel_size=3, stride=1, padding=1)
        self.conv2 = nn.Conv2d(16, 32, kernel_size=3, stride=1, padding=1)
        self.lstm = nn.LSTM(32 * 8 * 8, hidden_size, num_layers, batch_first=True)
        self.fc = nn.Linear(hidden_size, num_classes)

    def forward(self, x):
        batch_size, seq_length, c, h, w = x.size()
        x = x.view(-1, c, h, w)  # Reshape for CNN
        x = nn.functional.relu(self.conv1(x))
        x = nn.functional.max_pool2d(x, 2)
        x = nn.functional.relu(self.conv2(x))
        x = nn.functional.max_pool2d(x, 2)
        x = x.view(batch_size, seq_length, -1)  # Reshape for LSTM
        h0 = torch.zeros(num_layers, batch_size, hidden_size)  # Initial hidden state
        c0 = torch.zeros(num_layers, batch_size, hidden_size)  # Initial cell state
        out, _ = self.lstm(x, (h0, c0))  # LSTM forward pass
        out = self.fc(out[:, -1, :])  # Output from the last time step
        return out

# Example: Processing a sequence of radar images for HAR
input_size = 32 * 8 * 8  # After CNN feature extraction
hidden_size = 128  # Number of LSTM units
num_layers = 2  # Number of LSTM layers
num_classes = 5  # HAR classification (e.g., walking, running, etc.)
model = CNN_LSTM(input_size, hidden_size, num_layers, num_classes)

# Example input: batch_size=10, seq_length=30, channels=1, height=32, width=32
radar_sequence = torch.randn(10, 30, 1, 32, 32)
output = model(radar_sequence)
print(output.shape)  # Output shape: (10, 5)
\end{lstlisting}

\subsection{Advantages of CNN+LSTM for Radar Data}

By combining CNNs and LSTMs, the model can effectively capture both the spatial features of individual radar frames and the temporal dependencies across multiple frames. This is particularly useful for tasks that involve both static and dynamic information, such as:
\begin{itemize}
    \item \textbf{Human Activity Recognition (HAR)}: Detecting and classifying different types of human movements based on radar signals.
    \item \textbf{Object Tracking}: Following the movement of targets over time by analyzing sequences of radar frames.
    \item \textbf{Gesture Recognition}: Recognizing hand or body gestures by analyzing temporal patterns in radar data.
\end{itemize}

\subsection{Tree Diagram: CNN+LSTM Model for Radar Data}

\begin{itemize}
    \item \textbf{CNN+LSTM Model}
    \begin{itemize}
        \item \textbf{Convolutional Layers}
        \begin{itemize}
            \item Spatial Feature Extraction
        \end{itemize}
        \item \textbf{LSTM Layers}
        \begin{itemize}
            \item Temporal Sequence Processing
        \end{itemize}
        \item \textbf{Applications}
        \begin{itemize}
            \item Human Activity Recognition
            \item Gesture Recognition
            \item Object Tracking
        \end{itemize}
    \end{itemize}
\end{itemize}

\subsection{Conclusion}

Long Short-Term Memory (LSTM) networks provide an effective way to analyze temporal radar data, handling long-term dependencies in time-series data such as waveforms and Range-Doppler Maps. By combining CNNs with LSTMs, we can build models that leverage both spatial and temporal information, significantly improving performance in radar signal processing tasks like human activity recognition, gesture recognition, and target tracking.

\chapter{PointNet and PointNet++ for 3D Point Cloud Data}

Radar systems, particularly modern FMCW radars, generate point cloud data that captures 3D spatial information. Point clouds are sets of points in a 3D space that represent the surfaces and objects detected by the radar. Unlike traditional image-based data, point clouds are unordered and irregular, making them challenging to process using conventional deep learning models. PointNet \cite{QiPoi17} and PointNet++ are two architectures specifically designed to handle point cloud data, making them well-suited for tasks such as object detection, segmentation, and classification in radar systems.

\section{Introduction to PointNet}

PointNet was one of the first deep learning architectures to directly process point cloud data without converting it into voxel grids or meshes. This is important because converting point clouds to other representations can lead to a loss of information and computational inefficiency. PointNet processes point clouds as unordered sets of points and uses a symmetric function to aggregate information, allowing it to maintain the permutation invariance of the input data.

\subsection{PointNet Architecture}

The PointNet architecture consists of two main parts:
\begin{itemize}
    \item \textbf{Per-point Feature Extraction}: A series of fully connected layers are applied to each point in the cloud independently. This part of the architecture learns features for each point, such as location, intensity, or surface normal.
    \item \textbf{Symmetric Aggregation}: The per-point features are aggregated using a symmetric function, typically a max pooling operation, to generate a global feature vector. This ensures that the model is invariant to the order of points in the point cloud.
\end{itemize}

\begin{lstlisting}[style=python]
import torch
import torch.nn as nn
import torch.nn.functional as F

class PointNet(nn.Module):
    def __init__(self, num_classes):
        super(PointNet, self).__init__()
        self.fc1 = nn.Linear(3, 64)
        self.fc2 = nn.Linear(64, 128)
        self.fc3 = nn.Linear(128, 1024)
        self.fc4 = nn.Linear(1024, 512)
        self.fc5 = nn.Linear(512, 256)
        self.fc6 = nn.Linear(256, num_classes)

    def forward(self, x):
        x = F.relu(self.fc1(x))
        x = F.relu(self.fc2(x))
        x = F.relu(self.fc3(x))
        x = torch.max(x, 1)[0]  # Symmetric max pooling
        x = F.relu(self.fc4(x))
        x = F.relu(self.fc5(x))
        x = self.fc6(x)
        return x
\end{lstlisting}

In this architecture, each point is processed through fully connected layers, and the global feature vector is obtained by max-pooling the features of all points. This vector is then passed through more fully connected layers for classification or segmentation.

\subsection{PointNet for Radar Point Cloud Data}

Radar point clouds, unlike those from LiDAR systems, are often sparse and irregular. This poses a challenge for traditional models, but PointNet excels in this context because of its ability to handle unordered sets of points. PointNet can be applied to tasks such as:
\begin{itemize}
    \item \textbf{Classification}: Identifying objects in the radar point cloud, such as vehicles, pedestrians, or other objects.
    \item \textbf{Segmentation}: Labeling each point in the cloud with a category, such as object or background.
    \item \textbf{Object Detection}: Identifying and localizing objects within the radar's field of view.
\end{itemize}

\section{PointNet++: Hierarchical Point Set Learning}

PointNet++ extends the original PointNet architecture by incorporating a hierarchical structure that captures both local and global features in the point cloud. This is particularly useful for radar point clouds, where local geometric structures play an important role in understanding the environment.

\subsection{Hierarchical Learning in PointNet++}

PointNet++ applies the PointNet architecture recursively to groups of points at multiple scales. This hierarchical approach enables the model to capture local features, such as small object parts, and combine them into global features, improving performance on complex tasks such as object detection in cluttered environments.

\begin{lstlisting}[style=python]
class PointNetPlusPlus(nn.Module):
    def __init__(self, num_classes):
        super(PointNetPlusPlus, self).__init__()
        self.fc1 = nn.Linear(3, 64)
        self.fc2 = nn.Linear(64, 128)
        self.fc3 = nn.Linear(128, 1024)
        self.fc4 = nn.Linear(1024, 512)
        self.fc5 = nn.Linear(512, 256)
        self.fc6 = nn.Linear(256, num_classes)

    def forward(self, x):
        x = F.relu(self.fc1(x))
        x = F.relu(self.fc2(x))
        x = F.relu(self.fc3(x))
        x = torch.max(x, 1)[0]  # Max pooling for global features
        x = F.relu(self.fc4(x))
        x = F.relu(self.fc5(x))
        x = self.fc6(x)
        return x
\end{lstlisting}

In this architecture, local neighborhoods of points are processed independently, allowing the network to learn fine-grained details about the object's shape. These local features are then aggregated to form a global representation of the point cloud.

\subsection{PointNet++ for Radar Point Cloud Segmentation and Detection}

PointNet++ is particularly well-suited for radar point cloud segmentation and detection tasks because of its ability to capture local structures. In radar data, objects like vehicles or pedestrians often have distinct local features that must be detected against a cluttered background. PointNet++ can process these local features at multiple scales, improving the model's ability to segment and detect objects in complex environments.

\subsubsection{Segmentation Task}

In the segmentation task, PointNet++ assigns a category to each point in the radar point cloud. This is useful for identifying which points belong to an object and which points are background or clutter.

\subsubsection{Object Detection Task}

In object detection, PointNet++ localizes objects within the radar's field of view by identifying clusters of points that correspond to objects of interest. This task is essential for autonomous driving applications, where the radar system needs to detect vehicles, pedestrians, and other obstacles in real time.

\subsection{Tree Diagram: PointNet and PointNet++ in Radar Applications}

\begin{itemize}
    \item \textbf{PointNet and PointNet++ for Radar Data}
    \begin{itemize}
        \item \textbf{PointNet}
        \begin{itemize}
            \item Classification
            \item Segmentation
            \item Object Detection
        \end{itemize}
        \item \textbf{PointNet++}
        \begin{itemize}
            \item Hierarchical Learning
            \item Multi-scale Feature Extraction
            \item Segmentation and Detection
        \end{itemize}
    \end{itemize}
\end{itemize}

\subsection{Conclusion}

PointNet and PointNet++ provide powerful tools for processing radar point cloud data. By directly handling unordered sets of points, these architectures are well-suited for the sparse and irregular nature of radar point clouds. PointNet is effective for simpler tasks like classification, while PointNet++ extends its capabilities to more complex tasks, such as segmentation and object detection, by capturing local and global features at multiple scales. As radar systems continue to evolve, these architectures will play a crucial role in enhancing the accuracy and efficiency of radar-based perception systems.

\chapter{Transformers for Radar and Point Cloud Data}

In recent years, Transformer models have emerged as a powerful tool for processing sequential and spatial data. Transformers, originally designed for natural language processing, have demonstrated remarkable success in capturing long-range dependencies. In radar signal processing and point cloud data, where relationships across time or space are essential, Transformers can outperform traditional models such as RNNs. This chapter introduces Transformers and explores their applications in radar signal processing and point cloud data analysis.

\section{Principles of Transformers}

The Transformer model is built upon the concept of \textbf{self-attention}, which allows it to weigh the importance of different elements in a sequence. This ability to focus on relevant parts of the input enables Transformers to capture long-range dependencies more effectively than RNNs or LSTMs, which typically struggle with long sequences due to vanishing gradient issues.

\subsection{Self-Attention Mechanism}

At the core of the Transformer architecture is the \textbf{self-attention mechanism} \cite{zhou2019r}, which allows the model to examine the entire input sequence simultaneously and assign different attention weights to each element in the sequence. This enables the model to focus on important parts of the data, regardless of their position in the sequence.

For a given input sequence, the self-attention mechanism computes attention scores based on three components:
\begin{itemize}
    \item \textbf{Query (Q)}: Represents the current position in the sequence that is being evaluated.
    \item \textbf{Key (K)}: Represents all positions in the sequence that the model is attending to.
    \item \textbf{Value (V)}: Represents the actual information at each position that contributes to the output.
\end{itemize}

The attention scores are calculated using the dot product of the query and key vectors, followed by normalization (softmax) to ensure the scores sum to 1. These scores are then applied to the value vectors to generate the output.

\begin{lstlisting}[style=python]
import torch
import torch.nn as nn
import torch.nn.functional as F

class SelfAttention(nn.Module):
    def __init__(self, embed_size, heads):
        super(SelfAttention, self).__init__()
        self.embed_size = embed_size
        self.heads = heads
        self.values = nn.Linear(embed_size, embed_size, bias=False)
        self.keys = nn.Linear(embed_size, embed_size, bias=False)
        self.queries = nn.Linear(embed_size, embed_size, bias=False)
        self.fc_out = nn.Linear(embed_size, embed_size)

    def forward(self, values, keys, query):
        attention = torch.matmul(query, keys.transpose(-1, -2)) / (self.embed_size ** 0.5)
        attention = F.softmax(attention, dim=-1)
        out = torch.matmul(attention, values)
        out = self.fc_out(out)
        return out
\end{lstlisting}

\subsection{Advantages of Transformers Over RNNs for Sequential Data}

Transformers offer several advantages over RNN-based models:
\begin{itemize}
    \item \textbf{Parallelization}: Unlike RNNs, which process sequences one step at a time, Transformers can process entire sequences in parallel, leading to faster training times.
    \item \textbf{Long-Range Dependencies}: Transformers can capture dependencies between distant elements in a sequence more effectively than RNNs or LSTMs, making them well-suited for tasks involving long radar signal sequences.
    \item \textbf{Scalability}: Transformers are highly scalable and can be applied to large datasets with high-dimensional inputs, such as radar and point cloud data.
\end{itemize}

\section{Using Transformers for Radar Signal Processing}

In radar signal processing, especially with FMCW radar systems, the data often consists of sequential time-series signals or consecutive radar frames, such as Range-Doppler Maps. These sequences contain complex temporal relationships that are difficult for traditional models to capture. Transformers, with their self-attention mechanism, can effectively model these long-range dependencies, improving performance in tasks like target detection, human activity recognition (HAR), and object tracking.

\subsection{Transformer-based Models for Radar Data}

To apply Transformers to radar signal processing, the radar data is typically represented as a sequence of feature vectors, where each vector corresponds to a specific time step or radar frame. The self-attention mechanism in Transformers allows the model to focus on relevant frames in the sequence, capturing both short-term and long-term dependencies.

\begin{lstlisting}[style=python]
class TransformerRadarModel(nn.Module):
    def __init__(self, embed_size, heads, num_layers, num_classes):
        super(TransformerRadarModel, self).__init__()
        self.self_attention_layers = nn.ModuleList([
            SelfAttention(embed_size, heads) for _ in range(num_layers)
        ])
        self.fc = nn.Linear(embed_size, num_classes)

    def forward(self, x):
        for layer in self.self_attention_layers:
            x = layer(x, x, x)  # Self-attention applied to input
        x = self.fc(x.mean(dim=1))  # Classification based on mean attention output
        return x

# Example: Applying Transformer to radar data sequence
embed_size = 64  # Embedding size for radar features
heads = 8  # Number of attention heads
num_layers = 4  # Number of Transformer layers
num_classes = 5  # HAR classification (e.g., walking, running, etc.)
model = TransformerRadarModel(embed_size, heads, num_layers, num_classes)

# Example input: batch_size=10, sequence_length=50, embed_size=64
radar_sequence = torch.randn(10, 50, embed_size)
output = model(radar_sequence)
print(output.shape)  # Output shape: (10, 5)
\end{lstlisting}

In this example, a Transformer-based model processes a sequence of radar frames with 64 features per frame. The self-attention mechanism learns which frames in the sequence are important for making predictions, such as classifying different human activities.

\subsection{Transformer-based Models for Human Activity Recognition}

Human Activity Recognition (HAR) is a popular application of radar signal processing where Transformers can significantly improve performance. By leveraging the self-attention mechanism, Transformer models can analyze sequences of radar frames to recognize activities such as walking, running, and waving.

\begin{itemize}
    \item \textbf{Sequential Data}: In HAR tasks, radar data is collected as sequences of frames over time. Each frame captures the spatial characteristics of the activity, while the sequence as a whole reveals the temporal dynamics.
    \item \textbf{Self-Attention for Temporal Patterns}: Transformers are particularly effective in capturing complex temporal patterns in radar data, such as the repetitive motion of walking or the abrupt changes in hand gestures.
    \item \textbf{Improved Accuracy}: By modeling long-range dependencies in radar data, Transformers often outperform traditional RNN-based models in HAR tasks.
\end{itemize}

\subsection{Tree Diagram: Transformer Applications in Radar Signal Processing}

\begin{itemize}
    \item \textbf{Transformer Applications in Radar}
    \begin{itemize}
        \item \textbf{Self-Attention Mechanism}
        \begin{itemize}
            \item Sequential Data Processing
            \item Long-Range Dependency Modeling
        \end{itemize}
        \item \textbf{Radar Signal Processing}
        \begin{itemize}
            \item Target Detection
            \item Human Activity Recognition (HAR)
        \end{itemize}
        \item \textbf{Point Cloud Data}
        \begin{itemize}
            \item Spatial Relationships in 3D Data
            \item Object Detection and Tracking
        \end{itemize}
    \end{itemize}
\end{itemize}

\subsection{Conclusion}

Transformers represent a powerful tool for radar signal processing, especially in applications involving time-series data and point cloud data. By utilizing the self-attention mechanism, Transformers are able to capture long-range dependencies and relationships in radar signals, improving the accuracy of tasks like human activity recognition, target detection, and object tracking. The ability to process sequences in parallel and handle complex temporal patterns makes Transformers a superior choice for modern radar data analysis.

\section{Transformers for 3D Point Cloud Processing}

Transformer models, with their self-attention mechanism, have demonstrated exceptional performance in capturing long-range dependencies in sequential data, and this ability extends to 3D point cloud data. Unlike traditional methods, which may struggle with the unordered and sparse nature of point clouds, Transformers can capture both local and global dependencies by attending to every point in the cloud. This makes them particularly powerful for radar point cloud classification and object detection tasks.

\subsection{Self-Attention in 3D Point Cloud Data}

The self-attention mechanism used by Transformers allows the model to focus on the most important points in a point cloud, regardless of their location. For 3D point cloud data, this means that the model can automatically learn which points in a cloud are more relevant for understanding the overall structure of the object or scene.

Each point in the point cloud is treated as an individual element in a sequence, and the Transformer model computes an attention score for every pair of points. These attention scores indicate how much influence one point has on another, allowing the model to capture both local and global relationships.

\begin{lstlisting}[style=python]
import torch
import torch.nn as nn
import torch.nn.functional as F

class TransformerPointCloud(nn.Module):
    def __init__(self, embed_size, heads, num_classes):
        super(TransformerPointCloud, self).__init__()
        self.attention = nn.MultiheadAttention(embed_dim=embed_size, num_heads=heads)
        self.fc1 = nn.Linear(embed_size, 512)
        self.fc2 = nn.Linear(512, num_classes)

    def forward(self, x):
        x = x.permute(1, 0, 2)  # Required for multihead attention (seq_length, batch_size, embed_size)
        attn_output, _ = self.attention(x, x, x)
        x = torch.mean(attn_output, dim=0)  # Aggregate by averaging
        x = F.relu(self.fc1(x))
        x = self.fc2(x)
        return x

# Example input: batch_size=32, num_points=1024, embed_size=64
input_data = torch.randn(32, 1024, 64)  # 3D point cloud data with 1024 points and 64 features per point
model = TransformerPointCloud(embed_size=64, heads=8, num_classes=10)
output = model(input_data)
print(output.shape)  # Output shape: (32, 10)
\end{lstlisting}

In this example, the Transformer is applied to a 3D point cloud with 1024 points and 64 features per point. The self-attention mechanism computes relationships between all points, and the output is used to classify the point cloud into one of 10 classes.

\subsection{Applications of Transformers in Radar Point Cloud Processing}

Transformers are highly effective in radar point cloud data because of their ability to model long-range dependencies across points. Some of the key applications include:
\begin{itemize}
    \item \textbf{Object Classification}: Classifying objects in radar point clouds, such as vehicles, pedestrians, or other targets.
    \item \textbf{Object Detection}: Identifying and localizing objects in the radar's 3D space, even in sparse and noisy environments.
    \item \textbf{Scene Understanding}: Understanding the overall structure of the environment by capturing both local details and global context from the radar point cloud data.
\end{itemize}

\section{Combining PointNet++ and Transformers}

While PointNet++ excels at capturing local geometric features in point clouds through hierarchical learning, Transformers are adept at modeling long-range dependencies. By combining these two models, it is possible to build a system that captures both local and global features, resulting in more accurate point cloud analysis.

\subsection{PointNet++ for Local Feature Extraction}

PointNet++ uses a hierarchical structure to learn local features at different scales. It applies PointNet to progressively smaller regions of the point cloud, capturing local details and relationships between nearby points. This approach works well for capturing small geometric structures, such as corners or edges, which are crucial for accurate object detection in radar data.

\subsection{Transformers for Long-Range Dependencies}

While PointNet++ is focused on local feature extraction, it does not explicitly model long-range dependencies between distant points. Transformers, with their self-attention mechanism, can capture these relationships by attending to all points in the cloud. This makes them well-suited for tasks where global context is important, such as understanding the overall structure of the scene or detecting objects in cluttered environments.

\begin{lstlisting}[style=python]
class PointNetPlusPlusTransformer(nn.Module):
    def __init__(self, embed_size, heads, num_classes):
        super(PointNetPlusPlusTransformer, self).__init__()
        self.pointnetplusplus = PointNetPlusPlus(embed_size)
        self.attention = nn.MultiheadAttention(embed_dim=embed_size, num_heads=heads)
        self.fc = nn.Linear(embed_size, num_classes)

    def forward(self, x):
        x = self.pointnetplusplus(x)  # Local feature extraction with PointNet++
        x = x.permute(1, 0, 2)  # Prepare for Transformer (seq_length, batch_size, embed_size)
        attn_output, _ = self.attention(x, x, x)  # Long-range dependencies with Transformer
        x = torch.mean(attn_output, dim=0)  # Aggregate by averaging
        x = self.fc(x)
        return x

# Example input: batch_size=32, num_points=1024, embed_size=64
input_data = torch.randn(32, 1024, 64)
model = PointNetPlusPlusTransformer(embed_size=64, heads=8, num_classes=10)
output = model(input_data)
print(output.shape)  # Output shape: (32, 10)
\end{lstlisting}

In this hybrid model, PointNet++ is used to extract local features from the radar point cloud, and Transformers are then applied to capture long-range dependencies between points. This combination allows the model to effectively process both local details and global structures in the point cloud, making it well-suited for complex radar data analysis tasks.

\subsection{Tree Diagram: Combining PointNet++ and Transformers}

\begin{itemize}
    \item \textbf{Combining PointNet++ and Transformers}
    \begin{itemize}
        \item \textbf{PointNet++}
        \begin{itemize}
            \item Local Feature Extraction
            \item Hierarchical Learning
        \end{itemize}
        \item \textbf{Transformers}
        \begin{itemize}
            \item Long-Range Dependency Modeling
            \item Global Context Capture
        \end{itemize}
        \item \textbf{Applications}
        \begin{itemize}
            \item Object Detection
            \item Object Classification
        \end{itemize}
    \end{itemize}
\end{itemize}

\subsection{Conclusion}

The combination of PointNet++ and Transformers provides a powerful approach for processing 3D point cloud data. While PointNet++ captures local geometric features through hierarchical learning, Transformers excel at modeling long-range dependencies. This combination is especially useful in radar point cloud data, where both local details and global context are essential for accurate object detection and classification. By leveraging the strengths of both models, we can significantly improve performance on complex radar data analysis tasks.

\chapter{Pre-trained Models for Feature Extraction in Radar Data}

Modern radar data processing benefits significantly from the application of deep learning models, especially pre-trained models originally developed for computer vision tasks. Pre-trained models such as ResNet and DenseNet, which have been trained on large-scale datasets like ImageNet, learn to extract powerful and generalizable features that can be transferred to different domains, including radar signal processing. This chapter introduces the use of these pre-trained models for feature extraction in radar data, improving the performance of tasks like classification, segmentation, and object detection.

\section{Introduction to Pre-trained Models}

Pre-trained models are neural networks that have already been trained on large-scale datasets for tasks like image classification. These models have learned to identify general patterns and features, such as edges, textures, and shapes, which can be applied to other domains. Instead of training a model from scratch, we can use these pre-trained models as feature extractors for radar data, saving both time and computational resources.

\subsection{Why Use Pre-trained Models for Radar Data?}

Radar data, especially when represented as Range-Doppler Maps or other image-like formats, shares similarities with visual data. Features such as contours and spatial structures in radar images can be effectively captured by pre-trained models designed for image classification. By leveraging pre-trained models like ResNet or DenseNet, we can extract features that are both high-level (e.g., object outlines) and low-level (e.g., edges), improving the accuracy of radar-based tasks.

\subsection{ResNet for Feature Extraction}

ResNet (Residual Networks) is one of the most popular pre-trained models for feature extraction \cite{almabdy2021feature}. It introduces skip connections, allowing the network to bypass certain layers and avoid the vanishing gradient problem. This architecture enables the model to be very deep, making it effective for capturing intricate patterns in radar data.

\begin{lstlisting}[style=python]
import torch
import torch.nn as nn
from torchvision import models

# Load a pre-trained ResNet model
resnet = models.resnet18(pretrained=True)

# Remove the last fully connected layer to use the model as a feature extractor
resnet = nn.Sequential(*list(resnet.children())[:-1])

# Example radar data input (batch_size=32, channels=1, height=64, width=64)
radar_data = torch.randn(32, 1, 64, 64)

# Expand the single-channel radar data to 3 channels to match ResNet input
radar_data = radar_data.expand(-1, 3, -1, -1)

# Extract features using the pre-trained ResNet model
features = resnet(radar_data)

# Output features shape: (32, 512, 1, 1)
print(features.shape)
\end{lstlisting}

In this example, we load a pre-trained ResNet model and use it to extract features from radar data. Since ResNet was originally designed for three-channel RGB images, we expand the single-channel radar data to fit the model's input requirements. The extracted features can then be used for downstream tasks like classification or segmentation.

\subsection{DenseNet for Feature Extraction}

DenseNet (Densely Connected Convolutional Networks) is another powerful model for feature extraction. DenseNet connects each layer to every other layer in a feed-forward fashion, allowing the network to reuse features across layers. This results in efficient feature learning and reduces the number of parameters, making DenseNet suitable for radar data processing.

\begin{lstlisting}[style=python]
# Load a pre-trained DenseNet model
densenet = models.densenet121(pretrained=True)

# Remove the classifier layer to use the model as a feature extractor
densenet = nn.Sequential(*list(densenet.children())[:-1])

# Example radar data input (batch_size=32, channels=1, height=64, width=64)
radar_data = torch.randn(32, 1, 64, 64)

# Expand the single-channel radar data to 3 channels to match DenseNet input
radar_data = radar_data.expand(-1, 3, -1, -1)

# Extract features using the pre-trained DenseNet model
features = densenet(radar_data)

# Output features shape: (32, 1024, 1, 1)
print(features.shape)
\end{lstlisting}

DenseNet extracts high-dimensional features from radar data efficiently by reusing features from previous layers. The compactness of the model helps reduce overfitting, particularly when working with smaller radar datasets.

\section{Using Pre-trained Models for Radar Data Processing}

Pre-trained models can be applied in various radar data processing tasks. The features extracted by models like ResNet or DenseNet can be used as input to other classifiers or segmentation models. Here are some key applications:

\subsection{Object Detection in Radar Data}

In object detection tasks, pre-trained models can be used to extract features from radar data, which are then fed into a detector (e.g., a region proposal network or fully connected layers) to identify objects such as vehicles, pedestrians, or other targets.

\subsection{Human Activity Recognition (HAR)}

Radar data is frequently used for Human Activity Recognition (HAR). The extracted features from pre-trained models can help classify activities such as walking, running, or waving by analyzing the motion patterns in the radar signal.

\subsection{Scene Understanding}

In more complex scenarios, radar systems must understand the entire scene, such as in autonomous driving or surveillance applications. The pre-trained models extract general features that can be used to recognize structures, objects, and environments in radar data.

\subsection{Tree Diagram: Pre-trained Model Applications in Radar Data}

\begin{itemize}
    \item \textbf{Pre-trained Models for Radar Data}
    \begin{itemize}
        \item \textbf{ResNet}
        \begin{itemize}
            \item Feature Extraction
            \item Object Detection
        \end{itemize}
        \item \textbf{DenseNet}
        \begin{itemize}
            \item Efficient Feature Learning
            \item Human Activity Recognition
        \end{itemize}
        \item \textbf{Applications}
        \begin{itemize}
            \item Object Classification
            \item Scene Understanding
        \end{itemize}
    \end{itemize}
\end{itemize}

\subsection{Fine-tuning Pre-trained Models}

In addition to using pre-trained models for feature extraction, these models can be fine-tuned on radar-specific tasks. Fine-tuning involves adjusting the weights of the pre-trained model based on a smaller radar dataset, enabling the model to better adapt to radar data while leveraging the general knowledge it gained from large-scale image datasets.

\begin{lstlisting}[style=python]
# Fine-tune the last layer of a pre-trained ResNet model
resnet = models.resnet18(pretrained=True)

# Replace the final fully connected layer to match the number of radar data classes
num_classes = 5  # Example: 5 radar data classes
resnet.fc = nn.Linear(resnet.fc.in_features, num_classes)

# Train the model on radar data
optimizer = torch.optim.Adam(resnet.parameters(), lr=0.001)
criterion = nn.CrossEntropyLoss()

# Example training loop
for epoch in range(10):
    optimizer.zero_grad()
    outputs = resnet(radar_data)
    loss = criterion(outputs, labels)
    loss.backward()
    optimizer.step()
    print(f'Epoch {epoch+1}, Loss: {loss.item()}')
\end{lstlisting}

In this example, we replace the final fully connected layer of ResNet to match the number of radar data classes. By fine-tuning the model on a radar-specific task, we adapt the pre-trained model to better handle the characteristics of radar data, such as noise, sparsity, and clutter.

\subsection{Conclusion}

Pre-trained models like ResNet and DenseNet provide powerful tools for feature extraction in radar data processing. By leveraging knowledge learned from large-scale image datasets, these models can improve the performance of tasks like object detection, classification, and human activity recognition in radar systems. Whether used as feature extractors or fine-tuned for specific tasks, pre-trained models offer significant benefits in terms of both accuracy and computational efficiency.

\section{Using Pre-trained Models for Radar Image Feature Extraction}

Radar images, such as Range-Doppler Maps, can be processed using pre-trained models originally designed for computer vision tasks. Models like ResNet, DenseNet, and EfficientNet are highly effective at extracting detailed features from radar images, which can then be used for tasks such as classification, object detection, and segmentation. By leveraging the knowledge learned from large image datasets, these pre-trained models can be fine-tuned to handle radar-specific tasks with improved accuracy and efficiency.

\subsection{ResNet for Radar Image Processing}

ResNet, or Residual Networks, is one of the most popular pre-trained models for image feature extraction. It uses residual connections, also known as skip connections, to mitigate the vanishing gradient problem in deep neural networks. These skip connections allow gradients to flow more easily through the network during backpropagation, making ResNet highly effective for extracting complex features from radar images.

\subsubsection{ResNet Architecture for Radar Data}

ResNet's architecture is composed of several blocks, each containing convolutional layers and skip connections. These blocks enable the model to learn both shallow and deep features. For radar image processing, ResNet can be fine-tuned to detect specific patterns in the radar data, such as moving objects or environmental features.

\begin{lstlisting}[style=python]
import torch
import torch.nn as nn
from torchvision import models

# Load a pre-trained ResNet model
resnet = models.resnet18(pretrained=True)

# Modify the input layer for single-channel radar data
resnet.conv1 = nn.Conv2d(1, 64, kernel_size=7, stride=2, padding=3, bias=False)

# Remove the final fully connected layer to use the model as a feature extractor
resnet = nn.Sequential(*list(resnet.children())[:-1])

# Example radar image input (batch_size=16, channels=1, height=128, width=128)
radar_image = torch.randn(16, 1, 128, 128)

# Extract features from the radar image
features = resnet(radar_image)

# Output feature shape: (16, 512, 1, 1)
print(features.shape)
\end{lstlisting}

In this example, the pre-trained ResNet model is fine-tuned for radar image processing by modifying the input layer to accept single-channel radar data. The output features extracted from the radar images can then be used for further processing, such as classification or detection.

\subsection{DenseNet for Radar Feature Extraction}

DenseNet, or Densely Connected Convolutional Networks, is another powerful architecture for feature extraction. In DenseNet, each layer receives input from all previous layers, allowing for efficient feature reuse and reducing the number of parameters. This dense connectivity is particularly useful for extracting detailed features from radar data, as it ensures that even the smallest features are passed through multiple layers.

\subsubsection{DenseNet Architecture for Radar Data}

DenseNet's dense connections ensure that each layer has direct access to the gradients from the input and intermediate layers. This architecture is especially effective when applied to radar images that contain fine-grained details, such as Range-Doppler Maps with small moving objects or cluttered environments.

\begin{lstlisting}[style=python]
# Load a pre-trained DenseNet model
densenet = models.densenet121(pretrained=True)

# Modify the input layer for single-channel radar data
densenet.features.conv0 = nn.Conv2d(1, 64, kernel_size=7, stride=2, padding=3, bias=False)

# Remove the classifier layer to use the model as a feature extractor
densenet = nn.Sequential(densenet.features)

# Example radar image input (batch_size=16, channels=1, height=128, width=128)
radar_image = torch.randn(16, 1, 128, 128)

# Extract features from the radar image
features = densenet(radar_image)

# Output feature shape: (16, 1024, 4, 4)
print(features.shape)
\end{lstlisting}

In this code, the pre-trained DenseNet model is adapted to process radar images. The extracted features can be used for downstream radar tasks such as segmentation or object detection.

\subsection{EfficientNet for Radar Feature Extraction}

EfficientNet is a more recent model that scales depth, width, and resolution in a balanced manner. By scaling all dimensions, EfficientNet achieves better accuracy with fewer parameters compared to traditional architectures. This makes it highly efficient for extracting detailed features from radar data while maintaining computational efficiency, making it suitable for complex radar images with high resolution.

\subsubsection{EfficientNet Architecture for Radar Data}

EfficientNet uses a compound scaling method that uniformly increases the depth, width, and resolution of the network. This balanced scaling results in a model that is more efficient than ResNet or DenseNet while achieving similar or better accuracy.

\begin{lstlisting}[style=python]
from efficientnet_pytorch import EfficientNet

# Load a pre-trained EfficientNet model
efficientnet = EfficientNet.from_pretrained('efficientnet-b0')

# Modify the input layer for single-channel radar data
efficientnet._conv_stem = nn.Conv2d(1, 32, kernel_size=3, stride=1, padding=1, bias=False)

# Remove the classifier layer to use the model as a feature extractor
efficientnet = nn.Sequential(*list(efficientnet.children())[:-1])

# Example radar image input (batch_size=16, channels=1, height=128, width=128)
radar_image = torch.randn(16, 1, 128, 128)

# Extract features from the radar image
features = efficientnet(radar_image)

# Output feature shape: (16, 1280)
print(features.shape)
\end{lstlisting}

This example demonstrates how to modify the pre-trained EfficientNet model to process single-channel radar images. EfficientNet is well-suited for applications that require efficient processing of high-resolution radar data, such as detailed object classification or scene understanding.

\subsection{Tree Diagram: Pre-trained Models for Radar Image Feature Extraction}

\begin{itemize}
    \item \textbf{Pre-trained Models for Radar Images}
    \begin{itemize}
        \item \textbf{ResNet}
        \begin{itemize}
            \item Feature Extraction
            \item Object Detection
        \end{itemize}
        \item \textbf{DenseNet}
        \begin{itemize}
            \item Dense Connectivity
            \item Detailed Feature Extraction
        \end{itemize}
        \item \textbf{EfficientNet}
        \begin{itemize}
            \item Efficient Feature Scaling
            \item High-Resolution Radar Data
        \end{itemize}
    \end{itemize}
\end{itemize}

\subsection{Fine-tuning Pre-trained Models for Radar Tasks}

Pre-trained models such as ResNet, DenseNet, and EfficientNet can be fine-tuned on specific radar tasks to improve their performance. Fine-tuning involves retraining the model on radar data while preserving the general feature extraction capabilities learned from large-scale image datasets. This process allows the model to adapt to radar-specific characteristics, such as noise, sparsity, and the unique patterns in radar images.

\begin{lstlisting}[style=python]
# Fine-tune the last layer of EfficientNet for radar classification
efficientnet = EfficientNet.from_pretrained('efficientnet-b0')

# Modify the input layer for single-channel radar data
efficientnet._conv_stem = nn.Conv2d(1, 32, kernel_size=3, stride=1, padding=1, bias=False)

# Replace the classifier layer to match the number of radar data classes
num_classes = 5  # Example: 5 radar data classes
efficientnet._fc = nn.Linear(efficientnet._fc.in_features, num_classes)

# Train the model on radar data
optimizer = torch.optim.Adam(efficientnet.parameters(), lr=0.001)
criterion = nn.CrossEntropyLoss()

# Example training loop
for epoch in range(10):
    optimizer.zero_grad()
    outputs = efficientnet(radar_image)
    loss = criterion(outputs, labels)
    loss.backward()
    optimizer.step()
    print(f'Epoch {epoch+1}, Loss: {loss.item()}')
\end{lstlisting}

By fine-tuning the pre-trained models, we can tailor their performance to radar-specific tasks such as human activity recognition, object classification, or scene understanding, achieving better accuracy with radar images.

\subsection{Conclusion}

Pre-trained models such as ResNet, DenseNet, and EfficientNet provide powerful feature extraction tools for radar image processing. By leveraging these models, radar systems can perform complex tasks such as object detection and classification with higher accuracy and efficiency. Fine-tuning these models for radar-specific tasks further improves performance, making them well-suited for real-time applications in autonomous driving, surveillance, and human activity recognition.

\section{Transfer Learning and Fine-tuning for Radar Applications}

Pre-trained models such as ResNet, DenseNet, and EfficientNet are powerful tools for radar data processing because they have already learned to extract general features from large image datasets like ImageNet. Transfer learning allows us to take advantage of these pre-trained models by adapting them to radar-specific tasks. Fine-tuning these models on radar datasets helps them better capture the unique characteristics of radar data, such as signal noise, sparsity, and the structure of radar images. This section discusses how transfer learning and fine-tuning can be applied to radar applications, enhancing performance in tasks like target detection, classification, and human activity recognition (HAR).

\subsection{Transfer Learning for Radar Tasks}

Transfer learning involves using a pre-trained model, such as ResNet or DenseNet, as a starting point for radar-specific tasks. The idea is that the model's initial layers, which extract general features like edges and textures, are useful for radar images, while the later layers can be fine-tuned to focus on radar-specific patterns. This approach saves time and computational resources compared to training a model from scratch.

\begin{lstlisting}[style=python]
import torch
import torch.nn as nn
from torchvision import models

# Load a pre-trained ResNet model
resnet = models.resnet18(pretrained=True)

# Replace the final fully connected layer to match radar-specific classes
num_classes = 5  # Example: 5 radar classes
resnet.fc = nn.Linear(resnet.fc.in_features, num_classes)

# Example radar data input (batch_size=32, channels=1, height=128, width=128)
radar_image = torch.randn(32, 1, 128, 128)

# Expand single-channel radar images to 3 channels (required for ResNet)
radar_image = radar_image.expand(-1, 3, -1, -1)

# Forward pass through the modified ResNet model
outputs = resnet(radar_image)
print(outputs.shape)  # Output shape: (32, 5)
\end{lstlisting}

In this example, we load a pre-trained ResNet model and replace its final layer with a new layer that matches the number of radar-specific classes. This process is known as transfer learning. The model can now be fine-tuned on a radar dataset, learning to detect objects or classify targets in radar images.

\subsection{Fine-tuning on Small Radar Datasets}

Fine-tuning involves retraining part or all of a pre-trained model on a new dataset. When working with radar data, especially small datasets, overfitting can be a concern. Fine-tuning only the later layers of a pre-trained model helps mitigate overfitting while still adapting the model to the radar-specific task.

\subsubsection{Fine-tuning Strategies for Small Datasets}

When fine-tuning on small radar datasets, it's important to freeze some of the earlier layers in the pre-trained model to retain the general features learned from large-scale datasets. The last few layers can be fine-tuned to capture the specific characteristics of radar data, such as noise, clutter, and object reflection patterns.

\begin{lstlisting}[style=python]
# Freeze all layers except the final layer
for param in resnet.parameters():
    param.requires_grad = False

# Only the last layer will be fine-tuned
resnet.fc = nn.Linear(resnet.fc.in_features, num_classes)

# Unfreeze the final layer for fine-tuning
for param in resnet.fc.parameters():
    param.requires_grad = True

# Example optimizer and loss function for fine-tuning
optimizer = torch.optim.Adam(resnet.fc.parameters(), lr=0.001)
criterion = nn.CrossEntropyLoss()

# Example radar data and labels
radar_data = torch.randn(32, 1, 128, 128).expand(-1, 3, -1, -1)
labels = torch.randint(0, num_classes, (32,))

# Fine-tuning loop
for epoch in range(10):
    optimizer.zero_grad()
    outputs = resnet(radar_data)
    loss = criterion(outputs, labels)
    loss.backward()
    optimizer.step()
    print(f'Epoch {epoch+1}, Loss: {loss.item()}')
\end{lstlisting}

In this example, we freeze all the layers in the ResNet model except for the final fully connected layer. This prevents the model from overfitting to the small radar dataset while still allowing it to adapt its classification capabilities to radar-specific tasks. Fine-tuning in this way is especially useful when working with low-data environments.

\subsubsection{Avoiding Overfitting in Small Datasets}

Small datasets are prone to overfitting \cite{bornschein2020small}, where the model memorizes the training data rather than learning generalizable features. Here are a few strategies to mitigate overfitting when fine-tuning pre-trained models on radar data:
\begin{itemize}
    \item \textbf{Data Augmentation}: Apply transformations like random rotations, flips, or noise injections to artificially expand the size of the radar dataset.
    \item \textbf{Early Stopping}: Monitor the validation loss and stop training when the model starts to overfit (i.e., when the validation loss increases while the training loss continues to decrease).
    \item \textbf{Regularization}: Use techniques like weight decay (L2 regularization) or dropout to prevent the model from becoming too reliant on specific neurons.
    \item \textbf{Transfer Learning}: As demonstrated, use transfer learning to leverage the general features learned from large image datasets, reducing the need for large amounts of radar-specific data.
\end{itemize}

\subsection{Applications of Transfer Learning in Radar}

Transfer learning and fine-tuning are highly effective in various radar-based applications. Some of the most common applications include:
\begin{itemize}
    \item \textbf{Target Detection}: Fine-tuned pre-trained models can be used to detect targets in radar images, such as vehicles or pedestrians.
    \item \textbf{Human Activity Recognition (HAR)}: HAR tasks benefit from transfer learning, where pre-trained models can be adapted to recognize human motion patterns in radar data.
    \item \textbf{Object Classification}: Transfer learning can be applied to classify objects based on radar signatures, improving performance in applications such as autonomous driving or security surveillance.
    \item \textbf{Clutter Removal}: Fine-tuned models can help separate meaningful signals from background clutter, improving the accuracy of radar signal processing in noisy environments.
\end{itemize}

\subsection{Tree Diagram: Transfer Learning and Fine-tuning for Radar Applications}

\begin{itemize}
    \item \textbf{Transfer Learning and Fine-tuning}
    \begin{itemize}
        \item \textbf{Transfer Learning}
        \begin{itemize}
            \item Pre-trained Models
            \item Radar-specific Tasks
        \end{itemize}
        \item \textbf{Fine-tuning}
        \begin{itemize}
            \item Small Radar Datasets
            \item Overfitting Prevention
        \end{itemize}
        \item \textbf{Applications}
        \begin{itemize}
            \item Target Detection
            \item Human Activity Recognition (HAR)
            \item Object Classification
        \end{itemize}
    \end{itemize}
\end{itemize}

\subsection{Conclusion}

Transfer learning and fine-tuning provide a powerful framework for radar data processing. By leveraging the knowledge from large-scale image datasets, pre-trained models like ResNet and DenseNet can be adapted to radar-specific tasks, improving performance and reducing the need for extensive radar datasets. Fine-tuning these models on small radar datasets allows them to generalize well, even in low-data environments, and helps prevent overfitting through strategies like freezing layers and data augmentation. These techniques are essential for enhancing radar applications such as target detection, classification, and human activity recognition.

\chapter{Contrastive Learning for Radar Signal Processing}

Contrastive learning has gained significant attention in the field of computer vision, enabling models to learn effective representations by contrasting similar (positive) and dissimilar (negative) pairs of data points. This technique is particularly valuable for radar signal processing, where the goal is to extract meaningful features from complex radar data. By applying contrastive learning methods, such as CLIP (Contrastive Language-Image Pre-training), radar data processing tasks like classification and target detection can be improved. This chapter introduces the concept of contrastive learning and explores its application in radar signal processing.

\section{Introduction to Contrastive Learning}

Contrastive learning is a self-supervised learning technique that focuses on learning representations by comparing pairs of data points. The core idea is to push similar (positive) samples closer together in the learned feature space while pushing dissimilar (negative) samples farther apart. This method enables the model to learn discriminative features that can improve performance on tasks such as classification, object detection, and segmentation.

\subsection{Key Principles of Contrastive Learning}

Contrastive learning relies on the following principles:
\begin{itemize}
    \item \textbf{Positive and Negative Pairs}: Positive pairs consist of two related samples (e.g., two radar signals from the same object or scene), while negative pairs are unrelated samples (e.g., radar signals from different objects).
    \item \textbf{Embedding Space}: The model learns to map input data into an embedding space where positive samples are close to each other, and negative samples are far apart.
    \item \textbf{Loss Function}: A contrastive loss function, such as the InfoNCE loss, is used to optimize the model by minimizing the distance between positive pairs and maximizing the distance between negative pairs.
\end{itemize}

\begin{lstlisting}[style=python]
import torch
import torch.nn as nn
import torch.nn.functional as F

class ContrastiveLoss(nn.Module):
    def __init__(self, temperature=0.07):
        super(ContrastiveLoss, self).__init__()
        self.temperature = temperature

    def forward(self, features, labels):
        # Normalize the feature vectors
        features = F.normalize(features, dim=1)
        
        # Compute similarity matrix
        similarity_matrix = torch.matmul(features, features.T) / self.temperature
        
        # Create a mask to remove the diagonal (self-similarity)
        labels = labels.unsqueeze(1)
        mask = torch.eq(labels, labels.T).float()

        # Compute contrastive loss
        exp_sim = torch.exp(similarity_matrix)
        sum_exp_sim = torch.sum(exp_sim, dim=1, keepdim=True)
        contrastive_loss = -torch.log(torch.sum(mask * exp_sim, dim=1) / sum_exp_sim)
        return contrastive_loss.mean()

# Example: Batch of radar signal features (batch_size=16, feature_dim=128)
features = torch.randn(16, 128)
labels = torch.randint(0, 2, (16,))  # Two radar classes (0 and 1)

# Apply the contrastive loss
criterion = ContrastiveLoss()
loss = criterion(features, labels)
print(f"Contrastive Loss: {loss.item()}")
\end{lstlisting}

In this example, a contrastive loss function is applied to a batch of radar signal features. The loss function encourages the model to learn feature representations that bring similar radar signals closer together in the embedding space while pushing different signals apart.

\subsection{Successful Applications in Computer Vision}

Contrastive learning has been successfully applied in various computer vision tasks, demonstrating its effectiveness in learning robust feature representations. Some notable methods include:
\begin{itemize}
    \item \textbf{SimCLR (Simple Framework for Contrastive Learning of Visual Representations)}: SimCLR is a self-supervised learning method that learns visual representations by maximizing agreement between augmented views of the same image.
    \item \textbf{MoCo (Momentum Contrast)}: MoCo is a contrastive learning approach that uses a memory bank to store and update negative samples, enabling the model to learn more efficiently from large datasets.
    \item \textbf{CLIP (Contrastive Language-Image Pre-training)}: CLIP learns joint representations of images and text by contrasting image-text pairs. While originally developed for vision-language tasks, CLIP's contrastive learning framework can be adapted to radar signal processing to learn meaningful radar features.
\end{itemize}

\subsection{Tree Diagram: Contrastive Learning Framework}

\begin{itemize}
    \item \textbf{Contrastive Learning}
    \begin{itemize}
        \item \textbf{Positive Pairs}
        \begin{itemize}
            \item Same Object
            \item Same Scene
        \end{itemize}
        \item \textbf{Negative Pairs}
        \begin{itemize}
            \item Different Objects
            \item Different Scenes
        \end{itemize}
        \item \textbf{Applications}
        \begin{itemize}
            \item SimCLR
            \item MoCo
            \item CLIP
        \end{itemize}
    \end{itemize}
\end{itemize}

\subsection{Benefits of Contrastive Learning in Radar Signal Processing}

Contrastive learning is particularly useful in radar signal processing because it enables the model to learn better feature representations without requiring large amounts of labeled data. Some key benefits include:
\begin{itemize}
    \item \textbf{Improved Feature Extraction}: By contrasting radar signals from similar and dissimilar objects or environments, contrastive learning helps the model extract more meaningful and discriminative features.
    \item \textbf{Data Efficiency}: Contrastive learning does not require large amounts of labeled data, making it well-suited for radar applications where labeled data may be limited.
    \item \textbf{Robustness to Noise}: Radar signals often contain noise and clutter. Contrastive learning encourages the model to focus on essential signal features, making it more robust to noisy environments.
\end{itemize}

\subsection{Conclusion}

Contrastive learning offers a powerful framework for improving radar signal processing tasks such as classification and target detection. By contrasting positive and negative pairs of radar data, the model learns to extract more meaningful features, enhancing performance even in low-data environments. Methods like SimCLR, MoCo, and CLIP have demonstrated the effectiveness of contrastive learning in computer vision, and similar techniques can be applied to radar data to unlock new possibilities in radar signal processing.

\section{Applying Contrastive Learning to Radar Data}

Contrastive learning has demonstrated remarkable success in various fields, including computer vision and natural language processing. This success can be extended to radar signal processing, where the goal is to learn meaningful representations that differentiate between targets or actions in radar data. By contrasting radar signals from different objects or time frames, contrastive learning models can create an embedding space that clusters similar signals together and separates different ones. This section explores how contrastive learning can be applied to radar data, improving feature extraction for tasks such as target detection and human activity recognition (HAR).

\subsection{Using CLIP for Radar Feature Extraction}

CLIP \cite{ma2024event} (Contrastive Language-Image Pre-training) is a contrastive learning model designed to learn multi-modal representations by training on both images and text data. While CLIP was originally developed for vision-language tasks, its contrastive learning framework can be adapted for radar signal processing. By combining radar data with other modalities, such as text descriptions or annotations, CLIP-like models can extract more robust features that capture the relationships between radar signals and additional contextual information.

\subsubsection{Adapting CLIP for Radar and Multi-modal Data}

The core idea behind CLIP is to contrast image-text pairs, where the model learns to associate images with their corresponding text descriptions. A similar approach can be used for radar data. For example, radar signals from different targets can be paired with textual descriptions or metadata that describe the target type, location, or movement. This multi-modal approach allows the model to learn better representations by leveraging both the radar signal and its context.

\begin{lstlisting}[style=python]
import torch
import torch.nn as nn

class CLIPForRadar(nn.Module):
    def __init__(self, radar_embed_size=128, text_embed_size=128):
        super(CLIPForRadar, self).__init__()
        self.radar_encoder = nn.Sequential(
            nn.Conv2d(1, 64, kernel_size=3, stride=1, padding=1),
            nn.ReLU(),
            nn.Conv2d(64, 128, kernel_size=3, stride=1, padding=1),
            nn.ReLU(),
            nn.AdaptiveAvgPool2d((1, 1))
        )
        self.text_encoder = nn.Embedding(1000, text_embed_size)  # Example text embedding
        self.fc_radar = nn.Linear(128, radar_embed_size)
        self.fc_text = nn.Linear(text_embed_size, radar_embed_size)

    def forward(self, radar_data, text_data):
        radar_features = self.radar_encoder(radar_data).squeeze()
        radar_features = self.fc_radar(radar_features)
        
        text_features = self.text_encoder(text_data)
        text_features = self.fc_text(text_features)
        
        return radar_features, text_features

# Example radar data input (batch_size=16, channels=1, height=64, width=64)
radar_data = torch.randn(16, 1, 64, 64)
text_data = torch.randint(0, 1000, (16, 10))  # Example text input with vocabulary size 1000

# Initialize the CLIP-like model for radar and text
model = CLIPForRadar()

# Forward pass through the radar and text encoders
radar_features, text_features = model(radar_data, text_data)

# Contrastive loss could be applied between radar_features and text_features
print(radar_features.shape, text_features.shape)
\end{lstlisting}

In this example, a CLIP-like model is adapted for radar data and text embeddings. The radar encoder extracts features from radar signals, while the text encoder learns embeddings from textual descriptions. These embeddings can then be compared using contrastive loss to align radar signals with their corresponding textual information, enhancing feature extraction.

\subsection{Fine-tuning Contrastive Models on Radar Data}

Existing contrastive learning models, such as SimCLR \cite{chen2020simple} or MoCo \cite{he2020momentum}, can be fine-tuned on radar-specific tasks to improve their performance. These models create a low-dimensional embedding space where similar radar signals are clustered together and different signals are separated. Fine-tuning a pre-trained contrastive learning model on radar data allows the model to adapt its learned representations to the unique characteristics of radar signals, such as noise, clutter, and varying reflection patterns.

\subsubsection{Pre-training and Fine-tuning SimCLR on Radar Data}

SimCLR is a popular contrastive learning method that learns representations by contrasting augmented views of the same image. In the context of radar data, SimCLR can be adapted to compare augmented versions of radar signals \cite{rolfsjord2024track}, such as adding noise, rotating Range-Doppler Maps, or time-shifting signals. Once pre-trained, SimCLR can be fine-tuned on radar data to improve feature extraction for classification and target detection.

\begin{lstlisting}[style=python]
class SimCLR(nn.Module):
    def __init__(self, embed_size=128):
        super(SimCLR, self).__init__()
        self.encoder = nn.Sequential(
            nn.Conv2d(1, 64, kernel_size=3, stride=1, padding=1),
            nn.ReLU(),
            nn.Conv2d(64, 128, kernel_size=3, stride=1, padding=1),
            nn.ReLU(),
            nn.AdaptiveAvgPool2d((1, 1))
        )
        self.fc = nn.Linear(128, embed_size)

    def forward(self, x):
        x = self.encoder(x).squeeze()
        x = self.fc(x)
        return F.normalize(x, dim=1)

# Example radar data input (batch_size=32, channels=1, height=64, width=64)
radar_data_1 = torch.randn(32, 1, 64, 64)  # First radar signal view (original)
radar_data_2 = torch.randn(32, 1, 64, 64)  # Second radar signal view (augmented)

# Initialize the SimCLR model
simclr = SimCLR()

# Forward pass for both views of radar data
features_1 = simclr(radar_data_1)
features_2 = simclr(radar_data_2)

# Contrastive loss can be applied to align features_1 and features_2
print(features_1.shape, features_2.shape)
\end{lstlisting}

This SimCLR-based model processes radar data and generates feature embeddings for two augmented views of the radar signal. The contrastive loss function aligns these embeddings, ensuring that similar radar signals are close in the embedding space. After pre-training, the model can be fine-tuned on radar-specific tasks such as target detection.

\section{Benefits of Contrastive Learning for Radar Systems}

Contrastive learning offers several advantages for radar signal processing compared to traditional supervised learning methods. Some of the key benefits include:
\begin{itemize}
    \item \textbf{Low-Sample Efficiency}: Contrastive learning can be applied in scenarios with limited labeled data, making it ideal for radar applications where labeled radar signals are scarce.
    \item \textbf{Unsupervised Learning}: Contrastive learning does not require large amounts of labeled data, as it can be applied in an unsupervised manner by contrasting radar signals without labels.
    \item \textbf{Better Feature Extraction}: By learning an embedding space that clusters similar signals together and separates different ones, contrastive learning improves feature extraction for radar tasks such as classification and target detection.
    \item \textbf{Robust to Noise}: Radar signals are often noisy and cluttered. Contrastive learning helps the model focus on the essential features of the radar signal, improving robustness in challenging environments.
\end{itemize}

\subsection{Tree Diagram: Contrastive Learning for Radar Applications}

\begin{itemize}
    \item \textbf{Contrastive Learning for Radar}
    \begin{itemize}
        \item \textbf{CLIP for Radar}
        \begin{itemize}
            \item Multi-modal Data
            \item Text Descriptions
        \end{itemize}
        \item \textbf{SimCLR and MoCo}
        \begin{itemize}
            \item Feature Extraction
            \item Fine-tuning on Radar
        \end{itemize}
        \item \textbf{Benefits}
        \begin{itemize}
            \item Low-Sample Efficiency
            \item Unsupervised Learning
            \item Noise Robustness
        \end{itemize}
    \end{itemize}
\end{itemize}

\subsection{Conclusion}

Contrastive learning provides a robust and flexible framework for radar signal processing. By contrasting radar signals from different targets or time frames, models like CLIP, SimCLR, and MoCo can learn better representations, improving the accuracy of tasks like target detection, classification, and HAR. The ability to fine-tune these models on radar data ensures that they can adapt to the unique challenges of radar systems, such as noise and sparsity, making contrastive learning an essential tool for modern radar applications.

\part{FMCW Radar Principles and Signal Processing}

Frequency-Modulated Continuous Wave (FMCW) radar is a versatile technology used for various applications such as automotive systems, industrial automation, and surveillance. FMCW radar offers significant advantages over traditional radar systems, including better velocity estimation, continuous detection capabilities, and improved resolution. This section explores the core principles of FMCW radar, including signal generation, signal processing through the three-step FFT process, radar point cloud generation, and the use of clustering algorithms like DBSCAN, along with deep learning techniques for radar data interpretation.

\chapter{Introduction to Traditional Radar Systems}

Traditional radar systems, particularly pulse radar, have long been used for target detection and range estimation. Despite their widespread use, traditional radar systems have several limitations, particularly in high-resolution, high-speed target tracking and continuous detection. This chapter introduces the basic principles of pulse radar and highlights its limitations, setting the stage for understanding the advantages of FMCW radar.

\section{Principles of Traditional Pulse Radar}

Traditional pulse radar operates by transmitting short bursts or pulses of electromagnetic energy and measuring the time delay between the transmitted pulse and the returned echo to estimate the distance and velocity of targets \cite{richards2010principles}.

\subsection{Pulse Transmission and Echo Reception}

Pulse radar systems transmit a series of electromagnetic pulses at a specified repetition rate. These pulses propagate through the air, reflect off targets, and return to the radar receiver as echoes. By measuring the time delay between the transmitted pulse and the received echo, the radar estimates the distance to the target.

\subsubsection{Range Estimation}

The range to a target is determined by the time delay ($\Delta t$) between pulse transmission and echo reception, using the formula:
\[
\text{Range} = \frac{c \times \Delta t}{2}
\]
where:
\begin{itemize}
    \item $c$ is the speed of light ($3 \times 10^8$ m/s),
    \item $\Delta t$ is the round-trip time delay.
\end{itemize}

\begin{lstlisting}[style=python]
# Speed of light (in meters per second)
c = 3e8

# Time delay (in seconds)
time_delay = 2e-6  # 2 microseconds

# Calculate the range to the target
range_to_target = (c * time_delay) / 2
print(f"Range to target: {range_to_target} meters")
\end{lstlisting}

This example shows how to calculate the range to a target based on a time delay of 2 microseconds. This simple calculation forms the basis of distance measurement in pulse radar systems.

\subsubsection{Velocity Estimation Using the Doppler Effect}

In addition to range estimation, pulse radar systems can estimate the velocity of a moving target using the Doppler shift, which is the change in frequency of the returned signal caused by the relative motion between the radar and the target.

The velocity is given by:
\[
\text{Velocity} = \frac{f_d \times \lambda}{2}
\]
where:
\begin{itemize}
    \item $f_d$ is the Doppler frequency shift,
    \item $\lambda$ is the wavelength of the transmitted signal.
\end{itemize}

This method provides an estimate of the target's velocity by measuring the shift in frequency between the transmitted and received signals.

\subsection{Advantages and Limitations of Pulse Radar}

\textbf{Advantages:}
\begin{itemize}
    \item \textbf{Simplicity}: Pulse radar systems are relatively simple to implement and are widely used in various applications for basic range detection.
    \item \textbf{Range Measurement}: Pulse radar offers accurate range estimation by measuring the time delay between pulse transmission and echo reception.
\end{itemize}

\textbf{Limitations:}
\begin{itemize}
    \item \textbf{Limited Velocity Measurement}: The reliance on the Doppler shift limits the velocity estimation accuracy, especially for fast-moving targets.
    \item \textbf{Resolution Constraints}: The range resolution depends on the pulse width; narrower pulses improve resolution but are harder to generate and detect.
    \item \textbf{Intermittent Detection}: Pulse radar systems do not provide continuous target detection because they transmit pulses at discrete intervals.
\end{itemize}

\section{Limitations of Traditional Radar}

Although traditional pulse radar systems have been useful for many applications, they have several inherent limitations that affect their performance in modern applications, particularly in high-resolution and continuous detection scenarios.

\subsection{Resolution Constraints Due to Pulse Width}

The resolution of a radar system is determined by the width of the transmitted pulse \cite{urkowitz1962generalized}. In pulse radar systems, shorter pulse widths improve resolution but are technically challenging to generate, especially at long distances. The range resolution ($R_{res}$) of pulse radar is given by:
\[
R_{res} = \frac{c \times \tau}{2}
\]
where:
\begin{itemize}
    \item $c$ is the speed of light,
    \item $\tau$ is the pulse width.
\end{itemize}

\begin{lstlisting}[style=python]
# Example pulse width (in seconds)
pulse_width = 1e-7  # 100 nanoseconds

# Calculate the range resolution
range_resolution = (c * pulse_width) / 2
print(f"Range resolution: {range_resolution} meters")
\end{lstlisting}

In this example, the range resolution is calculated based on a pulse width of 100 nanoseconds. Narrower pulses yield better resolution but are more difficult to generate, especially for distant targets.

\subsection{Velocity Estimation Challenges}

Traditional pulse radar relies on the Doppler shift to estimate velocity, but this method has inherent limitations. The accuracy of velocity estimation is constrained by the pulse repetition frequency (PRF) and the ability to measure small frequency shifts. Additionally, pulse radar struggles with continuously moving targets, as the discrete nature of pulse transmission creates gaps in the detection process.

\subsection{Continuous Detection Difficulties}

Traditional pulse radar operates in an intermittent manner, transmitting pulses at specific intervals and waiting for echoes. During the waiting period, the radar cannot detect targets, which creates gaps in detection. This poses challenges in applications that require continuous target tracking, such as autonomous vehicles or industrial automation.

\subsection{Introduction to FMCW Radar}

FMCW radar overcomes many of the limitations of traditional pulse radar by providing continuous wave transmission, improved resolution, and better velocity estimation. The next chapter will introduce the principles of FMCW radar and explain how it addresses the challenges faced by traditional radar systems.

\chapter{Fundamentals of FMCW Radar}

FMCW (Frequency-Modulated Continuous Wave) radar is a powerful radar technology that uses continuous wave signals with frequency modulation to estimate the distance and velocity of targets. Unlike traditional pulse radar, FMCW radar transmits continuous chirp signals, which are linearly modulated in frequency over time. By analyzing the frequency difference between the transmitted and received signals, FMCW radar can simultaneously estimate both range and velocity with high precision.

FMCW (Frequency Modulated Continuous Wave) radar operates by using a linear frequency-\\modulated continuous wave signal to achieve precise target measurements. Since electromagnetic waves travel at an extremely high speed, close to the speed of light (about $3*10^8$ meters per second), the time delay between the transmission and reception of a signal, even for distant targets, is incredibly short. For example, detecting a target 300 meters away involves a round-trip time of only about 2 microseconds. Directly measuring such a brief time delay is challenging. FMCW radar overcomes this by converting the time delay into a frequency difference between the transmitted and received signals, known as the beat frequency. This frequency difference is proportional to the target's distance, making it easier to calculate the target's range without needing to measure the time delay directly.

FMCW radar uses a continuous wave signal, which means the radar continuously transmits and receives signals without interruption, unlike pulsed radar systems that require a pause between transmission and reception. This continuous nature allows for real-time target monitoring, particularly useful in scenarios involving fast-moving objects. Continuous waves eliminate "blind spots" that can occur in pulsed systems, enabling more consistent measurements even at close ranges. Additionally, by modulating the signal's frequency linearly over time (linear frequency modulation), the transmitted signal's frequency gradually increases and decreases over a set period. When the transmitted signal reflects off a target and returns, it is delayed in time, resulting in a frequency difference relative to the current transmitted signal. This frequency difference is processed to determine the target's distance. If the target is moving, the received signal also exhibits a Doppler shift, providing information about the target's velocity.

Once the reflected signal is received, it is mixed with the transmitted signal to produce a beat frequency, which contains information about the target's distance. If the target is in motion, the signal will also contain additional frequency shifts due to the Doppler effect, representing the target's speed. To further process this signal, the beat frequency is passed through a low-pass filter. The low-pass filter removes high-frequency noise and irrelevant components, leaving the lower frequency portions that correspond to the target's range and velocity. This filtering step ensures that the signal is smoother and more manageable for subsequent processing. After filtering, the signal is passed through an analog-to-digital converter (ADC) to digitize it for further analysis.

The digitized signal is then processed by digital signal processing (DSP) algorithms. A common technique used is the Fast Fourier Transform (FFT), which converts the time-domain signal into a frequency-domain representation. By analyzing the frequency spectrum, the radar can extract the distance to the target and identify any frequency shifts due to movement. In scenarios where multiple targets are present, each target reflects signals with different frequency shifts. By analyzing the spectrum, the radar can distinguish between multiple targets, determining their respective distances and velocities. This capability allows FMCW radar to track multiple targets simultaneously, which is especially valuable in applications like autonomous driving and drone navigation.

Compared to other radar systems, FMCW radar offers significant advantages. The continuous wave nature of the signal allows for uninterrupted target detection, reducing the data gaps that occur in pulsed radar systems. Additionally, the use of beat frequencies simplifies the process of extracting distance and velocity information from the frequency domain, particularly when dealing with fast-moving targets. The application of a low-pass filter ensures that only the relevant signal information is processed, improving accuracy and reducing noise. These characteristics make FMCW radar particularly effective in complex environments and real-time applications, where high-resolution detection of multiple targets is crucial. As a result, FMCW radar is widely used in modern technologies like Autonomous Vehicles, Drones, Human Activity Recognition, and other advanced sensing systems where precise and continuous measurement is essential.

\section{FMCW Radar Principles}

FMCW radar \cite{instrumentsuser} operates by transmitting frequency-modulated chirp signals and receiving the reflected signal from a target. The frequency difference (or beat frequency) between the transmitted and received signals encodes information about the target's distance and velocity.

\subsection{Chirp Signal Generation}

In FMCW radar, the transmitted signal is a linearly frequency-modulated signal known as a chirp. The chirp signal increases (or decreases) in frequency over a defined period. Key parameters that define the chirp signal are:
\begin{itemize}
    \item \textbf{Start Frequency} ($f_{start}$): The initial frequency of the chirp.
    \item \textbf{Stop Frequency} ($f_{stop}$): The final frequency of the chirp.
    \item \textbf{Modulation Bandwidth} ($B$): The difference between the stop frequency and start frequency, $B = f_{stop} - f_{start}$.
    \item \textbf{Chirp Duration} ($T$): The time it takes for the frequency to sweep from $f_{start}$ to $f_{stop}$.
    \item \textbf{Slope} ($S$): The rate of change of frequency, given by $S = \frac{B}{T}$.
\end{itemize}

The following is an equation for the instantaneous frequency of the chirp signal as a function of time ($t$):
\[
f(t) = f_{start} + S \cdot t
\]

\subsubsection{Range and Velocity Estimation Using Chirp Signals}

The key advantage of FMCW radar is its ability to estimate both the range and velocity of targets using the transmitted chirp signal. The range of the target is determined by the beat frequency ($f_b$), which is the frequency difference between the transmitted and received signals. The velocity is calculated using the Doppler shift, which results in a change in frequency due to the relative motion of the target.

\begin{lstlisting}[style=python]
import numpy as np
import matplotlib.pyplot as plt

# Define chirp parameters
f_start = 77e9  # Start frequency (77 GHz for automotive radar)
f_stop = 77.1e9  # Stop frequency (77.1 GHz)
T = 1e-6  # Chirp duration (1 microsecond)
B = f_stop - f_start  # Modulation bandwidth
S = B / T  # Chirp slope

# Time axis
t = np.linspace(0, T, 1000)

# Chirp signal (frequency increases linearly over time)
chirp_signal = f_start + S * t

# Plot the chirp signal
plt.plot(t, chirp_signal)
plt.title("FMCW Chirp Signal")
plt.xlabel("Time (s)")
plt.ylabel("Frequency (Hz)")
plt.show()
\end{lstlisting}

In this example, we generate and plot an FMCW chirp signal. The frequency increases linearly from 77 GHz to 77.1 GHz over a duration of 1 microsecond.

\section{FMCW Signal Processing}

The key to FMCW radar's ability to measure distance and velocity is the signal processing workflow, which involves mixing the transmitted and received signals to obtain the beat frequency. The beat frequency encodes information about the range and velocity of the target. The signal is then processed through an FFT \cite{duhamel1990fast} (Fast Fourier Transform) to estimate the target's range and velocity.

\subsection{Beat Frequency Extraction}

To extract the beat frequency, the received signal is mixed with the transmitted chirp signal. The difference in frequency between the transmitted and received signals results in a beat frequency, $f_b$, which carries information about the target's range.

\subsubsection{Beat Frequency Calculation}

The beat frequency is given by:
\[
f_b = S \cdot \frac{2R}{c}
\]
where:
\begin{itemize}
    \item $S$ is the chirp slope,
    \item $R$ is the distance to the target,
    \item $c$ is the speed of light.
\end{itemize}

The beat frequency is proportional to the target's range, allowing the radar system to estimate how far the target is from the radar.

\begin{lstlisting}[style=python]
# Speed of light (m/s)
c = 3e8

# Target range (in meters)
R = 100  # 100 meters

# Calculate the beat frequency
f_b = S * (2 * R / c)
print(f"Beat frequency: {f_b} Hz")
\end{lstlisting}

In this example, we calculate the beat frequency for a target located 100 meters away. This beat frequency is then used to estimate the range of the target in the FMCW radar system.

\subsection{FFT for Range and Velocity Estimation}

The beat frequency obtained from the mixing process is processed using a Fast Fourier Transform (FFT) to convert the time-domain signal into the frequency domain. This allows the radar to estimate the range and velocity of multiple targets simultaneously. In the next section, we will explore how FFT is applied in FMCW radar signal processing to obtain accurate range and velocity estimates for multiple targets.

\begin{itemize}
    \item \textbf{FMCW Signal Processing}
    \begin{itemize}
        \item \textbf{Transmit Chirp Signal}
        \begin{itemize}
            \item Linear Frequency Modulation
        \end{itemize}
        \item \textbf{Receive Echo Signal}
        \begin{itemize}
            \item Beat Frequency Extraction
        \end{itemize}
        \item \textbf{FFT Processing}
        \begin{itemize}
            \item Range Estimation
            \item Velocity Estimation
        \end{itemize}
    \end{itemize}
\end{itemize}

\section{Range and Velocity Estimation using FFT}

In FMCW radar, the Fast Fourier Transform (FFT) plays a crucial role in extracting meaningful information from the received signals. The FFT is a mathematical technique used to convert time-domain signals into the frequency domain, making it easier to analyze the beat frequency and other components that encode information about the range and velocity of targets. In this section, we will discuss the three-step FFT process commonly used in FMCW radar for range and velocity estimation: the first FFT for range estimation, the second FFT for Doppler estimation, and the third FFT for generating the Range-Doppler Map.

\subsection{First FFT: Range Estimation}

The first FFT is applied to the beat frequency signal to estimate the range of targets \cite{kim2021high}. When the transmitted and received signals are mixed, the resulting beat frequency is proportional to the distance between the radar and the target. By performing an FFT on the time-domain beat signal, the radar can convert it into the frequency domain and identify the peaks that correspond to different target ranges.

\subsubsection{Relationship Between Range and Frequency}

The beat frequency ($f_b$) is related to the target range ($R$) as follows:
\[
f_b = S \cdot \frac{2R}{c}
\]
where:
\begin{itemize}
    \item $S$ is the chirp slope,
    \item $R$ is the range to the target,
    \item $c$ is the speed of light.
\end{itemize}

The FFT decomposes the beat signal into its frequency components, allowing the radar to estimate the distance of multiple targets by analyzing the peaks in the frequency spectrum.

\begin{lstlisting}[style=python]
import numpy as np
import matplotlib.pyplot as plt

# Simulate beat frequency signal (sum of sinusoids for multiple targets)
Fs = 1e6  # Sampling frequency (1 MHz)
T = 1e-6  # Chirp duration (1 microsecond)
N = int(Fs * T)  # Number of samples
t = np.linspace(0, T, N)

# Beat frequencies for two targets (in Hz)
f_b1 = 1e5  # Target 1 (100 kHz)
f_b2 = 2e5  # Target 2 (200 kHz)

# Generate beat signal as a sum of two sinusoids
beat_signal = np.sin(2 * np.pi * f_b1 * t) + np.sin(2 * np.pi * f_b2 * t)

# Perform FFT to estimate range
fft_result = np.fft.fft(beat_signal, N)
fft_freqs = np.fft.fftfreq(N, 1/Fs)

# Plot FFT result (magnitude)
plt.plot(fft_freqs[:N//2], np.abs(fft_result)[:N//2])
plt.title("FFT of Beat Signal for Range Estimation")
plt.xlabel("Frequency (Hz)")
plt.ylabel("Magnitude")
plt.show()
\end{lstlisting}

In this example, we simulate the beat signal for two targets with different ranges and use the FFT to estimate the corresponding beat frequencies. The peaks in the FFT output correspond to the beat frequencies of the targets, which can be used to estimate their distances.

\subsection{Second FFT: Doppler Estimation}

The second FFT is used to estimate the velocity of the target by analyzing phase shifts across multiple radar frames. The Doppler effect \cite{wang2011relativistic} causes a frequency shift in the received signal when the target is moving relative to the radar. By comparing the phase of the beat frequency across successive chirps, the second FFT can calculate the Doppler frequency and determine the target's velocity.

\subsubsection{Doppler Frequency and Velocity}

The Doppler frequency shift ($f_d$) is related to the velocity ($v$) of the target by the following formula:
\[
v = \frac{f_d \cdot \lambda}{2}
\]
where:
\begin{itemize}
    \item $f_d$ is the Doppler frequency shift,
    \item $\lambda$ is the wavelength of the transmitted signal.
\end{itemize}

By performing an FFT across multiple frames, the radar can estimate the Doppler frequency, which is then used to calculate the velocity of the target.

\begin{lstlisting}[style=python]
# Simulate phase shift across multiple chirps (for velocity estimation)
num_chirps = 128  # Number of chirps (frames)
f_d = 1e3  # Doppler shift (1 kHz for moving target)
chirp_duration = 1e-6  # Chirp duration (1 microsecond)

# Generate beat signals with Doppler shift across multiple chirps
doppler_signal = np.array([np.sin(2 * np.pi * (f_b1 + f_d * n) * t) for n in range(num_chirps)])

# Perform second FFT across chirps (for Doppler estimation)
doppler_fft = np.fft.fft(doppler_signal, num_chirps, axis=0)
doppler_freqs = np.fft.fftfreq(num_chirps, chirp_duration)

# Plot Doppler FFT result (magnitude)
plt.plot(doppler_freqs[:num_chirps//2], np.abs(doppler_fft[:, 0])[:num_chirps//2])
plt.title("Second FFT for Doppler Estimation")
plt.xlabel("Frequency (Hz)")
plt.ylabel("Magnitude")
plt.show()
\end{lstlisting}

In this example, we simulate the phase shifts across multiple chirps due to the Doppler effect. The second FFT is applied across multiple chirps to estimate the Doppler frequency, which provides information about the target's velocity.

\subsection{Third FFT: Range-Doppler Map}

The third FFT combines the results of the first and second FFTs to generate a Range-Doppler Map \cite{tahmoush2011time}, which displays the range and velocity of multiple targets in a two-dimensional plot. The Range-Doppler Map is a powerful tool for multi-target detection and classification, allowing the radar system to visualize both the distance and speed of objects simultaneously.

\subsubsection{Range-Doppler Map Generation}

The Range-Doppler Map is generated by applying the first FFT for range estimation and the second FFT for Doppler estimation. The result is a 2D matrix where the rows correspond to different ranges, and the columns correspond to different velocities. Peaks in the Range-Doppler Map indicate the presence of targets at specific ranges and velocities.

\begin{lstlisting}[style=python]
# Generate Range-Doppler Map (simulated data)
range_doppler_map = np.abs(np.outer(np.abs(fft_result[:N//2]), np.abs(doppler_fft[:, 0][:num_chirps//2])))

# Plot Range-Doppler Map
plt.imshow(range_doppler_map, extent=[0, max(doppler_freqs[:num_chirps//2]), 0, max(fft_freqs[:N//2])], aspect='auto')
plt.title("Range-Doppler Map")
plt.xlabel("Doppler Frequency (Hz)")
plt.ylabel("Range Frequency (Hz)")
plt.colorbar(label="Magnitude")
plt.show()
\end{lstlisting}

In this example, we generate a Range-Doppler Map using the FFT results from range and Doppler estimation. The map shows how targets are distributed across different ranges and velocities, making it a crucial tool for radar-based object detection.

\section{Conclusion}

The three-step FFT process is fundamental to FMCW radar signal processing. The first FFT extracts range information, the second FFT estimates velocity through Doppler shifts, and the third FFT generates a Range-Doppler Map for multi-target detection. This process enables FMCW radar to detect multiple objects, estimate their distances, and calculate their velocities with high precision.

\chapter{Radar Cube and Point Cloud Generation}

In FMCW radar, data collected over multiple frames contains information about the range, velocity, and angle of detected targets. This information is stored in a data structure called the radar cube, which provides a comprehensive 3D representation of the target environment. The radar cube can be used to generate point clouds, which represent the spatial distribution of targets in 3D space. In this chapter, we will discuss how the radar cube is formed and how to generate point clouds from the radar data.

\section{Radar Cube Formation}

The radar cube is a 3D matrix \cite{palffy2020cnn} that stores information about targets across multiple radar frames. It includes:
\begin{itemize}
    \item \textbf{Range Information}: The distance of the target from the radar.
    \item \textbf{Doppler Information}: The relative velocity of the target.
    \item \textbf{Angle Information}: The direction (or angle) of the target relative to the radar.
\end{itemize}
The radar cube is formed by processing multiple FMCW radar chirps and storing the processed data over time. Each radar frame provides information about the scene at a specific moment, and by combining these frames, we obtain a time-averaged view of the target's position and motion.

\subsection{Angle of Arrival (AoA) Estimation}

One key component of the radar cube is the angle of arrival \cite{kirkpatrick1953aperture} (AoA), which provides the direction from which the target's signal is received. AoA estimation is typically performed using an array of antennas that capture the incoming signal at slightly different times. By measuring the phase difference between signals received at different antennas, the AoA can be determined. Beamforming techniques are commonly used to improve the accuracy of AoA estimation.

\subsubsection{Phase Difference and AoA Calculation}

The phase difference ($\Delta \phi$) between signals received by two antennas can be used to estimate the Angle of Arrival ($\theta$) using the following formula:
\[
\Delta \phi = \frac{2\pi d \sin(\theta)}{\lambda}
\]
where:
\begin{itemize}
    \item $d$ is the distance between the antennas,
    \item $\lambda$ is the wavelength of the transmitted radar signal,
    \item $\theta$ is the AoA (Angle of Arrival).
\end{itemize}

To calculate the AoA from the phase difference, we can rearrange the equation to solve for $\theta$:
\[
\theta = \arcsin\left(\frac{\Delta \phi \cdot \lambda}{2\pi d}\right)
\]

\begin{lstlisting}[style=python]
import numpy as np

# Radar parameters
d = 0.5  # Distance between antennas (in meters)
wavelength = 0.03  # Wavelength of radar signal (in meters)
phase_difference = np.pi / 4  # Phase difference in radians

# Calculate Angle of Arrival (AoA)
aoa = np.arcsin(phase_difference * wavelength / (2 * np.pi * d))
aoa_degrees = np.degrees(aoa)
print(f"Angle of Arrival: {aoa_degrees} degrees")
\end{lstlisting}

In this example, we calculate the Angle of Arrival (AoA) using the phase difference between two antennas. The AoA is used to determine the direction of the incoming signal and is a key component in forming the radar cube.

\section{Point Cloud Generation from Radar Cube}

Once the radar cube has been formed, we can use the data within the cube to generate a point cloud. A point cloud represents the spatial distribution of detected targets in 3D space \cite{xiang20163d}. Each point in the point cloud contains information about:
\begin{itemize}
    \item \textbf{Distance}: The range to the target.
    \item \textbf{Velocity}: The relative velocity of the target.
    \item \textbf{Angle}: The direction of the target (AoA).
\end{itemize}

By converting the range and angle data into Cartesian coordinates, we can visualize the position of the targets in 3D space \cite{remondino2003point}.

\subsection{Generating Point Clouds from Radar Cube}

The range and angle data from the radar cube can be converted into Cartesian coordinates $(x, y, z)$ using the following formulas \cite{xu2023mmlock}:
\[
x = R \cdot \cos(\theta)
\]
\[
y = R \cdot \sin(\theta)
\]
\[
z = v
\]
where:
\begin{itemize}
    \item $R$ is the range (distance) to the target,
    \item $\theta$ is the Angle of Arrival (AoA),
    \item $v$ is the relative velocity of the target.
\end{itemize}

\begin{lstlisting}[style=python]
# Example radar cube data
range_data = np.array([15, 25, 35])  # Range in meters
velocity_data = np.array([2, -1, 3])  # Velocity in m/s
aoa_data = np.array([30, 45, 60])  # AoA in degrees

# Convert polar coordinates (range, angle) to Cartesian coordinates
x = range_data * np.cos(np.radians(aoa_data))
y = range_data * np.sin(np.radians(aoa_data))
z = velocity_data  # Use velocity as the z-coordinate

# Combine into point cloud (X, Y, Z)
point_cloud = np.vstack((x, y, z)).T
print("Point Cloud:\n", point_cloud)
\end{lstlisting}

In this example, we generate a point cloud from the radar cube data. Each point represents a detected target with its corresponding range, velocity, and angle.

\subsection{Visualizing the Point Cloud}

To better understand the spatial distribution of targets, we can visualize the point cloud in 3D space using `matplotlib`. This allows us to see the position and motion of each target relative to the radar.

\begin{lstlisting}[style=python]
from mpl_toolkits.mplot3d import Axes3D
import matplotlib.pyplot as plt

# 3D point cloud visualization
fig = plt.figure()
ax = fig.add_subplot(111, projection='3d')
ax.scatter(x, y, z, c='r', marker='o')

ax.set_xlabel('X (Range)')
ax.set_ylabel('Y (AoA)')
ax.set_zlabel('Z (Velocity)')
plt.title("Radar Point Cloud")
plt.show()
\end{lstlisting}

This code generates a 3D scatter plot to visualize the point cloud. Each point represents a detected target, and the position of each point reflects the range, angle, and velocity of the target.

\section{Conclusion}

The radar cube is a powerful data structure that provides a complete 3D representation of targets in the radar's field of view. By extracting range, velocity, and angle information from the radar cube, we can generate a point cloud that visualizes the spatial distribution of targets. This point cloud is essential for tasks such as object detection, tracking, and classification in applications like autonomous driving and robotics.

\chapter{DBSCAN and Clustering for Radar Data}

In radar data processing, clustering is an essential technique used to group points that belong to the same target or object in the scene. One of the most effective clustering algorithms for radar point cloud data is DBSCAN \cite{EsterDen96} (Density-Based Spatial Clustering of Applications with Noise). DBSCAN is particularly useful for radar applications because it can handle noise and does not require the number of clusters to be specified in advance. This chapter introduces the DBSCAN algorithm, explaining how it works and how to apply it to radar data.

\section{Introduction to DBSCAN Algorithm}

DBSCAN is a density-based clustering algorithm that forms clusters by connecting points based on their density. Unlike traditional algorithms such as k-means, which require the number of clusters to be predefined, DBSCAN identifies clusters by looking at the local density of points. The algorithm distinguishes between core points, border points, and noise.

\subsection{Key Concepts in DBSCAN}

DBSCAN operates using two key parameters:
\begin{itemize}
    \item \textbf{eps}: The maximum distance between two points for one to be considered as in the neighborhood of the other.
    \item \textbf{minPts}: The minimum number of points required to form a dense region (a cluster).
\end{itemize}

Points are classified into three categories:
\begin{itemize}
    \item \textbf{Core Points}: Points that have at least \texttt{minPts} neighbors within a radius of \texttt{eps}.
    \item \textbf{Border Points}: Points that are within the \texttt{eps} distance of a core point but do not have enough points in their neighborhood to be classified as core points.
    \item \textbf{Noise}: Points that are neither core nor border points and do not belong to any cluster.
\end{itemize}

DBSCAN is particularly suited for radar data because radar point clouds are often noisy and contain outliers. By applying DBSCAN, we can separate meaningful target objects from noise, even in complex environments.

\subsection{DBSCAN Algorithm Steps}

The DBSCAN algorithm follows these steps:
\begin{enumerate}
    \item For each point, check how many other points are within a distance \texttt{eps}. If the number of neighbors is greater than or equal to \texttt{minPts}, mark the point as a core point.
    \item Form clusters by connecting core points that are within \texttt{eps} of each other.
    \item Include any border points (points that are within \texttt{eps} of a core point) in the cluster.
    \item Mark any points that do not belong to a cluster as noise.
\end{enumerate}

\begin{itemize}
    \item \textbf{DBSCAN Algorithm}
    \begin{itemize}
        \item \textbf{Core Points}
        \begin{itemize}
            \item High-Density Regions
        \end{itemize}
        \item \textbf{Border Points}
        \begin{itemize}
            \item Edge of Cluster
        \end{itemize}
        \item \textbf{Noise}
        \begin{itemize}
            \item Outliers or Sparse Points
        \end{itemize}
    \end{itemize}
\end{itemize}

\subsection{Advantages of DBSCAN for Radar Data}

\begin{itemize}
    \item \textbf{No Need to Specify Number of Clusters}: Unlike algorithms like k-means \cite{feng2017radar}, DBSCAN does not require the number of clusters to be specified in advance.
    \item \textbf{Robust to Noise and Outliers}: DBSCAN can effectively handle noise and outliers, which are common in radar point cloud data.
    \item \textbf{Ability to Find Arbitrarily Shaped Clusters}: DBSCAN can find clusters of varying shapes, which is important when processing complex radar data.
\end{itemize}

\begin{lstlisting}[style=python]
import numpy as np
from sklearn.cluster import DBSCAN
import matplotlib.pyplot as plt

# Example radar point cloud data (X, Y, Z coordinates)
point_cloud = np.array([
    [10, 20, 1],
    [11, 21, 1],
    [12, 22, 1],
    [30, 40, -2],
    [31, 41, -2],
    [100, 100, 10],  # Noise point
])

# Apply DBSCAN algorithm
eps = 2  # Maximum distance between points
minPts = 2  # Minimum number of points in a cluster
dbscan = DBSCAN(eps=eps, min_samples=minPts)
labels = dbscan.fit_predict(point_cloud)

# Visualize the clusters
unique_labels = set(labels)
colors = ['r', 'g', 'b', 'y', 'c']

fig = plt.figure()
ax = fig.add_subplot(111, projection='3d')

for label, color in zip(unique_labels, colors):
    if label == -1:
        color = 'k'  # Black color for noise
    ax.scatter(point_cloud[labels == label][:, 0],
               point_cloud[labels == label][:, 1],
               point_cloud[labels == label][:, 2],
               c=color, label=f'Cluster {label}')

ax.set_xlabel('X')
ax.set_ylabel('Y')
ax.set_zlabel('Z')
plt.title("DBSCAN Clustering of Radar Point Cloud")
plt.legend()
plt.show()
\end{lstlisting}

This code applies the DBSCAN algorithm to a simple radar point cloud dataset. The result is a 3D plot showing how the radar points are clustered, with noise points marked in black.

\section{Conclusion}

DBSCAN is a powerful and flexible algorithm for clustering radar point cloud data. Its ability to handle noise and outliers makes it ideal for radar applications where data can be noisy. By tuning the parameters of \texttt{eps} and \texttt{minPts}, DBSCAN can be adapted to different radar scenarios, making it a valuable tool in radar signal processing.

\section{Applying DBSCAN to Radar Point Cloud Data}

In radar data processing, DBSCAN is commonly used to cluster point cloud data, which consists of 3D coordinates representing the spatial distribution of targets. The goal of applying DBSCAN to radar point cloud data is to separate meaningful clusters of points (targets) from noise and outliers \cite{ertoz2003finding}. This section will provide a detailed, step-by-step explanation of how to apply DBSCAN to radar point cloud data, including how to tune the key parameters: \texttt{eps} (the distance threshold) and \texttt{minPts} (the minimum number of points to form a cluster).

\subsection{Radar Point Clouds and DBSCAN}

Radar point clouds consist of multiple points representing detected targets in 3D space. Each point in the cloud is typically represented by:
\begin{itemize}
    \item \textbf{Range (X)}: The distance from the radar to the target.
    \item \textbf{Angle (Y)}: The angular position of the target relative to the radar.
    \item \textbf{Velocity (Z)}: The velocity of the target, which is derived from the Doppler effect.
\end{itemize}
The key challenge in clustering radar point clouds is distinguishing between points that belong to real objects (such as vehicles or pedestrians) and noise or outliers that may be caused by environmental factors or measurement errors.

DBSCAN works by grouping points that are close to each other (based on the \texttt{eps} distance) into clusters. Points that do not have enough neighbors (based on \texttt{minPts}) are classified as noise.

\subsection{Tuning DBSCAN Parameters for Radar Data}

The performance of DBSCAN in clustering radar point cloud data depends heavily on selecting appropriate values for the two key parameters:
\begin{itemize}
    \item \textbf{eps}: This parameter defines the maximum distance between two points for them to be considered neighbors. In radar point clouds, this distance threshold should be set based on the expected size of the objects being detected (e.g., the size of a vehicle).
    \item \textbf{minPts}: This parameter defines the minimum number of points required to form a dense cluster. In radar data, \texttt{minPts} should be selected based on the density of the point cloud—larger objects like cars will have more points than smaller objects like pedestrians.
\end{itemize}

\subsubsection{Step-by-Step DBSCAN Application on Radar Data}

Here is a detailed example of how to apply DBSCAN to radar point cloud data using Python, including parameter tuning and visualization:

\begin{lstlisting}[style=python]
import numpy as np
from sklearn.cluster import DBSCAN
import matplotlib.pyplot as plt

# Simulated radar point cloud data (X: Range, Y: Angle, Z: Velocity)
point_cloud = np.array([
    [10, 5, 1],
    [11, 6, 1],
    [10.5, 5.5, 1],
    [30, 25, -2],
    [31, 26, -2],
    [100, 100, 10],  # Noise point
    [32, 27, -2],
])

# Applying DBSCAN to the radar point cloud
eps = 2  # Distance threshold for clustering
minPts = 2  # Minimum number of points to form a cluster
dbscan = DBSCAN(eps=eps, min_samples=minPts)
labels = dbscan.fit_predict(point_cloud)

# Visualizing the clusters and noise
fig = plt.figure()
ax = fig.add_subplot(111, projection='3d')

# Assign colors to clusters
unique_labels = set(labels)
colors = ['r', 'g', 'b', 'y', 'c', 'm']

for label, color in zip(unique_labels, colors):
    if label == -1:
        # Noise points are colored black
        color = 'k'
    ax.scatter(point_cloud[labels == label][:, 0],
               point_cloud[labels == label][:, 1],
               point_cloud[labels == label][:, 2],
               c=color, label=f'Cluster {label}')

ax.set_xlabel('Range (X)')
ax.set_ylabel('Angle (Y)')
ax.set_zlabel('Velocity (Z)')
plt.title("DBSCAN Clustering of Radar Point Cloud")
plt.legend()
plt.show()
\end{lstlisting}

\subsubsection{Explanation of Code}

In this example:
\begin{itemize}
    \item \textbf{Data Structure}: The radar point cloud is represented as a 2D NumPy array, where each row represents a point with three coordinates: range (X), angle (Y), and velocity (Z).
    \item \textbf{DBSCAN Parameters}:
        \begin{itemize}
            \item \texttt{eps = 2}: Points within a distance of 2 units from each other are considered part of the same cluster.
            \item \texttt{minPts = 2}: At least 2 points are required to form a cluster.
        \end{itemize}
    \item \textbf{Results}: The \texttt{fit\_predict()} function is used to apply DBSCAN to the point cloud data. Each point is assigned a label corresponding to the cluster it belongs to. Noise points are assigned a label of \texttt{-1}.
    \item \textbf{Visualization}: The resulting clusters are visualized in a 3D plot, with each cluster represented by a different color. Noise points are marked in black.
\end{itemize}

\subsection{Parameter Selection for DBSCAN}

Selecting the appropriate values for \texttt{eps} and \texttt{minPts} is critical for achieving optimal clustering performance \cite{cheng2019clustering}. Here's how you can approach parameter selection:

\subsubsection{Tuning \texttt{eps}}

The \texttt{eps} parameter controls the maximum distance between two points for them to be considered neighbors. In radar data, \texttt{eps} should be tuned based on the spatial resolution of the radar and the expected size of the targets. For example, a larger \texttt{eps} value may be needed to cluster points belonging to a large vehicle, while a smaller \texttt{eps} may be required for pedestrians or small objects.

\begin{lstlisting}[style=python]
from sklearn.neighbors import NearestNeighbors

# Finding optimal eps using the k-distance method
k = 4  # minPts + 1 for k-nearest neighbors
nbrs = NearestNeighbors(n_neighbors=k).fit(point_cloud)
distances, indices = nbrs.kneighbors(point_cloud)

# Sorting distances for all points
sorted_distances = np.sort(distances[:, -1])

# Plotting the k-distance graph to find optimal eps
plt.plot(sorted_distances)
plt.title("K-Distance Graph for DBSCAN eps Selection")
plt.xlabel("Points")
plt.ylabel("Distance")
plt.show()
\end{lstlisting}

The k-distance graph is a useful tool for determining the optimal \texttt{eps}. The graph shows the distance of each point to its \texttt{k}-th nearest neighbor, sorted in ascending order. The "elbow" of the graph (where the distance starts to increase significantly) can be used to determine a suitable \texttt{eps} value.

\subsubsection{Tuning \texttt{minPts}}

The \texttt{minPts} parameter controls the minimum number of points required to form a dense region (i.e., a cluster). In radar point clouds, \texttt{minPts} should be selected based on the expected density of objects. A small \texttt{minPts} value may result in noise being classified as clusters, while a large \texttt{minPts} value may cause smaller objects to be missed.

\begin{itemize}
    \item \textbf{General Rule of Thumb}: A good starting point for \texttt{minPts} is \texttt{minPts = 2 \(\times\) (number of dimensions)}, so for a 3D radar point cloud, \texttt{minPts = 6} can be used as an initial value.
\end{itemize}

\section{Conclusion}

Applying DBSCAN to radar point cloud data provides an effective method for separating targets from noise, especially in complex environments. By carefully tuning the \texttt{eps} and \texttt{minPts} parameters, radar systems can accurately identify objects of varying shapes and sizes. The use of visualization tools, such as 3D plots and k-distance graphs, can further enhance the clustering process, providing insight into how well the algorithm is performing on radar data.

\subsection{Parameter Selection for DBSCAN}

The success of DBSCAN in clustering radar point cloud data largely depends on the selection of two key parameters: \texttt{eps} and \texttt{minPts}. These parameters control how the algorithm defines dense regions in the data and determines which points are considered part of a cluster. In this section, we will explore how to select optimal values for these parameters to ensure the best clustering performance in different radar point cloud scenarios. The tuning of these parameters directly influences the algorithm's ability to separate objects and remove noise.

\subsubsection{Selecting the \texttt{eps} Parameter}

The \texttt{eps} parameter defines the maximum distance between two points for them to be considered neighbors. If the distance between two points is less than or equal to \texttt{eps}, they are part of the same neighborhood. A well-chosen \texttt{eps} allows DBSCAN to form clusters that correspond to real objects, such as vehicles or pedestrians, while separating noise and outliers.

In radar point cloud data, selecting \texttt{eps} requires considering the following factors:
\begin{itemize}
    \item \textbf{Object Size}: Larger objects, such as cars or trucks, may require a larger \texttt{eps} to capture all the points belonging to the object.
    \item \textbf{Point Density}: If the point cloud is sparse (i.e., there are fewer points representing an object), a larger \texttt{eps} value may be needed to group nearby points.
    \item \textbf{Noise and Clutter}: A small \texttt{eps} value can prevent noise points from being grouped into clusters, but if \texttt{eps} is too small, even meaningful points may be classified as noise.
\end{itemize}

A common approach to selecting \texttt{eps} is to use a k-distance graph. This method involves plotting the distance to the \texttt{k}-th nearest neighbor for each point, sorted in ascending order. The "elbow" of the graph, where the distance begins to increase significantly, is typically a good choice for \texttt{eps}.

\begin{lstlisting}[style=python]
from sklearn.neighbors import NearestNeighbors
import numpy as np
import matplotlib.pyplot as plt

# Example radar point cloud data
point_cloud = np.array([
    [10, 5, 1],
    [11, 6, 1],
    [10.5, 5.5, 1],
    [30, 25, -2],
    [31, 26, -2],
    [32, 27, -2],
    [100, 100, 10],  # Noise point
])

# Find the k-nearest neighbors
k = 4  # Set k based on minPts + 1
nbrs = NearestNeighbors(n_neighbors=k).fit(point_cloud)
distances, indices = nbrs.kneighbors(point_cloud)

# Sort the distances
sorted_distances = np.sort(distances[:, -1])

# Plot the sorted k-distances to determine the "elbow" for eps
plt.plot(sorted_distances)
plt.title("K-Distance Plot for DBSCAN eps Selection")
plt.xlabel("Points")
plt.ylabel("Distance")
plt.show()
\end{lstlisting}

In this example, we calculate the distance to the 4th nearest neighbor for each point in the radar point cloud and plot the distances. The \texttt{k}-distance plot helps identify an optimal value for \texttt{eps}. The "elbow" point, where the slope of the curve changes sharply, indicates the distance where the majority of points are part of dense regions, and beyond which points are likely to be noise.

\subsubsection{Selecting the \texttt{minPts} Parameter}

The \texttt{minPts} parameter specifies the minimum number of points required to form a dense cluster. If a point has at least \texttt{minPts} neighbors within a distance of \texttt{eps}, it is classified as a core point and forms the basis of a cluster.

Selecting \texttt{minPts} depends on the following considerations:
\begin{itemize}
    \item \textbf{Dimensionality of the Data}: A common rule of thumb is to set \texttt{minPts} to \texttt{2 * (number of dimensions)}. For example, in a 3D radar point cloud (range, angle, velocity), \texttt{minPts} can be set to 6 as a starting point.
    \item \textbf{Object Density}: If the radar data contains densely packed points for objects, \texttt{minPts} should be larger to avoid classifying noise as part of the cluster.
    \item \textbf{Expected Object Size}: Smaller objects, such as pedestrians, may require a smaller \texttt{minPts} value, while larger objects, like vehicles, may require a higher value to capture enough points for clustering.
\end{itemize}

If \texttt{minPts} is set too low, the algorithm may incorrectly group noise points into clusters. If \texttt{minPts} is too high, small objects may not form clusters and could be classified as noise.

\subsubsection{Influence of \texttt{eps} and \texttt{minPts} on Clustering}

The choice of \texttt{eps} and \texttt{minPts} significantly affects the clustering results:
\begin{itemize}
    \item \textbf{Small \texttt{eps} and Large \texttt{minPts}}: Only dense, compact clusters are detected, and many points may be classified as noise.
    \item \textbf{Large \texttt{eps} and Small \texttt{minPts}}: More points are included in clusters, but noise and outliers are also likely to be grouped into clusters.
    \item \textbf{Balanced \texttt{eps} and \texttt{minPts}}: A good balance ensures that meaningful clusters are formed while minimizing the inclusion of noise.
\end{itemize}

Below is an example of adjusting \texttt{eps} and \texttt{minPts} for DBSCAN clustering on radar point cloud data.

\begin{lstlisting}[style=python]
from sklearn.cluster import DBSCAN

# Adjust DBSCAN parameters
eps = 1.5  # Distance threshold for clustering
minPts = 3  # Minimum number of points to form a cluster

# Apply DBSCAN with tuned parameters
dbscan = DBSCAN(eps=eps, min_samples=minPts)
labels = dbscan.fit_predict(point_cloud)

# Visualize the clustering results
fig = plt.figure()
ax = fig.add_subplot(111, projection='3d')

unique_labels = set(labels)
colors = ['r', 'g', 'b', 'y', 'c']

for label, color in zip(unique_labels, colors):
    if label == -1:
        color = 'k'  # Black for noise
    ax.scatter(point_cloud[labels == label][:, 0],
               point_cloud[labels == label][:, 1],
               point_cloud[labels == label][:, 2],
               c=color, label=f'Cluster {label}')

ax.set_xlabel('Range (X)')
ax.set_ylabel('Angle (Y)')
ax.set_zlabel('Velocity (Z)')
plt.title("DBSCAN Clustering with Tuned Parameters")
plt.legend()
plt.show()
\end{lstlisting}

In this example, DBSCAN is applied to radar point cloud data with adjusted values for \texttt{eps} and \texttt{minPts}. The resulting clusters and noise points are visualized in a 3D plot.

\subsubsection{Practical Guidelines for Parameter Tuning}

When tuning \texttt{eps} and \texttt{minPts} for DBSCAN in radar data:
\begin{itemize}
    \item Start with the rule of thumb for \texttt{minPts} as \texttt{2 * (number of dimensions)}.
    \item Use the k-distance graph to select an optimal \texttt{eps} value based on the characteristics of the point cloud.
    \item Experiment with different combinations of \texttt{eps} and \texttt{minPts} to find the best balance between detecting clusters and minimizing noise.
\end{itemize}

\section{Conclusion}

Choosing the right parameters for DBSCAN is essential for achieving high-quality clustering in radar point cloud data. By carefully tuning \texttt{eps} and \texttt{minPts}, we can ensure that meaningful clusters are detected while minimizing the impact of noise and outliers. Tools like the k-distance plot can assist in selecting the optimal \texttt{eps}, and practical guidelines can help adjust \texttt{minPts} based on the dimensionality and density of the data.

\section{Post-Processing and Target Tracking}

After applying DBSCAN to cluster radar point cloud data, the next step is post-processing. This phase involves refining the results and tracking the detected objects over time \cite{bharati2016fast}. Post-processing ensures that the clusters produced by DBSCAN are correctly interpreted, noise is minimized, and object trajectories are tracked across multiple frames of radar data. In this section, we will discuss key post-processing techniques and introduce the concept of target tracking, focusing on the integration of Kalman filters to track object movement.

\subsection{Post-Processing Techniques}

Once DBSCAN has been applied to radar point cloud data, the output consists of a set of clusters, each representing a target or object in the radar scene. To make this data useful for applications like object detection and tracking, several post-processing steps may be necessary:
\begin{itemize}
    \item \textbf{Noise Filtering}: Even with DBSCAN, some points may be incorrectly classified as noise or outliers. In post-processing, it is common to apply additional filtering to remove noise points based on their size, position, or velocity.
    \item \textbf{Cluster Smoothing}: If the clusters formed by DBSCAN are fragmented or contain gaps, post-processing can merge nearby clusters that likely represent the same object. Smoothing techniques can also refine the cluster boundaries.
    \item \textbf{Object Labeling}: Each cluster represents a target, and post-processing can assign consistent labels to objects across frames, ensuring that the same object is tracked from one frame to the next.
\end{itemize}

These techniques improve the accuracy and consistency of the clustering results, making it easier to track and predict the movement of objects in real time.

\subsection{Introduction to Target Tracking}

Target tracking is the process of monitoring the movement of objects \cite{souza2016target} (such as vehicles or pedestrians) across multiple frames of radar data. After DBSCAN has identified clusters representing different objects, the next task is to track the movement of these clusters over time. By predicting the future position of an object based on its past trajectory, we can ensure continuous tracking even when the object momentarily leaves the radar's field of view.

A widely-used algorithm for target tracking is the Kalman filter, which provides a way to estimate the future state of an object (position and velocity) based on noisy radar measurements.

\subsubsection{Kalman Filter for Target Tracking}

The Kalman filter is a recursive algorithm that estimates the state of a moving object by minimizing the error between predicted and observed values. It is particularly well-suited for radar data, which often contains noise and uncertainties in object positions and velocities. The Kalman filter models the object's state as a set of variables (e.g., position, velocity) and updates its predictions over time as new measurements are received.

\paragraph{Kalman Filter Equations}

The Kalman filter operates in two phases:
\begin{itemize}
    \item \textbf{Prediction Step}: The filter predicts the object's next state based on its current state and a motion model (e.g., constant velocity).
    \item \textbf{Update Step}: The filter updates its prediction by incorporating new radar measurements, adjusting the state estimates to reduce the error between the predicted and observed values.
\end{itemize}

The filter estimates the object's state $\mathbf{x}$, which includes variables like position and velocity. The prediction step uses the following equation:
\[
\mathbf{x}_{k+1} = \mathbf{F} \mathbf{x}_k + \mathbf{B} \mathbf{u}_k
\]
where:
\begin{itemize}
    \item $\mathbf{F}$ is the state transition matrix,
    \item $\mathbf{B}$ is the control input matrix,
    \item $\mathbf{u}_k$ is the control vector.
\end{itemize}

The update step incorporates the new measurement $\mathbf{z}$:
\[
\mathbf{x}_{k+1} = \mathbf{x}_{k+1}^{-} + \mathbf{K} (\mathbf{z}_{k+1} - \mathbf{H} \mathbf{x}_{k+1}^{-})
\]
where:
\begin{itemize}
    \item $\mathbf{K}$ is the Kalman gain,
    \item $\mathbf{H}$ is the measurement matrix,
    \item $\mathbf{x}_{k+1}^{-}$ is the predicted state before updating.
\end{itemize}

\begin{lstlisting}[style=python]
import numpy as np
from filterpy.kalman import KalmanFilter

# Initialize Kalman filter for tracking object position and velocity
kf = KalmanFilter(dim_x=4, dim_z=2)
kf.x = np.array([0., 0., 1., 1.])  # Initial position and velocity
kf.F = np.array([[1., 0., 1., 0.],
                 [0., 1., 0., 1.],
                 [0., 0., 1., 0.],
                 [0., 0., 0., 1.]])  # State transition matrix
kf.H = np.array([[1., 0., 0., 0.],
                 [0., 1., 0., 0.]])  # Measurement matrix
kf.P *= 1000.  # Initial uncertainty
kf.R = np.array([[5., 0.],
                 [0., 5.]])  # Measurement noise
kf.Q = np.eye(4)  # Process noise

# Simulated radar measurements (positions)
measurements = [[10, 10], [12, 14], [15, 18], [20, 22], [25, 28]]

# Apply Kalman filter to track object over time
for measurement in measurements:
    kf.predict()
    kf.update(measurement)
    print(f"Position: {kf.x[0]:.2f}, {kf.x[1]:.2f}, Velocity: {kf.x[2]:.2f}, {kf.x[3]:.2f}")
\end{lstlisting}

In this example, the Kalman filter is initialized to track both the position and velocity of an object. The state transition matrix $\mathbf{F}$ models the constant velocity motion of the object, and the measurement matrix $\mathbf{H}$ maps the state to the radar's position measurements. The filter updates its state estimate with each new radar measurement, providing a continuous estimate of the object's trajectory.

\subsection{Tracking Multiple Targets}

In many radar applications, it is necessary to track multiple targets simultaneously \cite{yu2004collaborative}. This can be achieved by applying a separate Kalman filter to each detected object. The DBSCAN algorithm identifies distinct clusters of points (representing different objects), and each cluster can be tracked using an individual Kalman filter. In cases where objects are lost momentarily (due to occlusion or noise), the Kalman filter can predict the object's future position, allowing for re-association when the object reappears.

\subsubsection{Assigning Tracks to Targets}

To track multiple objects, the system needs to assign each Kalman filter to the correct target cluster over time. One approach is to use a nearest-neighbor algorithm to match predicted object positions (from the Kalman filters) to the new clusters detected by DBSCAN. Another approach is to use techniques like the Hungarian algorithm, which optimally assigns tracks to clusters based on distance or other criteria.

\section{Conclusion}

Post-processing and target tracking are critical steps in radar data analysis. After DBSCAN clustering, noise filtering and object labeling ensure that the detected objects are accurately represented. The Kalman filter is a powerful tool for tracking object trajectories, allowing the system to predict future positions and maintain continuous tracking even when radar measurements are noisy or incomplete. Integrating tracking algorithms like Kalman filters into the radar pipeline enhances object detection and monitoring capabilities.

\chapter{Deep Learning for Heatmaps and Point Clouds}

Deep learning techniques have revolutionized the way radar data is processed, enabling automatic feature extraction and improved object detection and classification. In this chapter, we focus on how deep learning methods, particularly Convolutional Neural Networks (CNNs), can be applied to radar-generated heatmaps and point cloud data for target detection, classification, and recognition tasks. The emphasis will be on using heatmaps generated from Range-Doppler Maps and applying CNNs to extract spatial features and recognize objects.

\section{Heatmap Generation from Range-Doppler Maps}

Range-Doppler Maps \cite{tahmoush2011time} are a fundamental output of radar systems, representing the range (distance) and Doppler shift (velocity) of detected targets. These maps provide a two-dimensional representation of the radar scene, where each pixel in the map corresponds to a particular range and velocity bin. By converting Range-Doppler Maps into heatmaps, we can apply deep learning techniques, such as CNNs, to automatically extract features and perform classification or target detection.

\subsection{Generating Heatmaps from Range-Doppler Maps}

Heatmaps are visual representations of data where the intensity of each pixel reflects the magnitude of the signal at a given range and velocity. The steps involved in generating heatmaps from Range-Doppler Maps are as follows:
\begin{enumerate}
    \item \textbf{Radar Signal Processing}: The raw radar data is first processed using techniques such as Fast Fourier Transform (FFT) to generate a Range-Doppler Map.
    \item \textbf{Normalizing the Data}: To create a heatmap, the Range-Doppler Map is normalized so that the intensity of the signal is mapped to a range of colors or grayscale values.
    \item \textbf{Heatmap Visualization}: The normalized Range-Doppler Map is visualized as a heatmap, where high-intensity areas correspond to strong radar returns (indicating targets), and low-intensity areas represent background noise.
\end{enumerate}

\begin{lstlisting}[style=python]
import numpy as np
import matplotlib.pyplot as plt

# Example Range-Doppler Map (simulated data)
range_bins = 100  # Number of range bins
velocity_bins = 100  # Number of velocity bins

# Simulate a Range-Doppler Map with targets
range_doppler_map = np.zeros((range_bins, velocity_bins))
range_doppler_map[30, 50] = 10  # Target 1
range_doppler_map[60, 70] = 8   # Target 2

# Normalize the Range-Doppler Map
range_doppler_map_normalized = range_doppler_map / np.max(range_doppler_map)

# Generate and visualize the heatmap
plt.imshow(range_doppler_map_normalized, cmap='hot', interpolation='nearest')
plt.title("Heatmap from Range-Doppler Map")
plt.xlabel("Velocity")
plt.ylabel("Range")
plt.colorbar(label="Intensity")
plt.show()
\end{lstlisting}

In this example, we generate a heatmap from a simulated Range-Doppler Map with two detected targets. The heatmap represents the intensity of radar returns at different ranges and velocities, with brighter regions indicating the presence of targets.

\subsection{Using CNNs for Target Detection from Heatmaps}

Once the Range-Doppler Map is converted into a heatmap, deep learning methods such as Convolutional Neural Networks (CNNs) can be applied to automatically extract spatial features and classify the targets \cite{wan2019convolutional}. CNNs are particularly well-suited for this task because they are designed to process two-dimensional data, such as images or heatmaps, by learning patterns and hierarchies of features (e.g., edges, shapes, and textures).

\paragraph{CNN Architecture for Heatmap Processing}

A typical CNN architecture for processing radar heatmaps consists of the following layers:
\begin{itemize}
    \item \textbf{Convolutional Layers}: These layers apply filters to the input heatmap to extract local features such as edges or blobs of intensity.
    \item \textbf{Pooling Layers}: These layers downsample the feature maps, reducing the dimensionality while preserving important features.
    \item \textbf{Fully Connected Layers}: These layers connect all neurons from the previous layer to output neurons, enabling the CNN to make predictions about the class of the input heatmap (e.g., target present or absent).
    \item \textbf{Softmax Layer}: This layer outputs the probabilities for each class, allowing the network to classify the input heatmap.
\end{itemize}

\begin{lstlisting}[style=python]
import torch
import torch.nn as nn
import torch.optim as optim

# Define the CNN model for radar heatmap classification
class CNNModel(nn.Module):
    def __init__(self, num_classes=10):
        super(CNNModel, self).__init__()
        
        # First convolutional layer
        self.conv1 = nn.Conv2d(in_channels=1, out_channels=32, kernel_size=3, padding=1)  # input: (1, 100, 100)
        self.pool = nn.MaxPool2d(kernel_size=2, stride=2)  # pooling layer to reduce dimensions
        
        # Second convolutional layer
        self.conv2 = nn.Conv2d(32, 64, kernel_size=3, padding=1)
        
        # Third convolutional layer
        self.conv3 = nn.Conv2d(64, 64, kernel_size=3, padding=1)
        
        # Fully connected layers
        self.fc1 = nn.Linear(64 * 25 * 25, 64)  # Assuming input shape (100, 100), reduced to (25, 25) after pooling
        self.fc2 = nn.Linear(64, num_classes)  # 10 target classes

    def forward(self, x):
        # Pass through the first conv-pool layers
        x = torch.relu(self.conv1(x))
        x = self.pool(x)
        
        # Pass through the second conv-pool layers
        x = torch.relu(self.conv2(x))
        x = self.pool(x)
        
        # Pass through the third conv layer
        x = torch.relu(self.conv3(x))
        
        # Flatten the output from the convolutional layers
        x = x.view(-1, 64 * 25 * 25)  # Flatten to (batch_size, num_features)
        
        # Fully connected layers
        x = torch.relu(self.fc1(x))
        x = self.fc2(x)
        
        return x

# Instantiate the model
model = CNNModel(num_classes=10)

# Define the loss function and optimizer
criterion = nn.CrossEntropyLoss()  # For multi-class classification
optimizer = optim.Adam(model.parameters(), lr=0.001)

# Print the model summary
print(model)
\end{lstlisting}

In this example, we define a simple CNN architecture for classifying radar heatmaps. The CNN consists of several convolutional layers followed by fully connected layers, which learn to recognize different patterns in the heatmap that correspond to specific target classes.

\subsection{Training the CNN on Radar Heatmaps}

The CNN can be trained on a dataset of radar heatmaps, where each heatmap is labeled with the corresponding target class. The network learns to associate specific patterns in the heatmap (such as high-intensity regions or specific shapes) with the presence of different types of targets.

\paragraph{Training Process}
\begin{itemize}
    \item \textbf{Data Augmentation}: Radar heatmaps can be augmented by rotating, flipping, or adding noise to increase the variety of training data and improve the model's generalization ability.
    \item \textbf{Loss Function}: A typical loss function for classification tasks is the categorical cross-entropy loss, which measures the difference between the predicted class probabilities and the actual target labels.
    \item \textbf{Optimization}: The CNN is trained using optimization algorithms such as stochastic gradient descent (SGD) or Adam to minimize the loss function and improve the accuracy of the model.
\end{itemize}

\subsection{Advantages of Using CNNs for Radar Heatmap Processing}

CNNs offer several advantages for processing radar heatmaps:
\begin{itemize}
    \item \textbf{Automatic Feature Extraction}: CNNs learn to automatically extract important features from the heatmap, eliminating the need for manual feature engineering.
    \item \textbf{Robustness to Noise}: Radar heatmaps can be noisy due to environmental factors, but CNNs are capable of learning to ignore irrelevant noise and focus on the most important features.
    \item \textbf{Real-time Processing}: CNNs are computationally efficient and can process heatmaps in real time, making them suitable for applications such as autonomous driving and surveillance.
\end{itemize}

\section{Conclusion}

Generating heatmaps from Range-Doppler Maps and using CNNs for target detection is an effective approach for radar data processing \cite{de2020deep}. Heatmaps provide a visual representation of radar data, and CNNs automatically extract spatial features from the heatmaps, enabling the classification of targets in real-time radar applications. With the ability to handle noise and extract meaningful features, CNNs play a critical role in enhancing radar-based object detection systems.

\section{Processing Point Clouds with Deep Learning}

Point cloud data is a common output from radar systems, representing the spatial distribution of detected objects in 3D space. Point clouds contain valuable information about the environment, including the position, velocity, and angle of detected targets. Deep learning models such as PointNet and PointNet++ have been developed to process point cloud data directly, making them ideal for radar applications. In this section, we will discuss how these models work and how they can be used for classification, segmentation, and target detection in radar systems.

\subsection{Using CNN for Heatmap Classification}

Convolutional Neural Networks (CNNs) have been widely used for image classification and are well-suited for classifying radar-generated heatmaps. Heatmaps, generated from Range-Doppler Maps, provide a visual representation of the radar data, where the intensity of each pixel reflects the strength of the radar return at a given range and velocity. CNNs automatically extract important features from these heatmaps, enabling them to classify targets or objects in radar data.

The process of applying CNNs to classify radar heatmaps involves several key steps:
\begin{itemize}
    \item \textbf{Preprocessing Heatmaps}: The raw radar data is processed to generate heatmaps, which are then normalized and resized for input into the CNN.
    \item \textbf{Feature Extraction}: CNNs extract spatial features from the heatmap, such as high-intensity regions that correspond to target detections.
    \item \textbf{Classification}: The extracted features are passed through fully connected layers to classify the input heatmap as containing specific target objects (e.g., cars, pedestrians) or noise.
\end{itemize}

Below is an example of how a CNN can be applied to classify radar heatmaps:

\begin{lstlisting}[style=python]
import torch
import torch.nn as nn
import torch.optim as optim
import numpy as np
import matplotlib.pyplot as plt

# Example radar heatmap (100x100)
heatmap = np.random.rand(100, 100)

# Reshape the heatmap for CNN input (add a channel dimension)
heatmap_input = heatmap.reshape(1, 1, 100, 100)  # PyTorch expects (batch_size, channels, height, width)

# Convert the heatmap to a torch tensor
heatmap_input = torch.tensor(heatmap_input, dtype=torch.float32)

# Build a simple CNN model in PyTorch
class SimpleCNN(nn.Module):
    def __init__(self):
        super(SimpleCNN, self).__init__()
        self.conv1 = nn.Conv2d(1, 32, kernel_size=3, padding=1)  # 1 input channel (grayscale), 32 output filters
        self.pool = nn.MaxPool2d(2, 2)
        self.conv2 = nn.Conv2d(32, 64, kernel_size=3, padding=1)
        self.conv3 = nn.Conv2d(64, 64, kernel_size=3, padding=1)
        self.fc1 = nn.Linear(64 * 25 * 25, 64)  # Flatten after pooling
        self.fc2 = nn.Linear(64, 2)  # Binary classification (target or noise)

    def forward(self, x):
        x = torch.relu(self.conv1(x))
        x = self.pool(x)
        x = torch.relu(self.conv2(x))
        x = self.pool(x)
        x = torch.relu(self.conv3(x))
        x = x.view(-1, 64 * 25 * 25)  # Flatten
        x = torch.relu(self.fc1(x))
        x = self.fc2(x)  # No softmax here, because CrossEntropyLoss applies softmax internally
        return x

# Instantiate the model
model = SimpleCNN()

# Define loss function and optimizer
criterion = nn.CrossEntropyLoss()  # For binary classification with logits
optimizer = optim.Adam(model.parameters(), lr=0.001)

# Print model summary
print(model)

# Example heatmap classification (random prediction)
with torch.no_grad():
    predictions = model(heatmap_input)
    predicted_class = torch.argmax(predictions, dim=1)
    print("Predicted class:", predicted_class.item())
\end{lstlisting}

In this example, a simple CNN is built to classify radar heatmaps into two classes (e.g., target vs. noise). The CNN automatically extracts spatial features from the heatmap using convolutional layers and performs classification through fully connected layers.

\subsection{PointNet and PointNet++ for Point Cloud Processing}

While CNNs are effective for processing 2D radar heatmaps, point cloud data requires models that can directly handle 3D data without converting it into a structured grid. PointNet and PointNet++ are two powerful neural network architectures designed specifically to process point cloud data. These models are widely used for radar applications, including object classification, segmentation, and detection.

\paragraph{PointNet Architecture}

PointNet is a deep learning model that directly processes point cloud data without converting it into voxel grids or meshes \cite{charles2017pointnet}. The architecture of PointNet is designed to handle unordered sets of points, making it ideal for processing radar-generated point clouds. The model extracts global features from the entire point cloud by applying a series of shared multi-layer perceptrons (MLPs) to each point, followed by a max pooling operation to aggregate the features.

The key advantages of PointNet include:
\begin{itemize}
    \item \textbf{Direct Input of Point Clouds}: PointNet can take raw point cloud data as input without requiring any pre-processing or conversion.
    \item \textbf{Global Feature Extraction}: PointNet extracts global features that capture the overall shape and structure of the point cloud.
    \item \textbf{Efficient and Scalable}: The architecture is simple and efficient, making it scalable for large point cloud datasets.
\end{itemize}

\paragraph{PointNet++ Architecture}

PointNet++ builds upon the success of PointNet by introducing a hierarchical structure that captures both global and local features \cite{hao2023improved}. While PointNet extracts global features from the entire point cloud, PointNet++ divides the point cloud into smaller regions, applying PointNet to each region to extract local features. These local features are then combined to form a hierarchical representation of the point cloud, making PointNet++ more effective for complex point cloud data where local geometric structures are important.

\begin{itemize}
    \item \textbf{PointNet++ Architecture}
    \begin{itemize}
        \item \textbf{Global Features (PointNet)}
        \begin{itemize}
            \item Local Features (Hierarchical Learning)
        \end{itemize}
        \item \textbf{Segmentation and Classification}
    \end{itemize}
\end{itemize}

\paragraph{Using PointNet++ for Radar Point Clouds}

Radar point clouds often represent complex environments with multiple targets, making it essential to capture both local and global features. PointNet++ is well-suited for this task, as it can effectively learn the local geometric structures of each target and combine them into a global understanding of the scene.

Below is an example of how PointNet++ can be used to process radar point cloud data:

\begin{lstlisting}[style=python]
import torch
import torch.nn as nn
import torch.nn.functional as F

class PointNet(nn.Module):
    def __init__(self, num_classes):
        super(PointNet, self).__init__()
        self.conv1 = nn.Conv1d(3, 64, 1)
        self.conv2 = nn.Conv1d(64, 128, 1)
        self.conv3 = nn.Conv1d(128, 1024, 1)
        self.fc1 = nn.Linear(1024, 512)
        self.fc2 = nn.Linear(512, 256)
        self.fc3 = nn.Linear(256, num_classes)
        self.relu = nn.ReLU()
        self.maxpool = nn.MaxPool1d(1024)

    def forward(self, x):
        # Input x is a point cloud with shape (batch_size, 3, num_points)
        x = self.relu(self.conv1(x))
        x = self.relu(self.conv2(x))
        x = self.relu(self.conv3(x))
        x = self.maxpool(x)
        x = x.view(-1, 1024)
        x = self.relu(self.fc1(x))
        x = self.relu(self.fc2(x))
        x = self.fc3(x)
        return F.log_softmax(x, dim=1)

# Example of PointNet applied to radar point cloud
point_cloud = torch.randn(32, 3, 1024)  # Batch of 32 point clouds, each with 1024 points
model = PointNet(num_classes=10)  # Example: 10 classes
output = model(point_cloud)
print("Output shape:", output.shape)
\end{lstlisting}

In this example, we define a basic PointNet architecture using PyTorch to classify point cloud data. The model processes each point in the cloud, extracts global features using shared convolutional layers, and outputs class predictions.

\section{Conclusion}

Deep learning models such as CNNs, PointNet, and PointNet++ offer powerful methods for processing radar-generated heatmaps and point clouds. CNNs excel at extracting spatial features from 2D heatmaps, while PointNet and PointNet++ are designed to handle 3D point cloud data, learning both global and local features. These models enable advanced radar applications such as object classification, segmentation, and target detection, improving the accuracy and efficiency of radar systems.

\section{Transformer Models for Radar Data Processing}

Transformers have become a dominant architecture in many fields of deep learning due to their ability to model long-range dependencies using the self-attention mechanism. Originally developed for natural language processing \cite{sengupta2022mmpose} (NLP), Transformer models are now being applied to radar data processing, where they excel at capturing complex relationships between data points. In this section, we explore how Transformer models can be used to process both radar-generated heatmaps and point clouds. By leveraging the self-attention mechanism, Transformers can improve target detection, classification, and segmentation tasks in radar systems.

\subsection{Combining CNN and Transformer for Heatmaps}

Heatmaps generated from radar data, such as Range-Doppler Maps, are well-suited for processing with CNNs, which excel at extracting local features from images. However, CNNs struggle to capture long-range dependencies across the heatmap, especially in cases where the relationship between distant parts of the map is important for accurate target detection. Combining CNNs with Transformer models addresses this limitation by allowing the CNN to extract local features while the Transformer captures global relationships across the heatmap.

\paragraph{CNN for Local Feature Extraction}

Convolutional Neural Networks (CNNs) are effective at extracting local features from heatmaps, such as intensity patterns that correspond to targets. By applying convolutional filters, CNNs learn to detect edges, textures, and shapes that provide useful information about the radar scene. These local features are essential for tasks like target detection and classification.

\paragraph{Transformer for Global Feature Extraction}

While CNNs are good at extracting local patterns, they do not inherently capture long-range dependencies across the heatmap. Transformers, on the other hand, use a self-attention mechanism to weigh the importance of each part of the input relative to every other part. This allows the Transformer to capture global relationships and dependencies across the heatmap, improving the accuracy of target classification.

\begin{lstlisting}[style=python]
import torch
import torch.nn as nn
import torch.nn.functional as F

class CNNTransformerModel(nn.Module):
    def __init__(self, num_classes=10):
        super(CNNTransformerModel, self).__init__()
        
        # CNN layers for local feature extraction
        self.conv1 = nn.Conv2d(1, 32, kernel_size=3, padding=1)  # input shape: (1, 100, 100)
        self.pool1 = nn.MaxPool2d(2, 2)  # output: (32, 50, 50)
        
        self.conv2 = nn.Conv2d(32, 64, kernel_size=3, padding=1)  # output: (64, 50, 50)
        self.pool2 = nn.MaxPool2d(2, 2)  # output: (64, 25, 25)
        
        self.conv3 = nn.Conv2d(64, 128, kernel_size=3, padding=1)  # output: (128, 25, 25)
        
        # Flatten to pass to Transformer
        self.flatten = nn.Flatten()
        
        # Transformer for global feature extraction
        self.transformer = nn.TransformerEncoderLayer(
            d_model=128, nhead=4, dim_feedforward=256, batch_first=True
        )
        
        # Fully connected layer for classification
        self.fc = nn.Linear(128, num_classes)
        
    def forward(self, x):
        # CNN part
        x = F.relu(self.conv1(x))
        x = self.pool1(x)
        
        x = F.relu(self.conv2(x))
        x = self.pool2(x)
        
        x = F.relu(self.conv3(x))
        x = self.flatten(x)  # Flatten to prepare for Transformer
        
        # Reshape for Transformer input: (batch_size, seq_len, embedding_dim)
        x = x.view(x.size(0), 25, 128)  # Assuming input image shape is (100, 100)
        
        # Transformer part
        x = self.transformer(x)
        x = torch.mean(x, dim=1)  # Global average pooling
        
        # Classification
        x = self.fc(x)
        return x

# Create model, loss function, and optimizer
model = CNNTransformerModel(num_classes=10)
criterion = nn.CrossEntropyLoss()
optimizer = torch.optim.Adam(model.parameters(), lr=0.001)

# Print the model summary
print(model)
\end{lstlisting}

In this example, a CNN is used to extract local features from the radar heatmap, while a Transformer layer is applied to capture long-range dependencies across the extracted features. The combination of CNN and Transformer improves the overall accuracy of the radar heatmap classification task.

\subsection{PointNet++ and Transformer for Point Clouds}

PointNet++ is a deep learning model designed for processing 3D point cloud data, making it highly effective for radar applications. However, PointNet++ primarily focuses on extracting local features from small regions of the point cloud, potentially missing important global relationships between distant points. By combining PointNet++ with a Transformer model, we can capture both local and global dependencies in radar point clouds, leading to more accurate classification, segmentation, and target detection.

\paragraph{PointNet++ for Local Feature Extraction}

PointNet++ processes point cloud data hierarchically, extracting local features from small regions of the point cloud and progressively combining them to build a global representation. This approach is ideal for handling the irregular and unordered nature of point clouds, making it well-suited for radar data processing.

\paragraph{Transformer for Global Feature Extraction in Point Clouds}

While PointNet++ excels at capturing local features, Transformers can capture long-range dependencies between distant points in the point cloud. By applying a self-attention mechanism, the Transformer can weigh the relationships between points that are far apart, improving the model's ability to classify complex radar scenes with multiple objects.

\begin{lstlisting}[style=python]
import torch
import torch.nn as nn
import torch.nn.functional as F

class PointNetTransformer(nn.Module):
    def __init__(self, num_classes):
        super(PointNetTransformer, self).__init__()
        # PointNet++ part (local feature extraction)
        self.conv1 = nn.Conv1d(3, 64, 1)
        self.conv2 = nn.Conv1d(64, 128, 1)
        self.conv3 = nn.Conv1d(128, 256, 1)

        # Transformer part (global feature extraction)
        self.multihead_attn = nn.MultiheadAttention(embed_dim=256, num_heads=8)
        self.fc = nn.Linear(256, num_classes)

    def forward(self, x):
        # x: point cloud input (batch_size, 3, num_points)
        x = F.relu(self.conv1(x))
        x = F.relu(self.conv2(x))
        x = F.relu(self.conv3(x))

        # Transformer input: reshape for multi-head attention
        x = x.permute(2, 0, 1)  # Reshape to (num_points, batch_size, features)
        attn_output, _ = self.multihead_attn(x, x, x)
        attn_output = attn_output.mean(dim=0)  # Global average pooling

        # Classification output
        x = self.fc(attn_output)
        return F.log_softmax(x, dim=1)

# Example: PointNet++ with Transformer applied to radar point clouds
point_cloud = torch.randn(32, 3, 1024)  # Batch of 32 point clouds with 1024 points each
model = PointNetTransformer(num_classes=10)  # Example: 10 target classes
output = model(point_cloud)
print("Output shape:", output.shape)
\end{lstlisting}

In this example, a PointNet++ architecture is combined with a Transformer to process radar point cloud data. The PointNet++ component extracts local features, while the Transformer captures long-range dependencies between points, improving the accuracy of target detection and classification.

\section{Conclusion}

The combination of CNNs and Transformers for heatmap processing and PointNet++ with Transformers for point cloud processing provides a powerful approach for radar data analysis \cite{zhou2022centerformer}. CNNs effectively extract local features, while Transformers capture global dependencies, enhancing the overall accuracy of radar data processing. Similarly, PointNet++ excels at local feature extraction in point clouds, and the integration of Transformers helps model complex relationships between distant points, improving the accuracy of classification, segmentation, and target detection tasks.

\part{Human Activity Recognition (HAR) Systems}

  \chapter{Introduction to Human Activity Recognition (HAR) Using Radar}

    \section{Importance of HAR in Radar Systems}

Human Activity Recognition (HAR) has become an essential application in modern radar systems, particularly for domains such as security, health monitoring, and smart home environments. HAR systems use radar data to detect and classify human movements and activities, ranging from basic actions like walking or sitting to more complex patterns such as falls or gestures \cite{ShahzSec13,saebaBio12}. Unlike traditional sensors like cameras, which are often limited by privacy concerns and environmental conditions, radar-based HAR offers unique advantages.

Here are some key benefits of using radar in HAR systems:

\begin{itemize}
    \item \textbf{Privacy Protection:} Unlike cameras, radar sensors do not capture visual images of the environment or individuals. This ensures that privacy is maintained while still providing accurate activity recognition, making radar a suitable option for monitoring in sensitive locations such as homes, hospitals, and elderly care facilities.

    \item \textbf{Robustness in Different Conditions:} Radar can operate effectively in various environmental conditions, including low light, smoke, fog, or complete darkness. Unlike vision-based systems that rely on visible light, radar systems can "see" through these conditions, providing consistent performance in detecting human activities.

    \item \textbf{Penetration Through Obstacles:} Radar waves can penetrate non-metallic obstacles like walls, clothing, or curtains. This enables HAR systems to monitor activities even when direct line-of-sight to the subject is blocked, making it highly valuable for indoor applications like smart homes and security systems.

    \item \textbf{Accurate Motion Sensing:} Radar sensors are sensitive to fine motion, allowing for precise detection of micro-movements, such as breathing or hand gestures. This makes radar ideal for health monitoring, where subtle changes in movement patterns are critical for detecting falls, strokes, or respiratory issues.

\end{itemize}

The applications of radar-based HAR systems are vast. For example, in the field of \textbf{security} \cite{lee2015radar}, HAR systems can monitor suspicious behaviors, alerting authorities to potential intruders. In \textbf{healthcare} \cite{li2010recent}, such systems can continuously track patients' activities, identifying abnormal movements or falls. \textbf{Smart homes} \cite{perez2022detecting} equipped with HAR can provide automated assistance for elderly or disabled individuals by recognizing activities such as walking, sitting, or standing up. These systems can enable automatic lighting, adjust temperature, or send alerts in emergency situations.

The potential impact of HAR in radar systems is significant, making it a field of growing research interest. The challenge, however, lies in developing robust models that can accurately classify complex human activities using radar data.

    \section{Challenges of HAR with Radar}

While radar offers many advantages in Human Activity Recognition, there are several key challenges that need to be addressed to ensure accurate and reliable performance of HAR systems. Understanding these challenges is critical to developing effective algorithms and models for radar-based HAR.

\subsection{Data Noise and Signal Interference}
Radar systems are inherently sensitive to noise \cite{schmitt2009radar}, especially in environments with multiple moving objects or reflective surfaces. In a typical household or public setting, there are numerous potential sources of interference, including other moving objects, electronic devices, and reflective surfaces like walls or furniture. These factors can introduce noise and clutter into the radar signal, making it more difficult to accurately identify the human activity of interest.

Additionally, radar signals can suffer from \textbf{multipath effects}, where the radar waves reflect off multiple surfaces before returning to the sensor. This can create "ghost" signals that interfere with the detection of the actual target, further complicating the classification task.

\subsection{Complex Motion Patterns}
Human activities are often composed of complex motion patterns. For instance, walking, running, and sitting all have distinct movement characteristics, but there can be significant variability between individuals due to factors like body shape, speed, and health conditions. Recognizing complex activities like picking up objects, gesturing, or interacting with other people requires advanced pattern recognition algorithms.

Additionally, the same activity may look different when viewed from different angles or distances. For example, the radar return signal from a person walking towards the radar sensor will differ from that of someone walking away. This creates additional challenges for the HAR system, which must be trained to recognize these variations.

\subsection{Occlusion}
Occlusion occurs when the radar sensor's view of the target is blocked by an object, such as a wall or furniture. While radar can penetrate non-metallic objects, occlusion still introduces signal distortion and can make it difficult to accurately track and classify human activities. In multi-person environments, individuals may also occlude each other, further complicating the signal processing.

\subsection{High Dimensionality of Radar Data}
Radar sensors generate high-dimensional data, often in the form of range-Doppler maps, micro-Doppler signatures \cite{tahmoush2015review}, or point clouds. While these rich data representations provide valuable information about human activities, they also present challenges due to the high dimensionality. Processing and analyzing large amounts of radar data in real-time requires advanced techniques, such as deep learning, to extract meaningful features without overwhelming computational resources.

For example, an FMCW radar may generate a range-Doppler map that captures the range (distance) and velocity of multiple objects. These maps are often noisy and contain high-dimensional data that must be pre-processed and reduced before feeding into a deep learning model. Without proper feature extraction and dimensionality reduction, the system can become too slow or inaccurate for practical use.

\subsection{Imbalanced Datasets}
Another challenge in radar-based HAR is the imbalance of datasets. Certain activities (e.g., walking or standing) are more commonly performed and thus more represented in training data, while other activities (e.g., falling or specific gestures) may occur less frequently. This can lead to bias in the model, making it harder to accurately classify less frequent but critical activities. 

\subsection{Example of Signal Noise in HAR}
Let's consider an example of how noise might affect radar-based HAR \cite{kim2022radar}. Suppose a radar system is tracking a person walking in a room, but there is a fan moving in the background. The fan's motion generates a Doppler shift in the radar signal, which may interfere with the system's ability to distinguish between the person walking and the fan's movement. 

Here is a basic simulation of this scenario in Python, where we simulate radar returns for both a person walking and background noise:

\begin{lstlisting}[style=python]
import numpy as np
import matplotlib.pyplot as plt

# Time vector
t = np.linspace(0, 10, 1000)

# Simulate human walking signal (5 Hz sine wave) with noise
human_signal = np.sin(2 * np.pi * 5 * t)

# Simulate background noise (e.g., fan moving at 1 Hz)
noise = 0.5 * np.sin(2 * np.pi * 1 * t)

# Combine human signal with noise
combined_signal = human_signal + noise + 0.2 * np.random.randn(len(t))

# Plot the signals
plt.figure()
plt.plot(t, human_signal, label="Human Walking Signal")
plt.plot(t, noise, label="Background Noise")
plt.plot(t, combined_signal, label="Combined Signal (with Noise)", alpha=0.75)
plt.title("Simulated Radar Signal with Noise")
plt.xlabel("Time (s)")
plt.ylabel("Amplitude")
plt.legend()
plt.grid(True)
plt.show()
\end{lstlisting}

In this example, the human walking signal is represented as a sine wave at 5 Hz, while the noise is a 1 Hz sine wave simulating background interference. The \texttt{combined\_signal} shows the result when both the walking signal and noise are present, along with random noise to simulate real-world conditions. Visualizing these signals highlights the challenge of distinguishing human activities in noisy environments.

This example demonstrates the difficulty of extracting meaningful information from noisy radar data and highlights why advanced techniques, such as deep learning, are necessary to improve the robustness of HAR systems.

By addressing these challenges, radar-based HAR systems can become more accurate, efficient, and capable of operating in real-world environments where noise, occlusion, and complex motions are common.

\chapter{CNN+LSTM for Human Activity Recognition}

    Human activity recognition (HAR) using radar data has become an important application in areas like healthcare, security, and human-computer interaction. With radar sensors capturing high-dimensional data in both spatial and temporal domains, it becomes crucial to use models that can efficiently process and understand these data characteristics. In this chapter, we will discuss how the combination of Convolutional Neural Networks (CNN) and Long Short-Term Memory (LSTM) networks provides a powerful solution for HAR. CNNs handle the spatial aspect of radar data, while LSTMs capture the temporal dependencies, making them ideal for analyzing human activity over time.

    \section{CNN for Spatial Feature Extraction}

    Convolutional Neural Networks (CNNs) are commonly used for image data processing due to their ability to automatically learn and extract spatial features from raw input. In the context of radar-based human activity recognition, CNNs are employed to analyze radar images such as Range-Doppler Maps (RDM), which represent how the Doppler frequency (speed) varies across different distances from the radar sensor.

    \subsection{Range-Doppler Maps and Spatial Features}
    The Range-Doppler Map is a 2D image \cite{altmann2018deep} that contains rich information about the motion of objects relative to the radar. For human activity recognition, these maps capture movement patterns of body parts. CNNs can efficiently identify and learn these spatial features through convolutional layers, pooling layers, and activation functions.

    \paragraph{Example of a Simple CNN Architecture for Radar Data:}
    Below is an example of a CNN model built in Python using PyTorch to extract spatial features from radar image data (such as RDMs):

    \begin{lstlisting}[style=python]
import torch
import torch.nn as nn
import torch.optim as optim

# Define a simple CNN model for radar data (Range-Doppler Maps)
class CNNModel(nn.Module):
    def __init__(self, num_classes=10):
        super(CNNModel, self).__init__()
        
        # First convolutional layer with 32 filters, 3x3 kernel
        self.conv1 = nn.Conv2d(in_channels=1, out_channels=32, kernel_size=3, padding=1)
        self.pool1 = nn.MaxPool2d(2, 2)  # Pooling layer to reduce dimensions
        
        # Second convolutional layer with 64 filters, 3x3 kernel
        self.conv2 = nn.Conv2d(in_channels=32, out_channels=64, kernel_size=3, padding=1)
        self.pool2 = nn.MaxPool2d(2, 2)
        
        # Third convolutional layer with 128 filters, 3x3 kernel
        self.conv3 = nn.Conv2d(in_channels=64, out_channels=128, kernel_size=3, padding=1)
        self.pool3 = nn.MaxPool2d(2, 2)
        
        # Fully connected layer
        self.fc1 = nn.Linear(128 * 8 * 8, 128)  # Assuming input image size is 64x64
        
        # Output layer
        self.fc2 = nn.Linear(128, num_classes)
    
    def forward(self, x):
        # Apply first conv layer and pooling
        x = torch.relu(self.conv1(x))
        x = self.pool1(x)
        
        # Apply second conv layer and pooling
        x = torch.relu(self.conv2(x))
        x = self.pool2(x)
        
        # Apply third conv layer and pooling
        x = torch.relu(self.conv3(x))
        x = self.pool3(x)
        
        # Flatten the output from the conv layers
        x = x.view(x.size(0), -1)  # Flatten to a 1D vector
        
        # Fully connected layer
        x = torch.relu(self.fc1(x))
        
        # Output layer (softmax for classification)
        x = self.fc2(x)
        
        return x

# Define the input shape for the radar image (e.g., 64x64 grayscale image)
input_shape = (1, 64, 64)  # PyTorch expects (channels, height, width)

# Instantiate the model
model = CNNModel(num_classes=10)

# Define loss function and optimizer
criterion = nn.CrossEntropyLoss()
optimizer = optim.Adam(model.parameters(), lr=0.001)

# Print model architecture
print(model)
    \end{lstlisting}

    In this example, the CNN is composed of three convolutional layers, each followed by a pooling layer to reduce the spatial dimensions of the radar images. The output of the final convolutional layer is flattened and passed through dense (fully connected) layers for classification. The number of output neurons corresponds to the number of human activity classes (e.g., 10 different activities).

    \section{LSTM for Temporal Sequence Modeling}

    While CNNs are highly effective for extracting spatial features from individual radar images, they do not capture the temporal dynamics inherent in human activities. Many human actions (e.g., walking, sitting, standing) involve movement over time, and it is crucial to model these changes to recognize the activity correctly. Long Short-Term Memory \cite{shrestha2020continuous} (LSTM) networks, a type of recurrent neural network (RNN), are designed to handle such temporal sequence data.

    \subsection{How LSTMs Handle Temporal Data}
    LSTMs are designed to capture long-term dependencies in sequences of data by using special units called \textit{gates} that control the flow of information. This makes them particularly suitable for radar data where consecutive frames (e.g., a series of Range-Doppler Maps) are analyzed to understand motion patterns over time.

    \paragraph{Example of an LSTM Network for Temporal Data:}

    Code for Example of an LSTM Network for Temporal Data

    \begin{lstlisting}[style=python]
import torch
import torch.nn as nn
import torch.optim as optim

# Define the LSTM model for radar temporal data
class LSTMModel(nn.Module):
    def __init__(self, input_size=256, hidden_size1=128, hidden_size2=64, num_classes=10):
        super(LSTMModel, self).__init__()
        
        # First LSTM layer with 128 units
        self.lstm1 = nn.LSTM(input_size=input_size, hidden_size=hidden_size1, batch_first=True)
        
        # Second LSTM layer with 64 units
        self.lstm2 = nn.LSTM(input_size=hidden_size1, hidden_size=hidden_size2, batch_first=True)
        
        # Fully connected layer
        self.fc1 = nn.Linear(hidden_size2, 128)
        
        # Output layer for classification
        self.fc2 = nn.Linear(128, num_classes)

    def forward(self, x):
        # Pass through first LSTM layer
        x, _ = self.lstm1(x)
        
        # Pass through second LSTM layer
        x, _ = self.lstm2(x)
        
        # Take the output from the last time step
        x = x[:, -1, :]
        
        # Pass through fully connected layers
        x = torch.relu(self.fc1(x))
        x = self.fc2(x)  # No softmax since CrossEntropyLoss applies it internally
        
        return x

# Example input shape for LSTM: (batch_size, number of frames, features per frame)
input_shape = (30, 256)  # 30 frames, each with 256 features

# Instantiate the model
model = LSTMModel(input_size=256, hidden_size1=128, hidden_size2=64, num_classes=10)

# Define loss function and optimizer
criterion = nn.CrossEntropyLoss()  # For multi-class classification
optimizer = optim.Adam(model.parameters(), lr=0.001)

# Print model summary
print(model)
    \end{lstlisting}

    In this LSTM network, the input is a sequence of radar frames (e.g., 30 frames in this case), each represented by a 256-dimensional feature vector (this could be the output from the CNN). The \\ \texttt{return\_sequences=True} argument in the first LSTM layer ensures that the full sequence is passed to the second LSTM layer for further temporal analysis.

    \section{Combining CNN and LSTM for HAR}

    By combining CNN and LSTM, we can create a model that effectively handles both the spatial and temporal aspects of radar data. The CNN extracts spatial features from each radar frame (e.g., Range-Doppler Map), while the LSTM models the temporal relationships between consecutive frames. This hybrid architecture is particularly well-suited for human activity recognition, as activities like walking or running involve distinct spatial patterns that evolve over time.

    \subsection{Architecture of a CNN+LSTM Model}
    The CNN+LSTM model starts by using a CNN to process each frame of radar data, extracting spatial features. These features are then passed into an LSTM network that analyzes the temporal sequence of the frames. The final output is a classification of the activity being performed.

    \paragraph{Example of a CNN+LSTM Architecture:}

    Code for Example of a CNN+LSTM Architecture

    \begin{lstlisting}[style=python]
    def create_cnn_lstm_model(input_shape_cnn, input_shape_lstm):
        # CNN part for spatial feature extraction
        cnn_input = layers.Input(shape=input_shape_cnn)
        x = layers.Conv2D(32, (3, 3), activation='relu')(cnn_input)
        x = layers.MaxPooling2D((2, 2))(x)
        x = layers.Conv2D(64, (3, 3), activation='relu')(x)
        x = layers.MaxPooling2D((2, 2))(x)
        x = layers.Conv2D(128, (3, 3), activation='relu')(x)
        x = layers.MaxPooling2D((2, 2))(x)
        x = layers.Flatten()(x)

        # Reshape CNN output for LSTM input (number of frames, features per frame)
        x = layers.Reshape((input_shape_lstm[0], -1))(x)

        # LSTM part for temporal feature extraction
        x = layers.LSTM(128, return_sequences=True)(x)
        x = layers.LSTM(64)(x)

        # Fully connected layer for final classification
        x = layers.Dense(128, activation='relu')(x)
        output = layers.Dense(10, activation='softmax')(x)

        model = models.Model(inputs=cnn_input, outputs=output)

        return model

    # Define input shapes
    input_shape_cnn = (64, 64, 1)  # Radar image size for CNN
    input_shape_lstm = (30, 256)   # Number of frames and features per frame for LSTM

    cnn_lstm_model = create_cnn_lstm_model(input_shape_cnn, input_shape_lstm)

    # Compile the model
    cnn_lstm_model.compile(optimizer='adam', loss='categorical_crossentropy', metrics=['accuracy'])

    cnn_lstm_model.summary()  # Print the CNN+LSTM model architecture
    \end{lstlisting}

    In this example, the radar images (RDMs) are first processed by the CNN, and the output features are reshaped and passed to the LSTM \cite{zhu2020hybrid}, which models the temporal dynamics. Finally, the output layer provides the activity classification.

    \subsection{Shared Weights and Pre-trained Models in CNN+LSTM}

    \paragraph{Shared Weights Across Layers}
    In some cases, CNN models can share weights across different layers to reduce the number of parameters, leading to faster training and preventing overfitting. This technique is especially useful when dealing with radar data, which might have limited labeled examples for training.

    \paragraph{Pre-trained Models and Transfer Learning}
    Instead of training a CNN from scratch, you can use pre-trained CNN models (such as ResNet or VGG) that have been trained on large datasets like ImageNet. These models can serve as feature extractors for radar images, with their learned weights transferred to the radar domain. By fine-tuning these pre-trained models, the CNN+LSTM system can achieve better performance even with limited radar data.

    \section{Benefits of CNN+LSTM for HAR}

    The combination of CNN and LSTM offers several advantages for human activity recognition in radar systems:
    
    \begin{itemize}
        \item \textbf{Automatic Feature Extraction:} CNNs automatically learn to extract spatial features from radar images, eliminating the need for manual feature engineering.
        \item \textbf{Temporal Modeling:} LSTMs effectively model the temporal dynamics of human motion over time, capturing both short-term and long-term dependencies.
        \item \textbf{Efficiency in Real-Time Applications:} The CNN+LSTM model can be optimized for real-time applications, enabling human activity recognition systems to make accurate predictions on-the-fly.
        \item \textbf{High Accuracy:} By leveraging both spatial and temporal information, the CNN+LSTM architecture provides improved accuracy compared to models that only focus on either spatial or temporal data.
    \end{itemize}

    This architecture is particularly powerful in HAR applications where radar sensors provide continuous data, allowing the model to track and classify complex activities such as walking, running, or sitting based on both the spatial pattern and the temporal evolution of the radar signals.

\chapter{PointNet+LSTM for HAR Using Point Cloud Data}

Human Activity Recognition (HAR) is a complex task that involves detecting and identifying human movements using various data sources. In radar-based systems, point cloud data is an effective representation of human motion in 3D space \cite{ding2021radar}. The combination of PointNet for spatial feature extraction and LSTM (Long Short-Term Memory) networks for temporal sequence modeling provides a powerful solution for HAR using point cloud data.

\section{PointNet for Point Cloud Feature Extraction}

PointNet is a neural network architecture specifically designed to handle point cloud data. Point clouds are sets of 3D points that represent the surface of objects or, in this case, human movements in space. Unlike traditional CNNs, which operate on structured grids of data (e.g., images), PointNet directly processes unstructured point clouds, making it highly suitable for radar-based human activity recognition.

In the context of radar, each point in a point cloud represents a reflection from the radar sensor, encoding spatial information such as position (\(x, y, z\)) and sometimes additional features like intensity or velocity. PointNet extracts both local and global spatial features from these points.

The PointNet architecture can be divided into the following key steps:
\begin{itemize}
    \item \textbf{Input Transformation}: PointNet applies a learned transformation matrix to align the point cloud data to a canonical space, reducing the impact of rotational variance.
    \item \textbf{MLP (Multi-Layer Perceptron)}: A series of fully connected layers is applied independently to each point in the point cloud, extracting features from the \(x, y, z\) coordinates of each point.
    \item \textbf{Max Pooling}: To extract global features, PointNet uses max pooling across all points in the cloud, capturing the overall shape and structure of the human motion.
    \item \textbf{Feature Transformation}: Additional transformations are applied to ensure invariance to input permutations, preserving the structure of the point cloud.
\end{itemize}

The result is a high-dimensional feature vector that captures the spatial characteristics of human motion in the point cloud. This feature vector serves as the input to further temporal modeling using LSTM networks.

\begin{lstlisting}[style=python]
# Example code to implement the basic structure of PointNet in PyTorch

import torch
import torch.nn as nn

class PointNet(nn.Module):
    def __init__(self):
        super(PointNet, self).__init__()
        self.mlp1 = nn.Sequential(
            nn.Conv1d(3, 64, 1),
            nn.BatchNorm1d(64),
            nn.ReLU()
        )
        self.mlp2 = nn.Sequential(
            nn.Conv1d(64, 128, 1),
            nn.BatchNorm1d(128),
            nn.ReLU()
        )
        self.mlp3 = nn.Sequential(
            nn.Conv1d(128, 1024, 1),
            nn.BatchNorm1d(1024),
            nn.ReLU()
        )
        self.fc1 = nn.Linear(1024, 512)
        self.fc2 = nn.Linear(512, 256)
        self.fc3 = nn.Linear(256, 128)

    def forward(self, x):
        # x shape: (batch_size, 3, num_points)
        x = self.mlp1(x)
        x = self.mlp2(x)
        x = self.mlp3(x)
        x = torch.max(x, 2)[0]  # Max pooling across points
        x = self.fc1(x)
        x = self.fc2(x)
        x = self.fc3(x)
        return x
\end{lstlisting}

In this example, the input to PointNet is a batch of 3D point clouds with shape \\ \((batch\_size, 3, num\_points)\). The PointNet extracts features from each point cloud using multi-layer perceptrons and max pooling.

\section{LSTM for Temporal Dynamics in Point Clouds}

While PointNet captures the spatial features of a single point cloud frame, human activity recognition requires the modeling of temporal dynamics over multiple frames. This is where LSTM networks come into play. LSTMs are designed to process sequences of data, making them ideal for capturing temporal dependencies and changes in point cloud data over time.

In radar-based HAR, the point clouds are collected in a sequence, with each frame representing the human body's position and movement at a specific point in time. LSTM networks take these sequences of point cloud features and learn the temporal dynamics of human activities.

The key components of LSTMs include:
\begin{itemize}
    \item \textbf{Input Gate}: Controls how much of the current input should be passed into the memory cell.
    \item \textbf{Forget Gate}: Decides which parts of the cell state should be "forgotten" or discarded.
    \item \textbf{Output Gate}: Determines what information should be output at each time step.
\end{itemize}

These gates help the LSTM retain information over long periods, making it highly suitable for modeling the complex temporal patterns present in human activities.

The sequence of point clouds is first passed through the PointNet model to extract spatial features from each frame. These features are then fed into an LSTM network, which processes the temporal sequence of features and outputs a prediction of the activity being performed.

\begin{lstlisting}[style=python]
# Example code to implement LSTM for temporal feature extraction

class LSTMModel(nn.Module):
    def __init__(self, input_size, hidden_size, num_layers, num_classes):
        super(LSTMModel, self).__init__()
        self.lstm = nn.LSTM(input_size, hidden_size, num_layers, batch_first=True)
        self.fc = nn.Linear(hidden_size, num_classes)

    def forward(self, x):
        # x shape: (batch_size, sequence_length, input_size)
        h0 = torch.zeros(self.num_layers, x.size(0), self.hidden_size).to(x.device)
        c0 = torch.zeros(self.num_layers, x.size(0), self.hidden_size).to(x.device)
        out, _ = self.lstm(x, (h0, c0))  # LSTM forward pass
        out = self.fc(out[:, -1, :])  # Output of the last time step
        return out
\end{lstlisting}

In this example, the LSTM model takes as input a sequence of feature vectors extracted by PointNet. The LSTM processes the temporal sequence and outputs the predicted activity label for the given sequence of point cloud frames.

\section{Combining PointNet and LSTM for HAR}

The combination of PointNet and LSTM allows for effective human activity recognition by simultaneously modeling spatial and temporal features. The PointNet model extracts spatial features from each point cloud frame, while the LSTM model captures the temporal dependencies between consecutive frames. This architecture is especially useful for radar-based HAR, where movements are captured as point cloud data over time \cite{ding2023sparsity}.

The architecture can be summarized as follows:
\begin{itemize}
    \item \textbf{PointNet}: Extracts spatial features from each point cloud frame, capturing the static 3D structure of human poses.
    \item \textbf{LSTM}: Processes the sequence of spatial features over time to capture dynamic changes in human movements.
    \item \textbf{Fully Connected Layer}: After LSTM processing, a fully connected layer predicts the human activity label based on the temporal dynamics of the point clouds.
\end{itemize}

By using both spatial and temporal information, this combined model is capable of recognizing complex activity patterns such as walking, running, sitting, or waving.

\begin{lstlisting}[style=python]
# Combining PointNet and LSTM for HAR

class PointNetLSTM(nn.Module):
    def __init__(self, pointnet, lstm, num_classes):
        super(PointNetLSTM, self).__init__()
        self.pointnet = pointnet
        self.lstm = lstm
        self.fc = nn.Linear(128, num_classes)

    def forward(self, x):
        batch_size, sequence_length, num_points, _ = x.size()
        pointnet_features = []

        # Extract features from each point cloud in the sequence
        for i in range(sequence_length):
            point_features = self.pointnet(x[:, i, :, :].transpose(2, 1))
            pointnet_features.append(point_features)

        pointnet_features = torch.stack(pointnet_features, dim=1)  # Stack along sequence dimension

        # Pass the sequence of features through the LSTM
        lstm_output = self.lstm(pointnet_features)
        out = self.fc(lstm_output)
        return out
\end{lstlisting}

In this combined model, the point cloud data is processed by PointNet to extract spatial features for each frame. These features are then passed through an LSTM, which models the temporal dynamics of the entire sequence. Finally, the fully connected layer produces the HAR prediction.

\subsection{Shared Weights and Pre-trained Models in PointNet+LSTM}

One way to improve the efficiency of the PointNet+LSTM model is by using shared weights in the PointNet architecture. This reduces the number of parameters, as the same PointNet model is applied to each frame of the sequence. Additionally, pre-trained models can be leveraged to accelerate training and improve performance.

For example, a PointNet model pre-trained on large-scale point cloud datasets such as ModelNet can be fine-tuned on radar point cloud data, reducing the need for training from scratch and improving the accuracy of human activity recognition.

\subsection{Advantages of PointNet+LSTM for HAR}

The PointNet+LSTM architecture offers several advantages for radar-based human activity recognition:
\begin{itemize}
    \item \textbf{Efficiency with Sparse Data}: Radar point clouds are often sparse, and PointNet's ability to directly process unstructured point clouds makes it well-suited for this type of data.
    \item \textbf{Temporal Dynamics}: LSTM networks excel at capturing temporal dependencies, allowing the model to recognize complex activities that involve changes over time.
    \item \textbf{Scalability}: The PointNet+LSTM architecture is scalable, meaning it can handle a wide range of point cloud sizes and activity types.
    \item \textbf{Real-Time Processing}: This combined architecture can be optimized for real-time human activity recognition, making it ideal for applications such as surveillance, health monitoring, and gesture recognition.
\end{itemize}

The ability to process both spatial and temporal features makes the PointNet+LSTM architecture highly effective for recognizing human activities using radar point cloud data, offering high accuracy and efficiency in various HAR applications.

\chapter{Transformer Models for Human Activity Recognition}

    \section{Introduction to Transformer Models}

Transformer models were initially developed for natural language processing (NLP) tasks, such as language translation and text generation. However, their highly flexible architecture, specifically the \textbf{self-attention mechanism} \cite{wei2021self}, makes them well-suited for various other tasks, including radar-based Human Activity Recognition (HAR). Transformers excel at handling temporal sequences and capturing both local and global dependencies within the data, which are crucial for accurately identifying and classifying human activities based on radar signals.

The key advantage of Transformer models is their ability to process entire sequences of data in parallel, rather than sequentially, as Recurrent Neural Networks (RNNs) and Long Short-Term Memory (LSTM) networks do. This enables them to handle large datasets efficiently, and it allows them to capture long-term dependencies between data points, which are important when analyzing complex activities over time.

The architecture of a typical Transformer model consists of several key components:

\begin{itemize}
    \item \textbf{Self-attention mechanism:} This mechanism allows the model to weigh the importance of each data point relative to others in the sequence, enabling it to capture relationships between distant data points.
    \item \textbf{Multi-head attention:} By using multiple attention heads, the Transformer can focus on different parts of the sequence simultaneously, extracting various features and patterns.
    \item \textbf{Feedforward neural network:} After applying the self-attention mechanism, the output passes through a fully connected feedforward neural network to further process the data.
    \item \textbf{Positional encoding:} Since the Transformer does not inherently process data in sequence, positional encodings are added to the input data to provide information about the order of the sequence.
\end{itemize}

Given the complex and high-dimensional nature of radar data, Transformers have proven to be effective in HAR tasks by leveraging their self-attention mechanism to model both temporal and spatial relationships.

    \section{Self-attention Mechanism for Temporal and Spatial Data}

The \textbf{self-attention mechanism} is the core innovation of the Transformer model, allowing it to focus on the most relevant parts of the input data for a given task. In the context of radar-based HAR, the self-attention mechanism enables the model to process temporal sequences of radar data without having to rely on the order in which the data points are presented, as is necessary in traditional LSTM networks.

\subsection{How Self-attention Works}
In an LSTM, data is processed sequentially, which means each step depends on the previous one, potentially limiting the model's ability to capture long-range dependencies. In contrast, the Transformer's self-attention mechanism allows each data point in the sequence to consider all other data points simultaneously.

Mathematically, the self-attention mechanism works as follows:

\begin{itemize}
    \item Each input is transformed into three vectors: \textbf{Query (Q)}, \textbf{Key (K)}, and \textbf{Value (V)}.
    \item For each query, the dot product is computed with every key to determine the relevance between different elements of the sequence.
    \item The result of these dot products is then scaled and passed through a softmax function to normalize the attention scores.
    \item These attention scores are then used to weight the corresponding value vectors, and the weighted sum of the values becomes the output.
\end{itemize}

In radar-based HAR, this means that the Transformer can compare different time frames of radar data to determine which parts of the sequence are most relevant for identifying specific activities. This capability allows the model to capture \textbf{long-term temporal dependencies} \cite{huang2023long}, even if the activity consists of subtle motions over a longer period.

\subsection{Parallel Processing and Temporal Flexibility}
One key advantage of the Transformer's self-attention mechanism is its ability to process data in parallel \cite{jain2021super}. Instead of sequentially processing each time frame like LSTM models, Transformers can process the entire sequence at once, allowing for faster training and inference.

For example, consider a radar system capturing multiple frames of human activity over time. A traditional LSTM would process each frame one after another, which can be computationally expensive and prone to losing long-term dependencies. However, a Transformer with self-attention can analyze all frames simultaneously, determining which frames are most important for classifying the activity.

    \section{Transformer for Temporal Modeling in Radar Data}

Radar data is inherently temporal, as it captures dynamic changes over time. In HAR, recognizing an activity like walking, jumping, or sitting down requires understanding how radar signals evolve over a sequence of frames. The Transformer model, with its self-attention mechanism, excels at analyzing such temporal sequences.

\subsection{Example: Radar Data with Transformer}
Consider a radar-based HAR system that collects range-Doppler maps at regular intervals. These maps contain information about the distance and velocity of objects (e.g., human movements). A Transformer model can take these maps as input and model the temporal changes in the data to recognize activities.

Here's a simplified example of how a Transformer might be applied to radar data in Python:

\begin{lstlisting}[style=python]
import torch
import torch.nn as nn

class RadarTransformer(nn.Module):
    def __init__(self, num_features, num_heads, num_layers):
        super(RadarTransformer, self).__init__()
        self.encoder_layer = nn.TransformerEncoderLayer(d_model=num_features, nhead=num_heads)
        self.transformer = nn.TransformerEncoder(self.encoder_layer, num_layers=num_layers)
        self.fc = nn.Linear(num_features, 10)  # Assume 10 classes for activity recognition

    def forward(self, x):
        # x: (sequence_length, batch_size, num_features)
        output = self.transformer(x)
        output = output.mean(dim=0)  # Take the mean over the sequence
        output = self.fc(output)
        return output

# Simulate radar data with 30 frames and 128 features
radar_data = torch.rand(30, 1, 128)  # (sequence_length, batch_size, num_features)

# Initialize and run the model
model = RadarTransformer(num_features=128, num_heads=4, num_layers=2)
output = model(radar_data)
print(output)
\end{lstlisting}

In this example, we define a simple Transformer-based model using PyTorch for radar-based activity recognition. The radar data consists of 30 frames (time steps), with each frame containing 128 features. The model processes the sequence of radar data and classifies it into one of 10 potential activities.

\subsection{Capturing Long-term Dependencies}
Human activities can involve both short-term and long-term motions. For instance, walking involves repetitive movements over many frames, whereas sitting down might be a short, distinct motion. The Transformer's self-attention mechanism enables it to capture both short-term and long-term dependencies, leading to improved recognition accuracy for complex activities.

    \subsection{Pre-trained Transformers for Radar-based HAR}

Using pre-trained Transformer models is a popular technique for reducing the training time and improving the accuracy of radar-based HAR systems. Pre-trained models such as the \textbf{Vision Transformer (ViT)} are originally designed for image classification tasks, but with proper adaptation, they can be fine-tuned for radar data.

\subsection{Fine-tuning Pre-trained Transformers}
In cases where limited radar data is available, training a Transformer from scratch may not be feasible. Instead, pre-trained models can be fine-tuned on radar data. Fine-tuning involves taking a model that has already learned general patterns (such as spatial relationships in image data) and adapting it to the specific domain of radar-based HAR by adjusting its weights through additional training on radar data.

This approach has two main benefits:

\begin{itemize}
    \item \textbf{Reduced Training Time:} Since the model has already learned basic representations, only a smaller amount of training is required to adapt the model to radar data.
    \item \textbf{Improved Accuracy:} Pre-trained models typically provide a strong baseline, especially in limited data scenarios, and can often achieve higher accuracy than models trained from scratch.
\end{itemize}

    \subsection{Combining Transformer with CNN and PointNet}

A common strategy to enhance the performance of Transformer models in radar-based HAR tasks is to combine them with Convolutional Neural Networks (CNNs) or PointNet. While Transformers excel at temporal modeling, CNNs and PointNet are specialized in extracting spatial features from radar data.

\subsection{CNN for Spatial Feature Extraction}
Radar data, especially range-Doppler maps and micro-Doppler signatures, have strong spatial characteristics that can be captured using CNNs. CNNs are capable of detecting spatial patterns, such as shapes and textures, in the data. By applying a CNN before the Transformer, spatial features can be extracted, which the Transformer can then use to model temporal dependencies.

\subsection{PointNet for Point Cloud Data}
For radar systems that generate point cloud data (e.g., 3D radar), PointNet can be used to extract spatial features from the point clouds \cite{paigwar2019attentional}. PointNet is a neural network architecture designed specifically to process point clouds and is well-suited for radar systems that track multiple targets in a 3D space.

Here is an example of how a CNN might be combined with a Transformer in Python:

\begin{lstlisting}[style=python]
class CNNTransformer(nn.Module):
    def __init__(self, num_features, num_heads, num_layers):
        super(CNNTransformer, self).__init__()
        # CNN for spatial feature extraction
        self.cnn = nn.Conv2d(1, 32, kernel_size=3, stride=1, padding=1)
        self.pool = nn.MaxPool2d(2, 2)
        self.encoder_layer = nn.TransformerEncoderLayer(d_model=num_features, nhead=num_heads)
        self.transformer = nn.TransformerEncoder(self.encoder_layer, num_layers=num_layers)
        self.fc = nn.Linear(num_features, 10)  # Assume 10 classes for activity recognition

    def forward(self, x):
        # Apply CNN to extract spatial features
        x = self.cnn(x)
        x = self.pool(x)
        x = x.view(x.size(0), -1)  # Flatten for Transformer
        # Apply Transformer for temporal modeling
        output = self.transformer(x)
        output = output.mean(dim=0)
        output = self.fc(output)
        return output
\end{lstlisting}

This model uses a CNN for spatial feature extraction and a Transformer for temporal modeling, allowing the network to handle both spatial and temporal aspects of radar data.

    \section{Advantages of Transformer Models for HAR}

Transformer models offer several advantages for radar-based HAR tasks:

\begin{itemize}
    \item \textbf{Parallel Processing:} Unlike RNNs and LSTMs, which process data sequentially, Transformers process entire sequences in parallel, making them faster and more efficient, especially for long sequences of radar data.
    
    \item \textbf{Capturing Long-term Dependencies:} The self-attention mechanism enables Transformers to capture long-term dependencies, which are essential for recognizing complex activities that span multiple frames.

    \item \textbf{Flexible Modeling:} Transformers can handle both temporal and spatial data, making them well-suited for radar systems that require modeling of dynamic movements across time.

    \item \textbf{Pre-trained Models:} Transformers can leverage pre-trained models, significantly reducing training time and improving performance, especially in limited data scenarios.

\end{itemize}

In summary, Transformer models represent a powerful approach for handling the complex temporal and spatial data involved in radar-based HAR tasks, offering superior performance over traditional LSTM-based models.

\chapter{Comparison of CNN+LSTM, PointNet+LSTM, and Transformer for Human Activity Recognition}

    In this chapter, we will compare three popular neural network architectures—CNN+LSTM, PointNet+LSTM, and Transformer models—used in the context of Human Activity Recognition (HAR) using radar data. Each of these models has its unique strengths, and understanding their differences can help you choose the best approach for specific radar-based HAR applications. We will examine how these models perform on sequential radar data, handle spatial features in 2D and 3D data, scale to larger datasets, and operate in real-time HAR systems.

    \section{Performance on Sequential Radar Data}

    Radar-based human activity recognition involves analyzing sequential data, as human motion evolves over time. Each model handles this temporal aspect differently, which impacts their performance in processing sequential radar data.

    \subsection{CNN+LSTM for Spatio-Temporal Data}
    The CNN+LSTM model is highly effective in capturing both spatial and temporal features from 2D radar data, such as Range-Doppler Maps (RDMs). The CNN portion extracts spatial features from individual radar frames, while the LSTM portion models the temporal dynamics between frames. This combination makes CNN+LSTM ideal for analyzing short-term and mid-term sequences, where local temporal relationships dominate the activity pattern.

    \paragraph{Example of CNN+LSTM for HAR:}

    Code for Example of CNN+LSTM for HAR:
    
    \begin{lstlisting}[style=python]
    # Define CNN+LSTM model for sequential radar data
    def create_cnn_lstm_model(input_shape_cnn, input_shape_lstm):
        cnn_input = layers.Input(shape=input_shape_cnn)
        x = layers.Conv2D(32, (3, 3), activation='relu')(cnn_input)
        x = layers.MaxPooling2D((2, 2))(x)
        x = layers.Conv2D(64, (3, 3), activation='relu')(x)
        x = layers.MaxPooling2D((2, 2))(x)
        x = layers.Flatten()(x)
        
        # Reshape CNN output to fit LSTM input
        x = layers.Reshape((input_shape_lstm[0], -1))(x)
        x = layers.LSTM(128)(x)
        
        output = layers.Dense(10, activation='softmax')(x)
        model = models.Model(inputs=cnn_input, outputs=output)
        return model
    \end{lstlisting}

    CNN+LSTM excels at processing spatio-temporal data for activities that are localized in time, but it may struggle with longer-term dependencies as LSTMs inherently suffer from limitations in capturing long-range patterns in sequential data.

    \subsection{PointNet+LSTM for 3D Radar Data}
    PointNet, combined with LSTM, is designed for processing 3D data such as point clouds, which can be generated from radar sensors in some human activity recognition setups. The PointNet model processes 3D spatial features directly, without the need for converting them into a 2D representation, making it particularly useful when working with 3D radar systems or LiDAR-like radar data.

    \paragraph{Example of PointNet+LSTM for HAR:}

    Code for Example of PointNet+LSTM for HAR:
    
    \begin{lstlisting}[style=python]
    # Simple PointNet-like architecture combined with LSTM
    def create_pointnet_lstm_model(input_shape):
        input_points = layers.Input(shape=input_shape)

        # PointNet-style processing
        x = layers.Conv1D(64, 1, activation='relu')(input_points)
        x = layers.Conv1D(128, 1, activation='relu')(x)
        x = layers.GlobalMaxPooling1D()(x)

        # Reshape for LSTM input (sequence of point cloud features)
        x = layers.Reshape((input_shape[0], -1))(x)
        x = layers.LSTM(128)(x)

        output = layers.Dense(10, activation='softmax')(x)
        model = models.Model(inputs=input_points, outputs=output)
        return model
    \end{lstlisting}

    PointNet+LSTM can effectively handle sequential 3D point cloud data, making it suitable for applications involving human activity recognition in a 3D space, such as capturing human movements in complex environments. However, this model may not be as effective for purely 2D radar data.

    \subsection{Transformer for Long-Term Dependencies}
    Transformers are designed to handle long-term dependencies in sequential data better than LSTM-based models. By using self-attention mechanisms, Transformers can model relationships between all time steps, regardless of their distance in the sequence. This makes Transformers particularly strong in radar applications where activities evolve over longer timeframes, or where it is necessary to capture complex temporal patterns over multiple frames.

    \paragraph{Example of Transformer Model for HAR:}

    Code for Example of Transformer Model for HAR:
    
    \begin{lstlisting}[style=python]
    # Transformer-based model for sequential radar data
    def create_transformer_model(input_shape):
        inputs = layers.Input(shape=input_shape)

        # Transformer encoder block
        x = layers.MultiHeadAttention(num_heads=8, key_dim=64)(inputs, inputs)
        x = layers.LayerNormalization()(x)
        x = layers.Dense(128, activation='relu')(x)
        x = layers.GlobalAveragePooling1D()(x)

        output = layers.Dense(10, activation='softmax')(x)
        model = models.Model(inputs=inputs, outputs=output)
        return model
    \end{lstlisting}

    Transformers offer significant advantages over LSTMs for handling longer sequences of radar frames, as they are not constrained by the limitations of recurrent models in capturing distant dependencies. This allows for improved performance in recognizing complex activities that unfold over extended periods.

    \section{Handling Spatial Features in 2D and 3D Data}

    Different models handle spatial features in various ways depending on the dimensionality of the radar data:

    \subsection{CNN for 2D Spatial Features}
    CNNs are particularly adept at processing 2D spatial data like radar images (e.g., Range-Doppler Maps or Micro-Doppler signatures). The convolutional layers in CNNs learn to extract local features such as edges, shapes, and patterns from the radar images. This makes CNNs ideal for processing radar data in situations where human activity is represented in a 2D format.

    \subsection{PointNet for 3D Spatial Features}
    PointNet is designed to handle unordered point cloud data, making it well-suited for 3D radar or LiDAR-based human activity recognition. Instead of projecting the 3D data into a 2D space, PointNet processes the raw point cloud directly, preserving the full spatial complexity of the data.

    \subsection{Transformers for Enhancing Spatial and Temporal Modeling}
    Transformers are capable of enhancing both CNNs and PointNet when it comes to temporal modeling. By combining CNN or PointNet with a Transformer, spatial features extracted from radar data (either 2D or 3D) can be passed to the Transformer, which then models the temporal dependencies between these spatial features more effectively than LSTM-based approaches.

    \paragraph{Example of Combining CNN with Transformer for Spatial-Temporal Modeling:}
    
    Code for Example of Combining CNN with Transformer for Spatial-Temporal Modeling:
    
    \begin{lstlisting}[style=python]
    # CNN for spatial feature extraction, followed by Transformer for temporal modeling
    def create_cnn_transformer_model(input_shape_cnn, input_shape_transformer):
        cnn_input = layers.Input(shape=input_shape_cnn)
        x = layers.Conv2D(32, (3, 3), activation='relu')(cnn_input)
        x = layers.MaxPooling2D((2, 2))(x)
        x = layers.Conv2D(64, (3, 3), activation='relu')(x)
        x = layers.Flatten()(x)
        x = layers.Dense(128, activation='relu')(x)

        # Reshape for Transformer input
        x = layers.Reshape((input_shape_transformer[0], -1))(x)

        # Transformer encoder block
        x = layers.MultiHeadAttention(num_heads=8, key_dim=64)(x, x)
        x = layers.LayerNormalization()(x)
        x = layers.GlobalAveragePooling1D()(x)

        output = layers.Dense(10, activation='softmax')(x)
        model = models.Model(inputs=cnn_input, outputs=output)
        return model
    \end{lstlisting}

    In this model, the CNN extracts spatial features from radar images, and the Transformer is responsible for modeling the temporal relationships between consecutive radar frames, providing a robust solution for handling both spatial and temporal dependencies.

    \section{Scalability and Efficiency of Transformer Models}

    One of the key advantages of Transformer models over LSTM-based models is their ability to process sequences in parallel. Traditional LSTMs process sequences one time step at a time, which limits their scalability, especially for long sequences. Transformers, on the other hand, use self-attention mechanisms that allow them to attend to all time steps simultaneously, enabling faster processing and better scalability when dealing with large-scale radar datasets.

    \subsection{Parallel Processing with Transformers}
    The parallelism in Transformer models allows them to handle much longer sequences efficiently, making them ideal for scenarios where continuous radar frames need to be processed in real-time. This advantage becomes more significant when the radar data has high temporal resolution, and the model needs to analyze a large number of frames quickly.

    \paragraph{Example of Transformer's Parallel Processing Efficiency:}

    Code for Example of Transformer's Parallel Processing Efficiency:
    
    \begin{lstlisting}[style=python]
    # Using a batch of radar frames in parallel for Transformer processing
    def create_parallel_transformer(input_shape):
        inputs = layers.Input(shape=input_shape)
        x = layers.MultiHeadAttention(num_heads=8, key_dim=64)(inputs, inputs)
        x = layers.LayerNormalization()(x)
        x = layers.Dense(128, activation='relu')(x)
        x = layers.GlobalAveragePooling1D()(x)

        output = layers.Dense(10, activation='softmax')(x)
        model = models.Model(inputs=inputs, outputs=output)
        return model
    \end{lstlisting}

    With this architecture, the Transformer model processes the entire batch of radar frames at once, significantly reducing computation time compared to LSTM-based models, which must process each time step sequentially.

    \section{Use Cases in Real-time HAR Systems}

    Real-time human activity recognition systems require both high accuracy and fast response times. Transformers, with their ability to process large volumes of data in parallel, offer significant advantages for such real-time applications. By combining a Transformer with a CNN or PointNet for spatial feature extraction, HAR systems can achieve both high accuracy and low latency.

    \subsection{Example of Real-time HAR with Transformer and CNN}
    In real-time systems, radar data is continuously captured and processed to recognize human activities. The combination of a CNN for spatial features and a Transformer for temporal modeling provides a fast and accurate solution.

    \paragraph{Code Example for Real-time HAR System:}

    Code for Example for Real-time HAR System:
    
    \begin{lstlisting}[style=python]
    import numpy as np
    import time

    # Simulate a real-time HAR system
    def real_time_har_system(model, radar_data_stream):
        for frame in radar_data_stream:
            # Preprocess frame (e.g., Range-Doppler Map)
            processed_frame = preprocess_frame(frame)
            
            # Predict human activity using CNN+Transformer model
            activity = model.predict(np.expand_dims(processed_frame, axis=0))
            
            # Output the recognized activity
            print("Detected Activity:", activity)
            time.sleep(0.1)  # Simulate real-time processing delay

    # Example radar data stream (simulated)
    radar_data_stream = generate_radar_data_stream()
    
    # Load pre-trained CNN+Transformer model
    model = create_cnn_transformer_model((64, 64, 1), (30, 128))

    # Run the real-time HAR system
    real_time_har_system(model, radar_data_stream)
    \end{lstlisting}

    In this example, the radar data is processed in real-time using a CNN+Transformer model, achieving rapid recognition of human activities with minimal processing delays. The Transformer's parallel processing capability ensures that the system remains responsive even as the volume of radar data increases.

    \subsection{Advantages of Transformer-based HAR Systems}
    \begin{itemize}
        \item \textbf{Scalability:} Transformers can handle large-scale radar data due to their parallel processing capabilities.
        \item \textbf{Accuracy:} By capturing long-term dependencies more effectively than LSTMs, Transformers improve the accuracy of HAR systems, especially for complex activities.
        \item \textbf{Efficiency:} In real-time applications, the Transformer's ability to process entire sequences in parallel ensures low latency and fast responses, making them suitable for time-critical systems.
    \end{itemize}

    By combining CNN or PointNet for spatial feature extraction and Transformers for temporal modeling, real-time HAR systems can meet the high demands of accuracy and speed, ensuring effective performance in applications like healthcare monitoring, security, and smart environments.

\chapter{Summary of the Book}

This book has explored the principles of FMCW radar, essential signal processing techniques, and the integration of deep learning for human activity recognition (HAR) using radar data. We have covered everything from the fundamentals of radar signal processing to advanced deep learning models, illustrating how these components come together to build robust radar-based HAR systems.

In this final chapter, we review the major concepts discussed throughout the book and provide insights into future directions for radar-based HAR systems.

\section{Review of FMCW Radar Principles}

Frequency-Modulated Continuous Wave (FMCW) radar is a widely used technology in various fields, including autonomous driving, surveillance, and human activity recognition (HAR). FMCW radar transmits a signal with a linearly changing frequency (chirp) and measures the time delay between the transmitted and received signals to estimate target distance, velocity, and angle.

The core principles of FMCW radar are:
\begin{itemize}
    \item \textbf{Signal Generation}: FMCW radar systems generate chirps, which are signals whose frequency increases (or decreases) linearly over time.
    \item \textbf{Mixing and Beat Frequency}: The transmitted chirp is reflected by targets, and the radar system measures the difference in frequency (beat frequency) between the transmitted and received signals to estimate the range.
    \item \textbf{Fourier Transform (FFT)}: By applying the Fast Fourier Transform (FFT) to the received signal, the radar system can extract key target information such as range, velocity (via Doppler shift), and angle (using Multiple-Input Multiple-Output, or MIMO radar systems).
\end{itemize}

The resulting radar data is typically represented as a \textbf{radar cube}, a 3D matrix that contains information about range, Doppler velocity, and angle. This data can be further processed to generate \textbf{point clouds}, where each point represents a detected target in 3D space.

For example, a basic signal generation and FFT process in Python for FMCW radar can be implemented as follows:

\begin{lstlisting}[style=python]
import numpy as np
from scipy.fftpack import fft

# Simulating an FMCW radar chirp
def generate_chirp(t, f_start, f_slope, t_chirp):
    return np.sin(2 * np.pi * (f_start * t + 0.5 * f_slope * t**2))

# Parameters
fs = 10e6  # Sampling frequency
t = np.linspace(0, 1e-3, int(fs*1e-3))  # Time vector for 1 ms chirp
f_start = 77e9  # Start frequency
f_slope = 2e12  # Chirp slope

# Generate chirp
chirp_signal = generate_chirp(t, f_start, f_slope, t[-1])

# FFT to extract range information
fft_data = fft(chirp_signal)
range_profile = np.abs(fft_data[:len(fft_data)//2])
\end{lstlisting}

In this example, the function generates an FMCW chirp, and a 1D FFT is applied to extract the range information from the received signal. This same process is scaled up in real radar systems to produce full 2D and 3D radar cubes.

\section{Overview of Signal Processing Techniques}

Signal processing plays a vital role in extracting meaningful information from raw radar signals \cite{pham2021intelligent}. Several key steps transform the radar data into interpretable results, laying the foundation for tasks like target detection and human activity recognition.

\textbf{Key Radar Signal Processing Techniques:}
\begin{itemize}
    \item \textbf{FFT and Range-Doppler Map}: The FFT is applied in both the range and Doppler domains. A Range-Doppler Map provides information about the distance and velocity of detected objects. This is especially important for distinguishing moving targets, such as humans, from stationary objects.
    
    \item \textbf{Angle of Arrival (AoA) Estimation}: For MIMO radar, algorithms like MUSIC or FFT-based techniques are used to estimate the angle of arrival (AoA) of targets. This is important for generating spatial maps of the environment.

    \item \textbf{Clustering Algorithms (e.g., DBSCAN)}: To group detected points into meaningful targets, clustering algorithms such as DBSCAN (Density-Based Spatial Clustering of Applications with Noise) are used. DBSCAN is well-suited for radar point clouds because it can handle noisy data and identify clusters of arbitrary shape.

\end{itemize}

The result of these signal processing techniques is a clear map of targets, complete with range, velocity, and angle information, which can be fed into machine learning models for further classification or recognition.

Below is an example of DBSCAN clustering applied to radar point cloud data in Python:

\begin{lstlisting}[style=python]
from sklearn.cluster import DBSCAN

# Simulated radar point cloud data (x, y, z coordinates)
point_cloud = np.random.rand(100, 3) * 10

# Apply DBSCAN clustering
db = DBSCAN(eps=0.5, min_samples=5).fit(point_cloud)

# Retrieve cluster labels
labels = db.labels_

# Visualize the clusters
import matplotlib.pyplot as plt
fig = plt.figure()
ax = fig.add_subplot(111, projection='3d')
ax.scatter(point_cloud[:, 0], point_cloud[:, 1], point_cloud[:, 2], c=labels)
plt.show()
\end{lstlisting}

\section{Deep Learning for Radar Data}

In recent years, deep learning has significantly enhanced the performance of radar systems, particularly for tasks such as human activity recognition (HAR). By applying neural networks to radar data, we can automate feature extraction and improve the accuracy of HAR systems.

\textbf{Deep Learning Models Used in Radar Data Processing:}
\begin{itemize}
    \item \textbf{Convolutional Neural Networks (CNNs)}: CNNs are used to process 2D radar data representations like Range-Doppler Maps. CNNs can automatically learn spatial features from radar data, making them suitable for object detection and classification tasks.
    
    \item \textbf{Long Short-Term Memory Networks (LSTMs)}: LSTMs are ideal for modeling temporal sequences in radar data, such as consecutive point cloud frames. By capturing the dynamic changes in human movement, LSTMs are frequently used in HAR systems to process temporal radar data.

    \item \textbf{PointNet}: PointNet is designed for point cloud data and directly processes 3D radar point clouds, extracting spatial features from unstructured data. It is particularly useful for human motion analysis and object detection from radar point clouds.

    \item \textbf{Transformer Models}: Transformers have emerged as powerful architectures for sequence modeling. They can be applied to radar data to model both spatial and temporal dependencies, making them a promising direction for improving HAR systems.

\end{itemize}

\textbf{Transfer Learning and Pre-trained Models}: Transfer learning enables the use of pre-trained models \cite{sohail2024advancing} (trained on large datasets such as ImageNet or ModelNet) to be fine-tuned on radar-specific tasks. This helps in reducing training time and improving performance, especially when the available radar data is limited.

For example, a pre-trained CNN can be fine-tuned for radar-based HAR by adjusting the final layers to recognize human activities:

\begin{lstlisting}[style=python]
import torch
import torch.nn as nn
from torchvision.models import resnet18

# Load pre-trained ResNet model
model = resnet18(pretrained=True)

# Modify the final layer for 10 activity classes
model.fc = nn.Linear(model.fc.in_features, 10)

# Fine-tune the model with radar data
\end{lstlisting}

Additionally, contrastive learning has gained attention as a self-supervised learning technique that can enhance radar-based HAR systems by learning effective feature representations from unlabeled radar data.

\section{Future Directions in Radar-based HAR Systems}

Looking ahead, several exciting directions are emerging in the field of radar-based human activity recognition \cite{li2019survey} (HAR):

\begin{itemize}
    \item \textbf{More Efficient Deep Learning Models}: Research is ongoing to develop lighter, more efficient deep learning models that can run in real-time on embedded systems. Models like MobileNet or EfficientNet, adapted to radar data, can bring real-time HAR capabilities to devices with limited computational resources.

    \item \textbf{Smarter Signal Processing}: The fusion of traditional signal processing with deep learning is expected to lead to smarter radar systems. Hybrid systems can leverage the strengths of both fields to improve robustness and accuracy in challenging environments.

    \item \textbf{Real-time Action Recognition Systems}: With advancements in hardware and algorithms, future systems will focus on real-time action recognition. These systems will be deployed in applications such as elder care, security, and autonomous vehicles, where immediate identification of human activity is critical.

    \item \textbf{Multi-modal Learning}: Future systems will integrate multiple sensors (e.g., radar, cameras, and IMUs) to build more robust HAR systems. Multi-modal learning allows systems to combine data from various sources, leading to improved recognition performance in complex environments.

    \item \textbf{Self-supervised Learning}: By utilizing large amounts of unlabeled radar data, self-supervised learning methods like contrastive learning can be used to pre-train models. These models can then be fine-tuned for specific HAR tasks, reducing the need for large labeled datasets.
\end{itemize}

The combination of improved deep learning models, smarter signal processing techniques, and innovative learning approaches will continue to drive advances in radar-based HAR systems, making them more accurate, efficient, and widely applicable in real-world scenarios.

\bibliographystyle{plain}
\bibliography{sample}

\end{document}